\documentclass[12pt]{article}
\usepackage[affil-it]{authblk}
\usepackage{multirow}
\usepackage{titling}
\usepackage{subcaption}
\usepackage{algorithm}
\usepackage[noend]{algpseudocode}
\usepackage{chngcntr}
\usepackage{apptools}
\usepackage{comment}
\usepackage{bm}

\usepackage{epigraph}
\usepackage{graphicx}
\usepackage{color}

\usepackage{setspace}
\usepackage{mdframed}

\definecolor{clemson-orange}{RGB}{234,106,32}
\definecolor{chicago-maroon}{RGB}{128,0,0}
\definecolor{northwestern-purple}{RGB}{82,0,99}
\definecolor{cornell-red}{RGB}{179,27,27}
\definecolor{sauder-green}{RGB}{171,180,0}
\definecolor{gray}{RGB}{192,192,192}
\definecolor{lawngreen}{RGB}{0,250,154}

\usepackage[round]{natbib}  

\makeatletter
\def\BState{\State\hskip-\ALG@thistlm}
\makeatother

\usepackage{amsmath}
\usepackage{amssymb}
\usepackage{amsthm}
\allowdisplaybreaks

\newcommand{\uh}{\overline{u}}

\theoremstyle{plain}
\newtheorem{theorem}{Theorem}
\newtheorem{lemma}{Lemma}
\newtheorem{corollary}{Corollary}
\newtheorem{proposition}{Proposition}

\theoremstyle{definition}

\newtheorem{defn}{Definition}

\AtAppendix{\counterwithin{lemma}{section}}
\AtAppendix{\counterwithin{proposition}{section}}
\AtAppendix{\counterwithin{corollary}{section}}

\usepackage[letterpaper,margin=1in]{geometry}
\usepackage{etoolbox}
\usepackage{lipsum}
\makeatletter
\patchcmd{\@addmarginpar}{\ifodd\c@page}{\ifodd\c@page\@tempcnta\m@ne}{}{}
\makeatother

\newcommand{\mar}[1]{}

\usepackage[colorlinks,citecolor=northwestern-purple,urlcolor=chicago-maroon,linkcolor=chicago-maroon]{hyperref}

\usepackage[nameinlink]{cleveref}
\crefname{assumption}{Assumption}{Assumptions}
\crefname{lemma}{Lemma}{Lemmas}
\crefname{theorem}{Theorem}{Theorems}
\crefname{corollary}{Corollary}{Corollaries}
\crefname{proposition}{Proposition}{Propositions}
\crefname{claim}{Claim}{Claims}
\crefname{procedure}{Procedure}{Procedures}
\crefname{algorithm}{Algorithm}{Algorithms}
\crefname{figure}{Figure}{Figures}
\crefname{remark}{Remark}{Remarks}
\crefname{section}{Section}{Sections}
\crefname{procedure}{Procedure}{Procedures}
\crefname{example}{Example}{Examples}
\crefname{definition}{Definition}{Definitions}
\crefname{table}{Table}{Tables}
\crefname{equation}{}{}
\crefname{enumi}{}{}
\crefname{conjecture}{Conjecture}{Conjectures}
\crefname{step}{Step}{Steps}
\crefname{appendix}{Appendix}{Appendices}
\crefname{footnote}{Footnote}{Footnotes}

\begin{document}

\global\long\def\vs{\varphi}%
\global\long\def\pl{\underline{p}}%
\global\long\def\ph{\overline{p}}%
\global\long\def\ql{\underline{q}}%
\global\long\def\qh{\overline{q}}%
\global\long\def\rl{\underline{r}}%
\global\long\def\rh{\overline{r}}%
\global\long\def\ch{\overline{c}}%
\global\long\def\cl{\underline{c}}%
\global\long\def\xih{\overline{\xi}}%
\global\long\def\e{\epsilon}
\global\long\def\s{\sigma}
\global\long\def\p{\pi}
\global\long\def\d{\delta}
\global\long\def\t{\theta}

 \newcommand{\yk}[1]{\textcolor{cornell-red}{[YK: #1]}}
 \newcommand{\sa}[1]{\textcolor{clemson-orange}{[SA: #1]}}
\newcommand{\ykk}[1]{\textcolor{blue}{#1}}
\newcommand{\saa}[1]{\textcolor{red}{#1}}

\title{\bf Pandora's Box Reopened:\\ Robust Search and Choice Overload\thanks{We thank Pak Hung Au, Ben Brooks, Laura Doval, Yingni Guo, Emir Kamenica, Teddy Kim, Andreas Kleiner, Suraj Malladi, Pietro Ortoleva, Ludvig Sinander, Bruno Strulovici, Leeat Yariv, Andy Zapechelnyuk, and audiences in various seminars and conferences for helpful comments and discussions. Sarah Auster gratefully acknowledges funding from the European Research Council (ERC)---grant No 101165999, the German Research Foundation (DFG) through Germany's Excellence Strategy---EXC 2126/1---390838866 and CRC TR 224 (Project B02).  Yeon-Koo Che is supported by the Ministry of Education of the Republic of Korea and the National Research Foundation of Korea (NRF-2024S1A5A2A0303850912)}
}
\author{Sarah Auster\qquad{}Yeon-Koo Che\thanks{Auster: Department of Economics, University of Bonn (email: auster.sarah@gmail.com); Che: Department of Economics, Columbia University (email: yeonkooche@gmail.com)}}

\maketitle

\begin{abstract}
This paper revisits the classic Pandora's box problem, studying a decision-maker (DM) who seeks to minimize her maximal ex-post regret. The DM decides how many options to explore and in what order, before choosing one or taking an outside option. We characterize the regret-minimizing search rule and show that the likelihood of opting out often increases as more options become available for exploration. 
 We show that this ``choice overload'' is driven by the DM's fear of \emph{selection error}---the regret from searching the wrong options---suggesting that steering choice via recommendations or cost heterogeneity can mitigate regret and encourage search.  
\end{abstract}

\section{Introduction}

The emergence of digital markets has revolutionized the way individuals make choices, offering unprecedented access to a vast array of products, services, and information. While this abundance empowers consumers, it also presents a significant challenge: navigating a complex choice environment where the value of alternatives is not immediately known and can only be discovered through costly effort. This challenge is starkly evidenced by the phenomenon of ``choice overload,'' where an overabundance of options can paradoxically discourage active selection and diminish satisfaction (see \citet{scheibehenne2010can} and \citet{chernev2015choice} for reviews).\footnote{This counterintuitive behavior has been documented in classic field experiments, such as shoppers being less likely to purchase jam from a selection of 24 varieties than from a selection of 6, and in high-stakes decisions, where employees are less likely to enroll in retirement plans with more fund options \citep{iyengar2000choice, sethi2004much}.} 

The canonical framework for analyzing such problems of costly discovery is the sequential search model pioneered by \citet{Weitzman1979}. This model provides an elegant and simple index-based rule for an expected utility-maximizing decision-maker (DM) who possesses a clear, {\it unambiguous} prior about the {\it independent} value distribution of different alternatives. In many real-world settings, however, individuals lack clear probabilistic perceptions of the merits of choices, particularly when they are complex and unfamiliar. Instead, they face ambiguity, also known as Knightian uncertainty. We therefore depart from the standard assumption by considering an individual facing such uncertainty.

Specifically, we adapt Weitzman's classic Pandora's box problem, where a DM faces \(n\) alternatives, or ``boxes,'' with unknown rewards. For most of the analysis, we focus on the binary case in which boxes either contain a high reward \(\overline u\) or nothing. Opening box \(i\) and observing its reward costs \(c_i\), with boxes indexed in increasing order of costs, \(c_1<\cdots<c_n\). The DM may inspect boxes sequentially and, after each inspection, decides whether to continue searching or to stop. Upon stopping, she chooses the best alternative found so far, or takes an outside option with payoff zero. Unlike in the standard Weitzman model, the DM does not know the joint distribution of rewards, \(F\in\Delta(\{0,\overline u\}^n)\), but considers all such distributions possible, allowing arbitrary correlation across the rewards of different boxes. Restrictions such as i.i.d. rewards are interpreted as additional information and considered separately below.

We assume that the DM confronts this ambiguity by seeking to minimize her {\bf maximal anticipated ex-post regret}. The minmax regret criterion is a compelling objective for individuals concerned with foregone opportunities.\footnote{Another robust approach is the maxmin criterion, under which the DM maximizes her payoff against the worst-case distribution of rewards. In the present setting, however, this solution concept does not capture a meaningful tradeoff: given the degenerate worst-case distribution of zero rewards, it yields a trivial ``universal no-search'' policy.} Rather than reflecting an emotional bias, this approach can be understood as a robust decision rule for an agent who cannot or will not adopt a single prior, effectively guiding her to hedge against the scenarios where her choices are most consequential. Analytically, the problem amounts to a zero-sum game between the DM and an adversarial ``Nature.'' For any search strategy the DM chooses, Nature responds by selecting the reward distribution that maximizes the DM's expected regret. The DM's regret is measured relative to an oracle---an ideal self who maximizes the payoff after observing realized rewards in all boxes. The DM's optimal search strategy is the one that minimizes the maximal regret.

Our first result shows that Nature's worst-case scenario takes a stark and tractable form. For any search rule chosen by the DM, there is a regret-maximizing distribution that places the high reward in {\it at most} one box. The robust search problem is therefore a ``needle-in-the-haystack'' problem. Nature need not make the world rich in opportunities; it is enough to make the DM worry that either there is no prize at all or a single prize is hidden in the box she is least likely to inspect. This reduction is important both economically and technically. Economically, it identifies the source of regret as a selection problem: with many boxes, the DM is not only concerned about searching too much or too little, but also about searching in the wrong place. Technically, it replaces the original ambiguity over all binary reward vectors by a game with \(n+1\) states.

Building on this insight, we characterize the DM's robustly optimal search rule. We show that the rule partitions the choice set into \emph{contiguous blocks} of boxes with similar search costs. The DM searches these blocks sequentially in order of increasing cost and treats each block as an all-or-nothing unit: upon reaching a block, she either stops or searches the entire block, randomizing over the order of boxes within any non-degenerate block. This structure equalizes regret across all boxes under consideration. Although cheaper boxes generate greater regret when the DM stops before reaching them, they are also searched earlier in expectation and therefore entail a lower expected search cost. When the cost difference between two adjacent boxes is sufficiently large, regret equalization requires the DM to open the cheaper box before the more expensive one with certainty, placing the two boxes in separate blocks. By contrast, when costs are similar, a deterministic ordering would impose excessive expected search costs when the prize is located in the later box. In that case, the boxes are grouped into the same block, and their search order is randomized.

Assuming that cost differences become smaller as costs increase, the general partition structure takes the form of a simple tail-bunching rule. Low-cost boxes are separated and searched sequentially, with random stopping between them, while high-cost boxes are bunched into a single tail block and searched exhaustively if the tail is reached. Behaviorally, the DM hesitates to continue search after each inspected option when the menu is large, but searches all remaining options once the set has been reduced to a manageable size.

The main technical challenge in the analysis is to characterize when regret equalization within a block can be implemented by a stochastic search order, formally a lottery over permutations of that block. After deriving a target vector of expected cumulative search costs for each location in a given block, we show that this vector is feasible if and only if no subset of boxes is assigned a lower weighted cumulative search cost than it could obtain by being searched first. The necessity of these conditions is immediate. Sufficiency follows by interpreting the set of allocations satisfying these inequalities as the core of a hypothetical cooperative game. We show that this game is supermodular and apply \citet{edmonds2003submodular,ichiishi1981super} to establish that every element of the core is implementable by some randomization over search orders.

The comparative statics link the model back to the challenges consumers face when navigating rich choice sets in digital markets. We show that adding new options always reduces the DM's willingness to initiate search, provided that the search costs of the new options are at least as high as that of the cheapest existing option. The model thus predicts a higher likelihood of opting out as the choice set expands, a key marker of choice overload. Holding fixed the true data-generating process, we also show that the DM's reduced inclination to initiate search generates the familiar hump-shaped relationship between menu size and the probability that the DM ultimately chooses an option from the menu (e.g., purchases a product from some seller): this probability increases for low values of \(n\), but falls for large values of \(n\). Intuitively, each additional option creates a new possible location for the best product and therefore expands the set of ways in which the DM may make a selection error. While such hump-shaped purchase patterns are observed both in experimental and online retail settings \citep{shah2007buying,long2025choice}, they are difficult to reconcile with Bayesian decision-making, where added options can be ignored and therefore weakly increase expected search.

The problem of not knowing where to start is exacerbated when boxes become more homogeneous ex ante, that is, when their search costs are similar. While a steep cost profile provides some guidance about where to begin, a flat profile leaves the DM facing a more symmetric and therefore more ambiguous navigation problem. The homogeneous limit makes this force stark. We show that, as costs converge to a common constant \(c\in(0,\uh)\), one of two possibilities arises. If \(n\) is sufficiently small, all boxes are grouped into a single block, which the DM searches exhaustively with positive probability. If, instead, \(n\) is sufficiently large, the probability of initiating search converges to zero as the cost profile becomes homogeneous. Importantly, in this second regime, the DM's limiting worst-case belief assigns probability one to the prize being in one of the boxes. Hence, the DM withdraws not because she thinks there is nothing to find, but because she lacks guidance about how to find it. This provides a clear interpretation of the digital-market paradox: online environments lower the cost of inspecting any single item, but they also create large, weakly differentiated menus in which the main burden is deciding how to navigate them.

Finally, we show that the main mechanism is robust to several extensions. We first relax the binary-reward assumption and allow rewards to take any value in the bounded interval \([0,\overline u]\). While the analysis becomes more complex, the core overload result continues to hold: sufficiently large homogeneous menus make immediate opt-out robustly optimal. We next show that, in the homogeneous-cost case, our results are robust when we allow dynamic reoptimization, thereby relaxing commitment, and when we add the restriction that rewards are {\it independently} and {\it identically} distributed across boxes.  

From an applied perspective, our work advances the understanding of how individuals navigate choice-rich environments with non-quantifiable uncertainty. The model's predictions align with empirical regularities in sequential search that are difficult to reconcile with the standard model---most notably under-searching (e.g., \citealp{SchotterBraunstein1981,CoxOaxaca1989,Sonnemans1998,Kogut1990,Einav2005}) and menu-size effects \citep{de2024rational}. Regret in our setting can arise along two margins: an \emph{intensity margin} (``how many options to inspect'') and a \emph{selection margin} (``in which order to inspect them''). Our results show that the selection margin plays a central role: the refusal to search can be traced to the DM's uncertainty about the ``right'' search order---an effect that is particularly strong when the options are ex ante homogeneous and offer little guidance on where to begin. This diagnosis yields concrete implications for \textit{assortment design}: platforms can mitigate overload by deliberately breaking symmetry, for example by creating cost heterogeneity or using recommendations to guide the search order.

Beyond these design implications, the analysis is interesting from a methodological point of view. While the optimal search rule in settings with independently distributed rewards is well understood, results for more general environments, allowing for arbitrary correlations across values of different choice options, have remained elusive. Our approach, based on worst-case regret minimization, proves highly tractable, paving the way for the study of new economic questions in sequential search.

\paragraph{Related Literature.}  Our work is situated at the intersection of the literatures on sequential search, decision-making under ambiguity, and the behavioral phenomenon of choice overload. The analysis directly revisits the canonical Pandora's box problem of \citet{Weitzman1979}, which provides the foundational model of optimal search for an expected-utility maximizer with known, independent priors. While this framework has generated a large and active literature in economics (e.g., \citet{olszewski2015more,doval2018whether,choi2020optimal}), existing studies typically adopt a Bayesian approach, in contrast to the present paper.  Meanwhile, computer science and operations research on the subject focus on the regret guarantees of given algorithms in rich settings rather than the optimal one in a more stylized environment like ours. Examples include \citet{beyhaghi2019pandora}, \citet{chawla2020pandora}, \citet{chawla2021approximating}, or \citet{gergatsouli2024weitzman} (see \citet{beyhaghi2024recent} for a recent overview).

Our primary departure is to introduce ambiguity, connecting our paper to other recent work that relaxes Weitzman's assumptions. Closest to our work, \citet{schlag2021robust} study minmax regret in an {\it infinite-horizon} sequential search problem. Their environment is not a finite-menu Pandora problem with option-specific inspection costs: search is undirected, and time is modeled through discounting rather than heterogeneous inspection costs. In a setting of \emph{spatial search}, \citet{malladi2025searching} studies a regret-minimizing DM who learns about rewards, knowing that nearby options have similar quality. \citet{malladi2025searching} shows that relaxing the restriction on how much rewards can vary across locations---thereby increasing the complexity of the problem---can eliminate search incentives, mirroring our search-breakdown result when rewards are arbitrarily correlated.\footnote{In his setting, the worst-case scenario is a quality distribution over the exploration space with a ``high" peak concentrated on a small interval of locations, which can be viewed as a spatial analogue of our “needle-in-a-haystack” structure. See also \citet{banchio2025rediscovery}.}

While our contribution is to the theory of search under ambiguity, our model provides a new theoretical foundation for the well-documented \textit{choice overload} phenomenon \citep{iyengar2000choice,sethi2004much,dean2022better}.\footnote{Other examples include \citet{dean2008status}, \citet{haynes2009testing}, \citet{iyengar2010choice}, and \citet{inbar2011decision}.}  Other theoretical explanations for this behavior have focused on consumer inference and steering under Bayesian choice \citep{kamenica2008contextual,nocke2024consumer} or reference dependence \citep{deb2018reference}. Our paper joins the stream of literature that links choice overload to anticipated regret, for which decision theorists have provided axiomatic foundations \citep{sarver2008anticipating, buturak2017choice}. We contribute a tractable search model with a standard, symmetric regret formulation that generates choice overload as a direct consequence of ambiguity about the menu of options. By contrast, \citet{de2024rational} study a rational expected-utility (EU) maximizer who is uncertain about the parameters of the reward distribution. This leads to a model of Bayesian learning and an effective positive correlation between options, whereas our regret-minimizing agent is guided by a worst-case belief that induces a negative correlation, generating distinct behavioral dynamics.

\section{Setting}\label{sec:setting}

We consider the following Pandora's box problem. A DM can open up to $n$ boxes before choosing one of them or taking an outside option with payoff zero (perfect recall). Boxes are indexed by $i\in[n]:=\{1,\ldots,n\}$ and deliver a reward $u_i\in \mathbb R$. We write a state as $\omega=(u_1,\ldots,u_n)$. For most of the analysis, we will focus on a binary reward setting where $u_i\in\{0,\uh\}$ with $\uh>0$---an option is a match or no match. The state space is thus
\[
\Omega=\{0,\uh\}^n.
\]
The restriction to binary rewards gives us sufficient tractability to characterize the structure of robust search. As we will see in \Cref{subsec:beyond-binary}, the main behavioral forces extend to the case where rewards are only restricted by an upper bound and, thus, may be continuously distributed.

To learn the reward of box $i$, the DM has to pay a search cost $c_i\in(0,\uh)$. We assume boxes have different inspection costs and index them in increasing order of their search costs:
\[
i<j\quad\Longleftrightarrow\quad c_i<c_j.
\]
The DM decides sequentially whether to open one remaining box or stop. If no high-reward box has been discovered, stopping yields payoff zero. If a high-reward box has been discovered, it is optimal to stop, since no remaining box can deliver a payoff above $\uh$ and search costs are positive.

\paragraph{EU benchmark.}
\citet{Weitzman1979} considered the case of independent rewards with a known distribution, showing that the optimization problem for a standard EU maximizer has a cutoff solution: assigning to each box a reservation value, representing the minimum value that justifies the opening cost, the optimal strategy is to open boxes in descending order of this value until the best reward found so far exceeds the reservation values of all unopened boxes. In the case of binary rewards, if $p_i$ denotes the probability that box $i$ delivers reward $\uh$, box $i$'s reservation value is
\[
z_i=\uh-\frac{c_i}{p_i}.
\]
Thus, before a high reward has been discovered, the DM opens boxes in descending order of $\uh-c_i/p_i$ as long as it is positive for some remaining box.

While the case of independently distributed rewards has a simple and well-understood solution, little is known about this fundamental problem in the general case allowing correlation.\footnote{Weitzman's index policy is a special case of the Gittins index policy. With correlated values, observing one box changes beliefs about the remaining boxes, so---absent additional structure---the simple state-independent index characterization is generally lost (see \citet{rothschild1974two}); economics treatments of such non-independent environments include \citet{keller2003branching,ke2020informational,bao2023search,gambato2024search,bizzarri2024pandora}.\label{ft:correlated}}

\paragraph{Robust search.} 
Our key assumption is that the DM does not know how rewards are distributed. Rather than maximizing expected utility against a known prior, she evaluates a search rule by its maximal ex-post regret. The DM considers all distributions over $\Omega$ possible, so the baseline ambiguity set is $\Delta(\Omega)$. This allows for arbitrary correlation across boxes and imposes no probabilistic model on the reward-generating process.

For any search rule $a$, the benchmark for regret is an {\it oracle} who knows the realized state $\omega\in\Omega$. Let $u(a,\omega)$ denote the DM's expected payoff from search rule $a$ (potentially stochastic) in state $\omega$, net of search costs, and let
\[
a^*(\omega)\in\arg\max_{a'}u(a',\omega)
\]
denote an oracle strategy. Since $c_i<\uh$ for every $i$, the oracle payoff is given by
\[
u(a^*(\omega),\omega)=
\begin{cases}
0, & \text{if } u_i=0 \text{ for all } i\in[n],\\[0.2cm]
\uh-\min\{c_i:u_i=\uh\}, & \text{otherwise.}
\end{cases}
\]
The regret of strategy $a$ in state $\omega$ is therefore
\[
u(a^*(\omega),\omega)-u(a,\omega).
\]

\paragraph{Strategies and payoffs.}
In the case of binary rewards, it is optimal for the DM to stop the search when discovering a high-reward box. Hence, it is without loss to describe the DM's strategy conditional on not having found the high reward so far. Let $\mathcal N:=2^{[n]}$ denote the set of all subsets of boxes. A pure contingent search plan is a mapping
\[
A=
\Bigl\{
\bar{a}:\mathcal N\to [n]\cup\{0\}
\;\Big|\;
\bar{a}(N)\in N\cup\{0\},\ \forall N\subseteq[n]
\Bigr\},
\]
where $\bar{a}(N)=i\in N$ means that the DM opens box $i$ when the set of remaining boxes is $N$, and $\bar{a}(N)=0$ means that she stops. A mixed search rule is an element of $\Delta(A)$.

The DM's goal is to minimize the maximum regret over all states in $\Omega$. A \textbf{robustly optimal search rule} thus solves the following problem:\footnote{This formulation of the problem assumes that the DM can commit to a dynamic search rule, which is not without loss under the minmax regret criterion. We return to this issue in \cref{subsec:dynamic-reoptimization}.}
\begin{eqnarray}\label{eq:opt-commitment}
\min_{a\in\Delta(A)}
\max_{\omega\in \Omega}
\Big[
u(a^*(\omega),\omega)-u(a,\omega)
\Big].
\end{eqnarray}
The solution to \eqref{eq:opt-commitment} is a Nash equilibrium---a \emph{saddle point}---of a zero-sum game between the DM, who chooses a search rule to minimize regret, and an adversarial ``Nature", who chooses a reward assignment to maximize regret.\footnote{Existence follows from the minmax theorem for finite zero-sum games: $A$ and $\Omega=\{0,\uh\}^n$ are finite, and the regret payoff is bilinear in mixed strategies.}
This saddle point is typically in mixed strategies. The set of mixed strategies for Nature is the simplex $\Delta(\Omega)$, i.e., the set of all reward distributions including those featuring arbitrary correlations. Defining $\mathcal{R}:\Delta(A)\times\Delta(\Omega)\to\mathbb R$ as the regret value for fixed strategies of the players in this hypothetical game, the pair $(a^*,F^*)$ is a saddle point if
\[
\mathcal{R}(a,F^*)\geq \mathcal{R}(a^*,F^*)\geq \mathcal{R}(a^*,F),\quad\text{for all }a\in\Delta(A),\ F\in\Delta(\Omega).
\]
The first inequality implies that, against the saddle-point belief $F^*$, the DM's saddle-point strategy $a^*$ is optimal in the standard sense: when fixing Nature's choice, the expected oracle payoff is fixed, so minimizing expected regret is equivalent to maximizing expected utility.

\section{Analysis and Main Results}\label{sec:analysis}

We now turn to the solution of the robust search problem. The analysis begins with the one-box case, which isolates the basic minmax-regret tradeoff. We then move to the multiple-box case, where a new force appears: the DM must decide not only how much to search, but also where to search and in what order.

\paragraph{Single box $n=1$:} We start with the simplest version of the problem, with a single box. Let  the cost of opening the box be \(c\), and let \(\alpha_1\in[0,1]\) denote the probability with which the DM opens it. In the saddle-point representation, write \(p\) for Nature's probability that the box contains the high reward.

There are two ways in which the DM can regret her decision. If the box contains no prize, the oracle takes the outside option and obtains zero. The DM regrets opening the box, because opening only wastes the search cost. This regret is
\[
R(\alpha_1;p=0)=\alpha_1 c.
\]
If instead the box contains the prize, the oracle opens it and obtains \(\uh-c\). The DM now regrets not opening the box. This regret is
\[
R(\alpha_1;p=1)=(1-\alpha_1)(\uh-c).
\]
Thus, increasing the search probability reduces regret in the prize state but increases regret in the no-prize state. Against a fixed \(p\), expected regret is
\[
R(\alpha_1;p)
=
(1-p)\alpha_1 c
+
p(1-\alpha_1)(\uh-c).
\]
\begin{figure}[t]
    \centering
    \includegraphics[width=.70\textwidth]{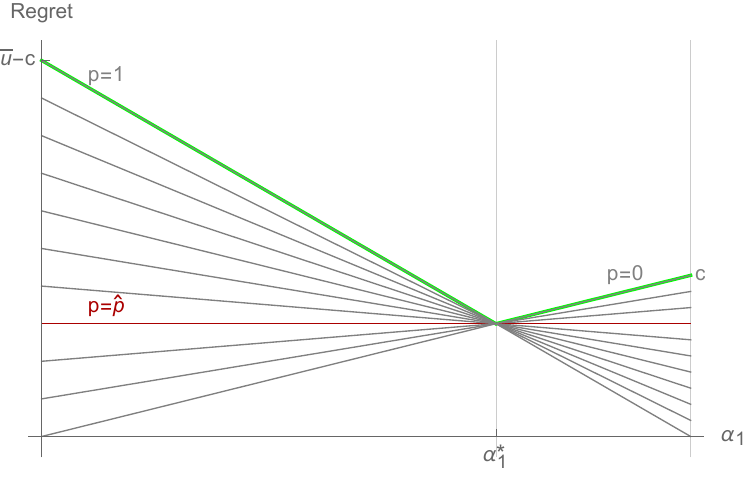}
    \caption{The one-box problem. The increasing line is regret when the box contains no prize; the decreasing line is regret when the box contains the prize. The opening probability \(\alpha_1^*\) that equalizes these two forms of regret minimizes the maximal regret, depicted by the green upper envelope.}
    \label{fig:one-box}
\end{figure}
Nature's pure states correspond to the two extreme lines in \cref{fig:one-box}: \(p=0\) generates the regret from searching in vain, while \(p=1\) generates the regret from missing the prize. The DM's minmax choice equalizes these two regrets:
\[
\alpha_1^*=\frac{\uh-c}{\uh}.
\]
Intuitively, the DM randomizes in order to hedge between two possible ex-post mistakes: searching when there is nothing to find and failing to search when the prize is present. The corresponding worst-case regret and Nature's saddle-point probability assigned to the prize state are
\[
R_1^*
=
\frac{c(\uh-c)}{\uh},\qquad \hat p=\frac{c}{\uh}.
\]
At the probability $\hat{p}$, the DM is indifferent between opening and not opening: opening risks paying \(c\) in the no-prize state, while not opening risks losing \(\uh-c\) in the prize state. In the figure, the line for \(p=\hat p\) is flat, and all expected-regret lines pass through the same minmax point.

\paragraph{Multiple boxes $n\geq2$:} With several boxes, the previous tradeoff remains, but a new margin appears. The DM must decide not only whether to search, but also in what order to search. A wrong order creates regret even when the DM eventually finds the prize. If the prize is in a cheap box but the DM first opens an expensive or unproductive box, the oracle would have gone directly to the cheap prize box and saved the extra cost. Thus the multi-box problem combines the one-box stopping tradeoff with a selection problem: the DM must guard against searching too much, searching too little, and searching the wrong boxes.

From a technical point of view, the key complication with multiple boxes is that rewards may be arbitrarily correlated, rendering the standard Weitzman index rule inapplicable and the resulting problem seemingly intractable. Crucially, imposing the robustness criterion restores tractability: our first result establishes that, despite the large underlying state space, Nature’s worst-case distribution takes a particularly simple form.

\begin{lemma}[Needle in the haystack]\label{lem:hc-one-prize}
Fix any search rule \(a\). Then there is a regret-maximizing distribution that assigns \(\uh\) to at most one box.
\end{lemma}

The intuition is simple. Consider any state with several high-reward boxes, and let \(i^*(\omega)\) be the cheapest box among those containing the high reward. Now remove all other high rewards, leaving only the prize in box \(i^*(\omega)\). The oracle's payoff is unchanged, because the oracle would open the cheapest prize box in either state. The DM's payoff, however, can only fall. Removing prizes cannot make the DM find a prize earlier, and may make her search longer or fail to find a prize at all. Hence regret weakly increases after all but the cheapest prize are removed. Nature can therefore restrict attention to states with no prize or exactly one prize.

As a consequence, we can replace the original state space \(\{0,\uh\}^n\) by the reduced state space
\[
\{0,1,\ldots,n\},
\]
where state \(0\) means that no box contains a prize and state \(i\in[n]\) means that box \(i\) is the unique prize box. Nature's mixed strategy is then denoted by
\[
\mathbf p=(p_0,p_1,\ldots,p_n)\in\Delta(\{0,1,\ldots,n\}).
\]
In reduced state \(i\geq 1\), the oracle opens box \(i\) immediately, so its payoff is \(\uh-c_i\). The oracle's expected payoff under \(\mathbf p\) is therefore
\[
\sum_{i=1}^n p_i(\uh-c_i).
\]

The one-prize reduction is economically important. It says that the worst case is not a world with many hidden opportunities. It is a needle-in-the-haystack environment.\footnote{The same prize structure is studied in \citet{kamenica2008contextual}; in his setting, however, it is assumed exogenously rather than arising endogenously as the worst-case scenario, as it does in ours.} Nature makes the DM worry that there may be no prize at all, and, if there is one, that it may be hidden in the box the DM is least inclined to search. The rest of the analysis characterizes how the DM optimally hedges against this possibility.

\subsection{The Block Structure of the Saddle Point}\label{subsec:structure-theorem}

We now establish the main structure of the saddle points of the zero-sum game. We refer to \emph{active boxes} as the boxes to which Nature, under her saddle-point strategy, assigns the prize with positive probability.

We say that a rule searches a block of boxes \emph{exhaustively} if, conditional on reaching the block without having found a prize, it continues until either a prize is found or every box in the block has been opened.

\begin{theorem}\label{thm:hc-block-structure}
In any saddle point, the active boxes form an initial segment \([m]\subseteq[n]\), which can be partitioned into maximal contiguous blocks
\[
B^{(1)},B^{(2)},\ldots,B^{(K)}
\]
such that
\[
\frac{p_i}{c_i}=\lambda^{(k)}
\quad\text{for all } i\in B^{(k)},
\qquad
\lambda^{(1)}>\lambda^{(2)}>\cdots>\lambda^{(K)}.
\]
Moreover, every pure rule in the support of the DM's mixed strategy searches blocks in the order
\[
B^{(1)},B^{(2)},\ldots,B^{(K)},
\]
and, for each block, either does not enter the block or searches it exhaustively.
\end{theorem}

The theorem says that robust search has a simple architecture. Boxes are ordered according to their costs and then grouped into blocks. If the active set is \([m]\), then there are endpoints \(0=n_0<n_1<\cdots<n_K=m\) such that
\[
B^{(k)}=\{n_{k-1}+1,\ldots,n_k\},
\qquad k=1,\ldots,K.
\]
The DM searches block by block, possibly exiting between blocks and randomizing the order within non-degenerate blocks. Under a saddle-point strategy, any active block is searched with positive probability and contains the prize with positive probability. High-cost boxes outside \([m]\), if any, are ignored by both the DM and Nature.

\paragraph{Outline of the proof.} The proof of \cref{thm:hc-block-structure} has two main steps. In the first step, we characterize the DM's best reply to an arbitrary one-prize structure.

\begin{lemma}\label{lem:hc-best-reply}
Fix \(\mathbf p=(p_0,p_1,\ldots,p_n)\in\Delta(\{0,1,\ldots,n\})\). Any optimal search rule is such that:
\begin{enumerate}
\item Boxes are opened in decreasing order of \(p_i/c_i\).
\item Whenever $B\subseteq[n]$ is a maximal class of boxes satisfying $p_i/c_i=\lambda>0$ for all $i\in B$, the rule either does not enter $B$ or searches it exhaustively.
\end{enumerate}
\end{lemma}

The resemblance of part 1 to Weitzman’s policy may seem surprising given the strong negative dependence induced by Nature’s one-prize distribution. The source---and limits---of this similarity become clearer once Weitzman’s policy is decomposed into a selection and a stopping rule. The selection rule survives: conditional on continuing search, the DM opens boxes in decreasing order of the probability-to-cost ratio $p_i/c_i$, or equivalently in decreasing order of the binary Weitzman reservation value $z_i=\uh-c_i/p_i$ when $p_i>0$. To see why, compare two boxes \(i\) and \(j\). Searching \(i\) before \(j\) is beneficial in the state where \(i\) contains the prize, because it saves the cost of opening \(j\). Searching \(j\) before \(i\) is beneficial in the state where \(j\) contains the prize, because it saves the cost of opening \(i\). Hence \(i\) should precede \(j\) exactly when
\[
p_i c_j\geq p_j c_i,
\]
or equivalently when \(p_i/c_i\geq p_j/c_j\).

The stopping rule does not survive, however. Since rewards are correlated in the reduced problem, opening a box has experimentation value: finding no reward in one box makes it more likely that the prize is hidden elsewhere. As a result, the DM may be willing to open a box even when opening it and then stopping would yield a negative expected payoff. This feature is reflected in part 2 of \cref{lem:hc-best-reply}. If a group of boxes has the same probability-to-cost ratio, the optimal rule treats the group as a unit: the DM either stops upon reaching the block or searches the block exhaustively. Searching only part of the block may yield a loss on its own, but this loss can be compensated by the information that an unsuccessful search provides about the remaining boxes in the block. The same principle applies across blocks. Exploring a block is optimal if the likelihood of finding the prize in that block, together with the continuation value generated by not finding it there, justifies the cost of exhaustively searching the block. Thus, the Weitzman index survives as a selection device, but not as a box-by-box stopping index.

The second step of the proof shows that, at a saddle point, these probability-to-cost ratios must be ordered by the cost ranking.

\begin{lemma}\label{lem:hc-monotone-ratio}
Let \((a,\mathbf p)\) be a saddle point of the reduced game. Then,
\[
\frac{p_i}{c_i}\geq \frac{p_j}{c_j}
\qquad\text{for all }1\leq i<j\leq n.
\]
\end{lemma}

Suppose instead that \(p_i/c_i<p_j/c_j\) for some cheaper box \(i\) and more expensive box \(j\). By \cref{lem:hc-best-reply}, the DM searches \(j\) before \(i\). Conditional on this search rule, the DM would do weakly worse if the prize is in \(i\) than it is in \(j\), because the rule reaches \(i\) later. Moreover, the oracle does strictly better in state \(i\), since \(c_i<c_j\). Regret is therefore strictly higher when the prize is in \(i\) than when it is in \(j\). Nature could then profitably move probability from state \(j\) to state \(i\), contradicting optimality.

The structure theorem follows by combining these properties. \cref{lem:hc-one-prize} gives the reduced one-prize representation. \cref{lem:hc-monotone-ratio} implies that if a more expensive box is active, then every cheaper box must also be active; hence the active prize states form an initial segment. The maximal sets on which \(p_i/c_i\) is constant are contiguous blocks, and \cref{lem:hc-best-reply} implies that the DM searches these blocks in order, treating each block as an all-or-nothing unit.

\paragraph{Saddle-point requirements.} \Cref{thm:hc-block-structure} gives the architecture of any saddle point but leaves open the question of \emph{how exactly} boxes are partitioned under the robustly optimal search rule. Starting from a block profile, the two support conditions of the zero-sum game must be satisfied: every state used by Nature must generate the same maximal regret, and every search path used by the DM must be optimal under Nature's assignment of rewards. Operationally, the DM's continuation probabilities between blocks equalize regret across active blocks, her lottery over within-block orders equalizes regret across prize locations within a block, and Nature's reward distribution makes the DM willing to follow the prescribed block order and indifferent exactly at histories where she randomizes between stopping and continuing. While we postpone the formal treatment of these requirements to \cref{subsec:supporting-block-profile}, let us explain the high-level idea here.

$\square$ \emph{Regret equalization.} To hedge against uncertainty over where the prize is hidden, the DM's search rule equalizes regret \emph{across} and \emph{within} all active blocks. To gain intuition for which boxes are grouped and which are separated, consider two consecutive boxes $i$ and $j$ with $c_i<c_j$. The cheaper box $i$ yields a higher oracle payoff than $j$, since $\uh-c_i>\uh-c_j$. Nature's indifference requires that $j$'s disadvantage in oracle payoff be compensated by a higher expected search cost when Nature hides the prize in $j$. Hence, in expectation, the cheaper box $i$ must be searched before the more expensive box $j$. If the cost differential between the two boxes is sufficiently large, the required expected-search-cost difference cannot be implemented by a lottery over orders within one block. Box $i$ must then be placed deterministically before $j$, so the two boxes belong to separate blocks. If, instead, costs are close to each other, the oracle payoffs are similar, and a deterministic ranking would create too large a difference in cumulative search costs across the two prize states. The DM must therefore randomize between opening box $i$ or box $j$ first, so the two boxes are placed in the same block.

$\square$  \emph{Worst-case belief.} Optimality of the block-search rule for the DM requires that the probability-to-cost ratios $(\lambda^{(k)})_{k=1}^K$ are strictly decreasing in $k$ and that, upon reaching an active block $B^{(k)}$, the DM weakly prefers entering it over opting out---with exact indifference when her strategy prescribes randomized stopping upon reaching $B^{(k)}$. Hence, conditional on reaching any active block, the chance of finding the prize must be sufficiently high to justify the average search cost. This average search cost is formalized as follows. For each $S\subseteq[n]$, let
\[
\Phi(S):=\frac{C(S)^2+\Sigma(S)}{2C(S)},
\qquad
\text{where } C(S):=\sum_{i\in S}c_i
\text{ and } \Sigma(S):=\sum_{i\in S}c_i^2 .
\]
The object $\Phi(S)$ has a simple interpretation. Suppose the prize is known to lie somewhere in $S$ and Nature's probability-to-cost ratio is constant on $S$, so that the conditional probability of box $i$ is $c_i/C(S)$. Then $\Phi(S)$ is the expected cumulative cost of locating the prize when the DM searches $S$ exhaustively. A key feature of this expectation is that it is independent of the exhaustive order used by the DM.\footnote{The formal accounting identity is stated and proved in \Cref{lem:accounting}. In particular, for any exhaustive order of $S$, if $E_i$ denotes the cumulative cost incurred when the prize is in box $i$, then $\sum_{i\in S}c_iE_i=C(S)\Phi(S)$. Dividing by $C(S)$ gives the stated cost-proportional expectation.} Optimality of entering block $B^{(k)}$ then requires that the conditional expected reward outweigh the average cost of exhaustively searching $B^{(k)}$, given by $\Phi(B^{(k)})$. If $\Phi(B^{(k)})$ is sufficiently small, $\lambda^{(k)}$ can be calibrated to keep the DM indifferent between stopping and entering block $B^{(k)}$.\footnote{Of course, the aggregate prize probability across active blocks cannot exceed one, which is why some boxes may have to be ignored by both the DM and Nature.} Since $\Phi(B^{(k)})$ increases as boxes are added to $B^{(k)}$, this limits how many boxes can be grouped together under the saddle-point block profile.

\paragraph{Global profitability.} Applied to the whole set $[n]$, the term $\Phi([n])$ has a useful interpretation: conditional on the prize being somewhere, $\Phi([n])$ is the average cost of exhaustive search when all boxes have the same probability-to-cost ratio, so there is no inference that allows the DM to target more promising options first. Now suppose
\begin{equation}\label{eq:global-profitability-main}
\uh\geq \Phi([n]).
\end{equation}
Under this ``global profitability'' condition, exhaustive search remains worthwhile even in the least informative case, provided a prize is known to exist. We next show that this condition, and its violation, have direct implications for the properties of the saddle-point block profile. To state the result, we abstract from uninteresting boundary cases and impose the following genericity condition:
\begin{equation}\label{eq:genericity}
\uh\neq \Phi(S)
\qquad\text{for every nonempty }S\subseteq[n].
\end{equation}

\begin{proposition}\label{thm:global-profitability}
Suppose genericity condition \eqref{eq:genericity} holds.
\begin{enumerate}
\item If \(\uh>\Phi([n])\), then, at every saddle point, Nature's support is
\[
\{0,1,\ldots,n\},
\]
and the DM inspects every box with positive probability.
\item If \(\uh<\Phi([n])\), then every saddle point has at least two active blocks.
\end{enumerate}
\end{proposition}

\Cref{thm:global-profitability} gives the sense in which $\Phi([n])$ disciplines the saddle-point structure. Suppose first that $\uh>\Phi([n])$. Then, even if Nature gives the DM no guidance within the menu---in the sense that the probability-to-cost ratio is constant across all boxes---the full menu is still worth searching when the prize is known to be present. To keep the DM willing to randomize over whether to begin, Nature must therefore put positive probability on the no-prize state. Once the no-prize state is in support, every prize location must also be in support; otherwise Nature could move the prize to an ignored box and strictly raise regret. The worst-case belief thus has full support, and the DM's optimal search rule inspects every box with positive probability.

The case $\uh<\Phi([n])$ is different. If all boxes were pooled into a single undifferentiated block, then exhaustive search of the whole menu would not be profitable in the cost-proportional benchmark. A single block would therefore provide too little direction: the DM would either want to stop immediately or, if induced to search on a subset of boxes, Nature would have a profitable deviation to boxes outside the block. Any saddle point must instead split the active boxes into at least two blocks. The point is not that adjacent costs must be far apart. Rather, when the menu as a whole is too costly relative to $\uh$, Nature can sustain search only by giving the DM some direction about where to begin and by allowing her to stop between groups of boxes.

\subsection{Tail-Block Search under Concave Costs}\label{subsec:concave-costs}

Before moving to the formal treatment of the saddle-point requirements laid out in the previous subsection, we consider a shape restriction on the cost profile that gives us a sharp characterization. In particular, in addition to the maintained ordering \(c_1<\cdots<c_n\), we assume that cost differences are weakly decreasing along the sequence:
\begin{equation}\label{eq:concave-costs}
c_{i+1}-c_i\leq c_i-c_{i-1}
\quad\text{for all }i=2,\ldots,n-1.
\end{equation}

We will show that under this restriction the saddle point has a tail-block structure:
\[
\{1\},\{2\},\ldots,\{\ell-1\},[\ell,n].
\]

The economic intuition for this pattern follows from a simple saddle-point tradeoff. Consider two consecutive boxes, $i$ and $i+1$. Equalizing regret requires the marginal oracle advantage of box $i$ to be offset by the higher search priority assigned to it. The magnitude of the oracle effect is determined by the cost difference $c_{i+1}-c_i$, whereas the magnitude of the priority effect is determined by $c_i$, the cost wasted when the DM searches box $i$ in state $j$. Concavity weakens the first force relative to the second as costs rise: the adjacent oracle advantage, $c_{i+1}-c_i$, weakly decreases, whereas the regret generated by searching the earlier box in vain, $c_i$, strictly increases. This suggests using a deterministic ranking among low-cost boxes, where the priority effect remains weak, and randomization in the upper tail, where a deterministic ordering would give Nature too strong an incentive to place the prize in a later box.

The theorem below establishes that this local force can be reconciled with the remaining saddle-point conditions. The singleton blocks are searched in increasing order of costs, with the DM potentially stopping between them. Once the DM reaches the tail, a fixed order would create too much regret imbalance across possible prize locations. The DM therefore randomizes over orders within the tail and, conditional on entering it, searches the tail exhaustively.

 To state the result, write
 \[
 C_{\ell:n}:=\sum_{i=\ell}^n c_i,
 \qquad
 \Sigma_{\ell:n}:=\sum_{i=\ell}^n c_i^2.
 \]

\begin{theorem}\label{thm:hc-concave-tail}
Assume concave costs and genericity condition \eqref{eq:genericity}. There exists a saddle point whose active prize states are \([n]\) and whose block partition is
\[
\{1\},\{2\},\ldots,\{\ell^*-1\},[\ell^*,n],
\]
where
\begin{equation}\label{eq:hc-ell-star-main}
\ell^*
:=
\min\Bigl\{
\ell\in[n]:
\uh>\Phi([\ell,n])
\text{ and }
\uh\geq \bar u_{\ell:n}
\Bigr\},
\end{equation}
and
\begin{equation}\label{eq:hc-tail-cutoff-main}
\bar u_{\ell:n}
:=
c_\ell+
\frac{2C_{\ell:n}^2\bigl(\Sigma_{\ell:n}/C_{\ell:n}-c_\ell\bigr)}
{C_{\ell:n}^2-\Sigma_{\ell:n}},\quad \ell<n;\qquad \bar u_{n:n}:=c_n.
\end{equation}

\end{theorem}

The theorem turns the abstract block structure into an explicit search procedure. The DM begins with the low-cost boxes, which are searched one by one in increasing order of cost. After an unsuccessful search of a singleton block, she may stop before entering the next block. This is the region of \emph{hesitant exploration}: the menu is still large, and the DM is concerned both about searching in vain and about searching the wrong part of the menu. Because cost differences provide relatively sharp guidance among low-cost boxes, these boxes provide a natural direction for search, but the DM still randomizes over how far to continue.

If the tail block is reached, the nature of the problem changes. The remaining boxes are relatively more similar, and the menu has become more manageable. Conditional on entering the tail, the DM searches it exhaustively, using a randomized order within the block. Thus the same robust rule that generates hesitation early in the search path also generates full exploration toward the end: once the DM has narrowed the problem to a sufficiently homogeneous tail, the value of continuing dominates the cost of protecting against further selection mistakes.

The threshold \(\ell^*\) is the first tail boundary for which two requirements are met. The profitability requirement, \(\uh>\Phi([\ell,n])\), says that, once the DM reaches the tail and the prize is known to lie somewhere in it under Nature's one-prize distribution, the prize is valuable enough to justify exhaustive tail search. The second requirement, \(\uh\geq \bar u_{\ell:n}\), implies that regret equalization within the tail block is implementable by some stochastic search order over the tail. Why implementability reduces to a single inequality on $\uh$ is explained in the next subsection.

\subsection{Saddle-Point Requirements: Formal Conditions}\label{subsec:supporting-block-profile}

This subsection provides the verification machinery behind the block construction. This machinery is useful for checking candidate block profiles for arbitrary cost structures and for proving \cref{thm:hc-concave-tail}, but it is not needed for the main economic message derived below. Readers interested primarily in that message can skip to \cref{sec:comparative-statics} on a first reading.

Fix a candidate partition of the active boxes into blocks $B^{(1)}, \dots, B^{(K)}$. By \cref{thm:hc-block-structure}, Nature's probabilities must satisfy $p_i/c_i = \lambda^{(k)}$ for all $i \in B^{(k)}$, with $\lambda^{(1)} > \dots > \lambda^{(K)}$ chosen to make the DM indifferent between continuing and stopping at the start of every non-terminal block. The DM's strategy can be described by two objects. The first is a sequence of reach probabilities $Q^{(1)}, \dots, Q^{(K)}$, where $Q^{(k)}$ is the probability that the DM enters block $B^{(k)}$ in a state in which no earlier block contains the prize. The second is, for each non-singleton block, a lottery over the deterministic orders of boxes within that block. Reach probabilities are calibrated to make Nature indifferent across blocks; the lottery over search orders is chosen to make Nature indifferent within a block.

\paragraph{DM's stopping indifference.} Write the probability that the prize is in $B^{(k)}$ as
\[
P^{(k)}=\sum_{i\in B^{(k)}}p_i=\lambda^{(k)}C(B^{(k)}).
\]
The DM's expected payoff from exhaustively searching block $B^{(k)}$ and then stopping, conditional on the prize not being in one of the earlier blocks, is
\[
V^{(k)}:=\frac{P^{(k)}}{p_0+\sum_{\ell=k}^KP^{(\ell)}}(\uh-\Phi(B^{(k)}))-\frac{p_0+\sum_{\ell=k+1}^KP^{(\ell)}}{p_0+\sum_{\ell=k}^KP^{(\ell)}}C(B^{(k)}).
\]
The first term captures the event where the prize lies in block $k$, which implies that the DM obtains $\uh$ and pays the average cost $\Phi(B^{(k)})$. The second term captures the opposite case---there is no prize or it is located in a later block---so the DM pays the cost of searching the entire block without finding the prize.

We distinguish two cases according to whether the DM randomizes at the start of every block---this occurs when $p_0>0$---or only at the start of every non-terminal block---this occurs when $p_0=0$. If the DM opts out with positive probability at the start of every active block, the continuation value is always zero, so $V^{(k)}=0$ for all $k=1,\ldots,K$. These indifference conditions together with $p_0+\sum_kP^{(k)}=1$ pin down the vector $(P^{(k)})_{k\in\{1,\ldots,K\}}$ and $p_0$. One then has to verify that this solution is a valid probability vector and that the resulting ratios
\[
\lambda^{(k)}=\frac{P^{(k)}}{C(B^{(k)})}.
\]
are indeed declining in $k$. A necessary condition for the solution to be a valid probability system, in particular for $P^{(k)}\geq0$ for all $k$, is that each block is sufficiently inexpensive as a group, so that $\uh>\Phi(B^{(k)})$ for all $k$; that is, if the prize were guaranteed to be in the block, the reward would justify the average cost of finding it.

The other case is where, conditional on reaching it, the DM enters the final block $B^{(K)}$ with probability one and $p_0=0$.   Conditional on having eliminated earlier blocks as possible prize locations, the DM is then certain to find the prize and obtains payoff
\[
V^{(K)}=\uh-\Phi(B^{(K)})>0.
\]
To offset the positive continuation value, $V^{(K-1)}$ must now be negative.\footnote{This does not imply that the DM would skip $B^{(K-1)}$: because $\lambda^{(K-1)}>\lambda^{(K)}$, \Cref{lem:hc-best-reply} requires the former block to be searched first.} For all preceding blocks $k<K-1$, the payoff $V^{(k)}$ continues to equal zero.

\paragraph{Nature's across-block indifference.} Just as Nature's prize probabilities keep the DM indifferent between exiting and continuing between blocks, the DM's continuation probabilities between blocks must be calibrated to make Nature indifferent between assigning the prize to any given block and, potentially, not assigning a prize at all. Given the DM's reach probabilities, the average regret generated by hiding the prize in block $B^{(k)}$ is
\begin{equation}\label{eq:hc-block-regret}
R^{(k)}
=
(1-Q^{(k)})\left(\uh-\frac{\Sigma(B^{(k)})}{C(B^{(k)})}\right)
+
\sum_{m<k} Q^{(m)}C(B^{(m)})
+
Q^{(k)}\left(\Phi(B^{(k)}) - \frac{\Sigma(B^{(k)})}{C(B^{(k)})}\right).
\end{equation}
The first term is the oracle payoff forgone when the DM stops before reaching block $B^{(k)}$---the reward $\uh$ minus the expected cost of finding the prize immediately, which, given Nature's random prize allocation, equals
\[
\sum_{i\in B^{(k)}}\frac{c_i}{C(B^{(k)})}\cdot c_i=\frac{\Sigma(B^{(k)})}{C(B^{(k)})}.
\] 
The second term is the cost wasted on earlier blocks, which do not contain the prize in this state. The third term is the within-block selection cost: conditional on entering the block, the DM finds the prize but pays the average cumulative cost $\Phi(B^{(k)})$, while the oracle pays only the average cost of directly opening the ``right'' box, $\Sigma(B^{(k)})/C(B^{(k)})$. This last term is zero for singleton blocks.

At a saddle point, $R^{(k)}$ must be the same for all active blocks. The no-prize state generates regret
\begin{equation}\label{eq:hc-no-prize-regret}
    R^{(0)} = \sum_{m=1}^{K} Q^{(m)} C(B^{(m)}).
\end{equation}
This value must be no larger than the common active-prize regret, and it is equal to that regret whenever Nature assigns positive probability to the no-prize state. These equalities and inequalities determine the reach probabilities.

\paragraph{Nature's within-block indifference.} Finally, the DM's selection rule within blocks must equalize Nature's regret across all boxes contained in a given block. To formalize this requirement, fix an active block $B^{(k)}$ and its reach probability $Q^{(k)}>0$. Let $E_i$ be the expected cumulative cost incurred inside the block under the DM's lottery over exhaustive search orders, conditional on entering the block and conditional on the prize being in box $i \in B^{(k)}$. Equalizing regret between two boxes $i, j \in B^{(k)}$ requires:
\begin{equation}\label{eq:within-block-ind}
c_j-c_i
=
Q^{(k)}(E_j-E_i).
\end{equation}
The left-hand side represents the oracle's absolute payoff advantage from finding the prize in the cheaper box $i$ rather than the more expensive box $j$. For Nature to remain indifferent, this oracle advantage must be offset by the right-hand side, which captures the DM's additional expected search costs from prioritizing the cheaper box $i$ over $j$ in the randomized search order when the prize is in fact in $j$. Condition \eqref{eq:within-block-ind}, together with an accounting identity from \Cref{lem:accounting} in the Appendix, pins down the regret-equalizing target vector\footnote{See \cref{app:accounting-identity} for details.}
\begin{equation}\label{eq:target-vector-main}
E_i
=
\frac{c_i}{Q^{(k)}}
+
\Phi(B^{(k)})
-
\frac{\Sigma(B^{(k)})}{Q^{(k)}C(B^{(k)})},
\qquad i\in B^{(k)}.
\end{equation}

\paragraph{The implementability problem.}

Equation \eqref{eq:target-vector-main} gives the expected cumulative search costs that would render Nature indifferent across all boxes in a block. The key question is feasibility: can some lottery over within-block search orders generate this vector $(E_i)_{i\in B^{(k)}}$? The following result gives the answer.

\begin{proposition}\label{lem:core-implementability}
A vector \(E=(E_i)_{i\in B}\) is implementable by a lottery over permutations of \(B\) if and only if
\begin{equation}\label{eq:core-inequality-main}
\sum_{i\in S}c_iE_i
\geq
\frac{C(S)^2+\Sigma(S)}{2}
\qquad
\text{for every }S\subseteq B,
\end{equation}
with equality at \(S=B\).
\end{proposition}

The inequality has a simple interpretation. The term
\[
v(S):=\frac{C(S)^2+\Sigma(S)}{2}
\]
is the smallest cost-weighted cumulative search cost that the boxes in $S$ can generate: it is attained when all boxes in $S$ are searched before any box outside $S$. Therefore, no lottery over orders can assign the boxes in $S$ a lower total cost-weighted cumulative search cost than $v(S)$. This proves necessity. When $S=B$, the inequality is an equality because the total cost-weighted cumulative search cost of exhaustively searching the whole block is invariant to the order.

For sufficiency, consider a hypothetical {\it cooperative game} with {\it characteristic function} $v(S)$. The inequalities in \eqref{eq:core-inequality-main}, together with equality at $B$, say that the vector $(c_iE_i)_{i\in B}$ lies in the core of this game. Moreover, the marginal contribution of any box $j$,
\[
v(S\cup\{j\})-v(S)
=
c_j\bigl(C(S)+c_j\bigr),
\]
becomes larger when the coalition $S$ that is already present becomes larger. The game is therefore convex, or supermodular. By the core characterization for convex games---equivalently, the polymatroid characterization---the core is the convex hull of its extreme points, each consisting of marginal contributions of boxes along a sequence of nested subsets within $B$ \citep{shapley1971cores,edmonds2003submodular,ichiishi1981super}. Each sequence of nested subsets induces a permutation of boxes in $B$, and, importantly, the associated marginal contributions coincide with the DM's cost-weighted cumulative search costs when she searches according to that order.\footnote{To see this, let $\pi=(\pi_1,\ldots,\pi_m)$ be the induced permutation and let $S_k=\{\pi_1,\ldots,\pi_k\}$. Searching according to $\pi$ yields cumulative cost
$E_{\pi_k}^{\pi}=C(S_k)$
when $\pi_k$ is the prize box, while the corresponding marginal contribution is
\[
v(S_k)-v(S_{k-1})
=
c_{\pi_k}C(S_k)
=
c_{\pi_k}E_{\pi_k}^{\pi}.
\]
Hence the marginal vector associated with the nested sequence is exactly the cost-weighted cumulative-search-cost vector $(c_iE_i^\pi)_{i\in B}$ generated by the order $\pi$.} This means that if the regret-equalizing cost-weighted expenditure vector $(c_iE_i)_{i\in B}$ lies in the core, satisfying \eqref{eq:core-inequality-main}, it can be implemented by a lottery over deterministic search orders. We thus conclude that \eqref{eq:core-inequality-main} guarantees the implementability of $E$ via some randomized search order.

The core test in \Cref{lem:core-implementability} requires checking every subset of \(B\). The target vector in \eqref{eq:target-vector-main}, however, has additional structure. Because the weighted target \(c_iE_i\) contains a positive quadratic term in $c_i$ and a common linear term, coalitions that contain expensive boxes receive relatively high weighted cumulative costs. The constraints that can bind are therefore the lower-cost coalitions. To formalize this, write the boxes in \(B\) in increasing cost order as \(b_1,\ldots,b_{|B|}\), and define the prefix sets
\[
S_k:=\{b_1,\ldots,b_k\},
\qquad k=1,\ldots,|B|.
\]

\begin{proposition}\label{prop:prefix-sufficiency}
Fix a block \(B\) and the target vector \((E_i)_{i\in B}\) defined in \eqref{eq:target-vector-main}. Then \eqref{eq:core-inequality-main} holds for all sets \(S\subseteq B\) if and only if it holds for all prefix sets \(S_k\subseteq B\).
\end{proposition}

\cref{prop:prefix-sufficiency} shows that implementability is hardest for the cheap prefixes. These boxes are precisely the ones that the target tries to reach early. If the lottery over search orders can give enough priority to every cheap prefix, then coalitions containing more expensive boxes automatically satisfy the core inequalities.  The high-dimensional implementability problem thus collapses to a nested sequence of prefix constraints.

In the case of concave costs, the simplification goes one step further. Not only do we need to check only prefixes; among the prefixes of a tail, the binding one is the singleton prefix consisting of the cheapest tail box.

\begin{lemma}\label{lem:hc-concave-block-single-prefix}
Assume costs are concave and fix a tail block \(T=[\ell,n]\). If \(T\) is a singleton, within-block implementability is automatic. If \(\ell<n\), then within-block implementability is satisfied if and only if the core condition \eqref{eq:core-inequality-main} holds for the singleton prefix \(\{\ell\}\). Letting \(Q^T\) be the DM's reach probability of tail \(T\), this condition can be written as
\begin{equation}\label{eq:block-prefix-threshold-main}
Q^T
\geq
\frac{\Sigma(T)/C(T)-c_\ell}{\Phi(T)-c_\ell}.
\end{equation}
\end{lemma}

In the case of concave costs, among all prefix constraints for $T$, the most demanding one is the singleton prefix \(\{\ell\}\), the cheapest box in the tail. If the lottery over search orders can give sufficient priority to this cheapest box, i.e., if \eqref{eq:block-prefix-threshold-main} holds, then concavity guarantees that all longer prefixes are also feasible. Noting that the DM's probability of entering the tail $Q^T$ increases with the size of the potential reward $\uh$, the implementability requirement reduces to the simple cutoff condition $\uh\geq\bar u_{\ell:n}$.

\section{Comparative Statics}\label{sec:comparative-statics}

We now turn to some of the questions motivating our analysis. Digital platforms have changed consumer search in two ways. On the one hand, they have made the inspection of any given option substantially cheaper.  On the other hand, they have dramatically expanded the number of available options. It has been noted that this may lead to paradoxical outcomes, where, even though each individual search is easier, consumers may feel overwhelmed by the size of the menu and disengage from search altogether. 

Our model provides a simple mechanism for this phenomenon. When the menu grows, or when inspection costs become similar, so that available options become harder to rank ex ante, the DM's concern over mistakes from exploring the wrong options becomes more severe, reducing her inclination to search at all. The comparative statics below make this force precise: focusing on the saddle-point construction under concave costs from \cref{thm:hc-concave-tail}, we consider the DM's probability of initiating search, $Q^{(1)}$, under the robustly optimal search rule and study how it changes when more options become available and when options become more similar ex ante.

\subsection{Adding options}\label{subsec:cs-adding-options}

We first ask what happens when the menu expands. In a standard Bayesian search
model, adding an option to an existing menu cannot hurt the DM: she can always ignore it. In the
robust problem, however, an additional option gives Nature a new possible
location for the prize. Since the worst case is a one-prize environment, the
added option increases the scope for selection mistakes. The DM must now hedge
against a larger set of possible ways in which she may search the wrong boxes.
The next result shows that this force unambiguously reduces the probability of
initiating search.

\begin{theorem}\label{thm:cs-adding-option}
Let \(c=(c_1,\ldots,c_n)\) be strictly increasing and concave, and suppose a new
box with cost \(c_{\mathrm{new}}>c_1\) is inserted at position
\(k\in\{2,\ldots,n+1\}\), yielding a strictly increasing concave sequence
\(\tilde c=(\tilde c_1,\ldots,\tilde c_{n+1})\). Let \(Q^{(1)}\) and
\(\tilde Q^{(1)}\) be the probabilities of initiating search in the saddle-point
construction of \cref{thm:hc-concave-tail} under \(c\) and \(\tilde c\),
respectively. Then
\[
\tilde Q^{(1)}\le Q^{(1)}.
\]
\end{theorem}

\cref{thm:cs-adding-option} captures a stark form of choice overload. As the choice set grows, the DM becomes more likely to opt out before inspecting any option. This effect is driven not by the cost of inspecting each additional option in isolation, but by the increasing difficulty of navigating a larger menu:  if there is only one prize, a larger menu provides more places for it to be hidden and, consequently, more ways for the DM to begin searching in the wrong place.

\begin{figure}[htb]
    \centering
    \includegraphics[width=0.65\textwidth]{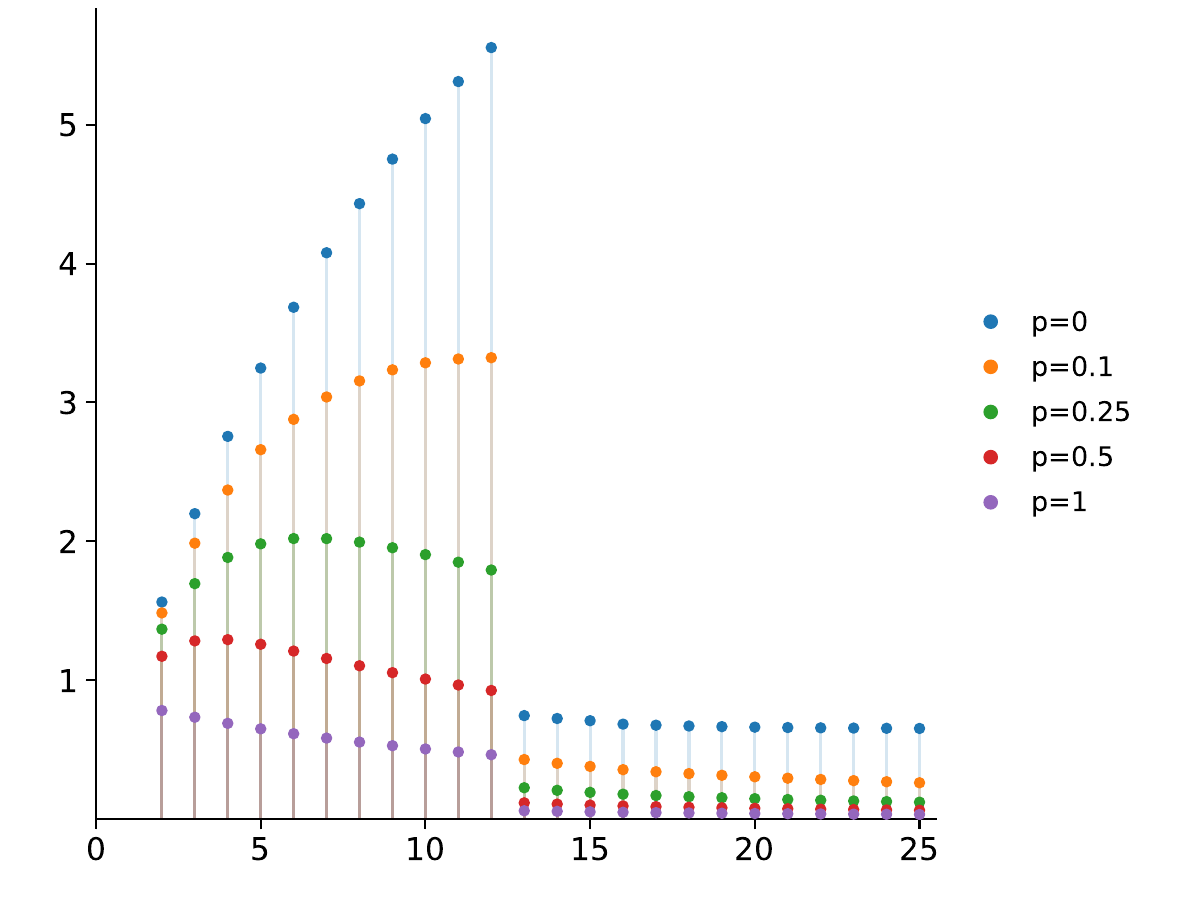}
    \caption{Expected number of opened boxes when rewards are i.i.d. Bernoulli
    with \(p\in\{0,0.1,0.25,0.5,1\}\), as a function of \(n\). For each \(n\),
    costs vary between \(0.1\) and \(0.2\) with constant increments. The drops
    correspond to points at which the saddle-point structure transitions from a
    single block to multiple blocks.}
    \label{fig:total}
\end{figure}

The theorem concerns the extensive margin of search. Another natural variable of interest is the expected number of options inspected. In contrast to the initial search probability, this object depends not only on the
robust search rule, but also on the true data-generating process (DGP) for rewards. \cref{fig:total} fixes this process to be i.i.d. Bernoulli, with probability \(p\) of a high reward, and plots the expected number of opened boxes as the menu size varies. Of course, not knowing this true DGP, the DM employs the robust search rule characterized earlier.

The figure shows a hump-shaped pattern: for small menus, adding options can
increase the total amount of search, while for large menus it decreases it. This
is consistent with a standard theme in the choice-overload literature:  additional
options may initially be attractive, but sufficiently large assortments can reduce
choice, purchase, and satisfaction
\citep{iyengar2000choice,ReutskajaHogarth2009}. Recent evidence from online
recommender systems documents a closely related pattern: increasing the number of
recommended products first raises and then lowers purchase, with much of the
decline coming from a lower probability of starting search
\citep{long2025choice}. At the same time, the empirical literature emphasizes that choice overload is not a universal phenomenon; rather, its emergence depends on factors such as choice-set complexity, decision difficulty, preference uncertainty, and the consumer’s decision goal \citep{scheibehenne2010can,chernev2015choice}.

In our robust model, the hump shape comes from two opposing forces.  As options are added, the DM becomes less likely to initiate search but, conditional on starting and finding nothing in the early boxes, may search further. For small menus, the continuation effect dominates; for large menus, the decline in initiation more than offsets it. Thus choice overload is not modeled as a psychological anomaly. It is a rational response to a harder robust-search problem.

\subsection{Flattening the cost profile}\label{subsec:cs-flattening}

The previous result shows that expanding the menu can reduce search. Our interpretation is that the key friction is the absence of guidance. To substantiate this reading of the model, we next consider the effect of changes in the cost profile. When costs differ sharply, the DM has a natural way to organize search: start with the cheap boxes, and only later move to the expensive ones. When the cost profile becomes flatter, this guidance weakens. The boxes become more similar from the point of view of search, and the DM has fewer reasons to choose one path over another.

The following comparative static formalizes this idea. It compares two concave cost profiles, where the second profile is both higher and flatter.

\begin{theorem}\label{thm:cs-flatter-costs}
Let \(c=(c_1,\ldots,c_n)\) and \(\tilde c=(\tilde c_1,\ldots,\tilde c_n)\) be strictly increasing
concave cost sequences such that \(\tilde c_i\ge c_i\) for all \(i\), and
\(\tilde c_i-\tilde c_{i-1}\le c_i-c_{i-1}\) for all \(i=2,\ldots,n\). Let \(Q^{(1)}\) and
\(\tilde Q^{(1)}\) be the probabilities of initiating search in the saddle-point construction of \cref{thm:hc-concave-tail}
under \(c\) and \(\tilde c\), respectively. The probability of initiating search weakly decreases:
\[
\tilde Q^{(1)}\le Q^{(1)}.
\]
\end{theorem}

There are two forces in the theorem. The first is direct: higher search costs make search less attractive. The second is informational: a flatter profile makes it harder to use costs as a guide. It is important to emphasize that higher search costs alone do not guarantee a lower probability of initiating search. Indeed, it is easy to construct pairs of cost profiles in which one has uniformly lower but flatter costs, yet induces a lower probability of initiating search. A stark illustration arises in the homogeneous limit below, where costs become entirely flat. Whereas a steep profile gives the DM direction, a flat profile deprives her of guidance, leaving her with a more symmetric and therefore more ambiguous navigation problem.

\paragraph{Homogeneous limit.} The cleanest way to isolate the role of flattening is to consider the limiting case in which cost differences vanish. Fix $c\in(0,\uh)$. For each flatness parameter $\nu>0$, let
\[
c^\nu=(c^\nu_1,\ldots,c^\nu_n)
\]
be a strictly increasing, concave cost profile satisfying $\max_{i\in[n]}|c_i^\nu-c|\le \nu$. For each $\nu$, let $(a_\nu,\mathbf p_\nu)$ be the saddle point constructed in \cref{thm:hc-concave-tail}, and let $\ell_\nu$ be its tail boundary. Thus the blocks are
\[
B_\nu^{(1)}=\{1\},\ldots,B_\nu^{(\ell_\nu-1)}=\{\ell_\nu-1\},
\qquad
B_\nu^{(\ell_\nu)}=[\ell_\nu,n],
\]
with the convention that there is only the tail block when $\ell_\nu=1$. Let $Q_\nu^{(k)}$ denote the probability of reaching block $B_\nu^{(k)}$. We study the limit as $\nu\downarrow0$.

For a flat profile, the global profitability condition simplifies substantially:
\[
\uh>\Phi([n])
=
\frac{n^2c^2+nc^2}{2nc}
=
\frac{n+1}{2}c.
\]
Fixing $\uh$ and $c$, we observe that this condition becomes violated when $n$ becomes sufficiently large. Let
\[
\bar{n}:=\max\Bigl\{k\in[n]:\ \frac{k+1}{2}c\leq\uh\Bigr\}
\]
be the largest size of a homogeneous menu that can be searched profitably when the prize is known to be present.

\begin{theorem}\label{thm:hc-flat-limit}
Fix $c\in(0,\uh)$. For each $\nu>0$, let
$c^\nu=(c^\nu_1,\dots,c^\nu_n)$ be a strictly increasing, concave cost profile satisfying
\[
\max_{i\in[n]} |c_i^\nu-c|\le \nu.
\]
Let $\ell_\nu$ be the tail boundary in the saddle-point construction of
\cref{thm:hc-concave-tail}, and let $Q_\nu^{(k)}$ be the probability of reaching the
$k$th block in that construction. Then, as $\nu\downarrow0$:
\begin{enumerate}
\item[\textup{(i)}] If \(\uh>\frac{n+1}{2}c\), then $\ell_\nu=1$ for all sufficiently small $\nu$, and
\[
Q_\nu^{(1)}\to Q^*:=
\frac{\uh-c}{\uh+\frac{n-1}{2}c}\in(0,1).
\]

\item[\textup{(ii)}] If \(\uh<\frac{n+1}{2}c\), then
\[
Q_\nu^{(k)}\to0
\qquad\text{for every }k=1,\dots,n-\bar n+1.
\]
In particular, $Q_\nu^{(1)}\to0$.
\end{enumerate}
\end{theorem}

The theorem has two regimes. If \(\uh>\frac{n+1}{2}c\), the whole flat menu is globally profitable. In the limit, Nature assigns equal mass to all prize states and positive mass to the no-prize state. The DM initiates search with probability $Q^*=\frac{\uh-c}{\uh+\frac{n-1}{2}c}$ and, conditional on searching, explores all options. In the approximating constructions, $\ell_\nu=1$ eventually, so the whole menu is treated as a single block. In the limit, Nature and the DM treat boxes symmetrically: each box has the same probability of containing the prize, and, conditional on searching, the DM uses a uniformly random order. Consistent with \cref{thm:cs-adding-option}, the probability of initiating search strictly declines with \(n\): a larger homogeneous menu makes search less attractive because it increases the expected cost of finding the single prize.

If \(\uh<\frac{n+1}{2}c\), global profitability fails. In this regime, the probability of initiating search converges to zero as boxes become homogeneous. This is the starkest form of choice overload in the model: as the cost profile becomes flat, the DM disengages from search altogether. The implication is that lower search costs need not induce more search. Fix any strictly increasing concave profile $\hat c$ satisfying the maintained assumptions and $\hat c_i>c$ for every $i$, and let $\hat Q^{(1)}>0$ be the initiation probability generated by \cref{thm:hc-concave-tail} under $\hat c$. Then, for all sufficiently small $\nu$, the profile $c^\nu$ is uniformly cheaper than $\hat c$, yet $Q_\nu^{(1)}<\hat Q^{(1)}$. Thus the DM searches more under the uniformly higher, but more informative, cost profile than under the uniformly lower, sufficiently flat one.

The limiting belief in the second regime is revealing: Nature places no mass on the no-prize state ($p_0^*=0$). The DM’s disengagement is therefore not driven by the possibility that there is nothing to find, but by the difficulty of locating the prize in a large, homogeneous menu. Under the worst-case scenario, the prize is always hidden somewhere, yet the DM has no indication of where to search first. When the menu is sufficiently large relative to the value of the prize, this absence of direction is enough to eliminate search.

The homogeneous limit therefore sharpens the interpretation of the previous comparative statics. Adding options reduces search because it expands the set of possible mistakes. Flattening the cost profile reduces search because it removes the cost-based ranking that helps the DM avoid those mistakes. Digital platforms combine both forces: they make individual inspection cheap, but they also create large and often weakly differentiated menus. The robust-search friction is precisely that, in such environments, the consumer's problem is not only whether to search, but how to begin.

\section{Discussion and Robustness}\label{sec:discussion}

We conclude by examining three robustness issues. The binary-reward assumption is the clearest limitation of the baseline model: although it yields a clean characterization, it leaves open the extent to which the overload result is driven by this simplification. We address this concern first. We then return to the binary benchmark to examine two further dimensions: dynamic reoptimization and restricting Nature’s ambiguity set to i.i.d. reward distributions.

To streamline the exposition, we restrict attention to the homogeneous case in which \((c_1,\ldots,c_n)=(c,\ldots,c)\), with \(c\in(0,\uh)\). This benchmark corresponds to the flat limit characterized in \cref{thm:hc-flat-limit}. In particular, if global profitability fails---that is, if \(\uh<\frac{n+1}{2}c\)---the DM opts out with probability one. By contrast, if \(\uh>\frac{n+1}{2}c\), the robustly optimal commitment rule initiates search with probability
\begin{equation}\label{eq}
Q^*(n)=\frac{\uh-c}{\uh+\frac{n-1}{2}c},
\end{equation}
where we make the dependence on \(n\) explicit. Conditional on initiating search, the DM searches exhaustively, selecting each remaining box uniformly at random at every stage.

\subsection{Beyond Binary Rewards}\label{subsec:beyond-binary}
The binary-reward assumption is the main tractability device in the baseline analysis: it renders the state space finite and makes stopping automatic once a high reward is found.
We now relax this restriction and set
\[
 \Omega=[0,\uh]^n, 
\]
with Nature choosing any distribution from \(\Delta([0,\uh]^n)\).\footnote{Without the upper bound \(\uh\), the minmax regret problem need not have a finite solution.}

The DM's strategy must now condition not only on the set of remaining boxes, but also on the rewards observed so far. In particular, the current best reward \(z\in[0,\uh]\) becomes a state variable: after stopping, the DM takes the best currently available payoff, including the outside option. The following simple observation allows us to focus on rules that treat the unopened boxes symmetrically.

\begin{lemma}\label{lem:rich-symmetric}
 Suppose \(\Omega=[0,\uh]^n\) and costs are homogeneous.  If the  robust search  problem has a saddle point, then it has one in which Nature's distribution is exchangeable and the DM's rule is symmetric: at every history, conditional on continuing search, each unopened box is selected with equal probability.
\end{lemma}

 The lemma removes labels from the choice of which unopened box to inspect. The description of the search rule can thus be reduced to a continuation probability as a function of the previously observed rewards. 

For a sufficiently small upper bound  $\uh$, the DM employs a threshold rule.  In particular, we show in \cref{sec:beyond-n2} that, in the two-box problem, after opening the first box and observing reward \(z\), she opens the second box if \(z<\uh-c\) and stops otherwise. When the upper bound $\uh$ is large, however, a fixed acceptance threshold exposes the DM to two opposite forms of regret. A low threshold risks stopping too early after a mediocre draw, while a high threshold risks excessive search after a sufficiently good draw. The optimal rule then requires a randomized acceptance threshold (see \cref{sec:beyond-n2}).

The additional state dependence makes the general problem substantially less tractable than the binary model. The main behavioral implication, however, is unchanged: sufficiently large homogeneous menus lead to complete disengagement.

\begin{corollary}\label{co:rich-optout}
 Let \(\Omega=[0,\uh]^n\), and suppose search costs are homogeneous. If
\[
\uh\leq \frac{n+1}{2}c,
\]
then immediate opt-out is robustly optimal.
\end{corollary}

The proof uses the binary model as a lower bound on Nature's power. Restrict Nature to the binary one-prize distributions used in the homogeneous benchmark. When \(\uh\leq (n+1)c/2\), the binary problem already allows Nature to impose regret \(\uh-c\), which is exactly the regret from immediate opt-out. Enlarging Nature's feasible set from binary distributions to all distributions on \([0,\uh]^n\) cannot reduce the maximal regret of any search rule. At the same time, opting out immediately guarantees regret of at most \(\uh-c\), since the oracle’s net payoff cannot exceed \(\uh-c\). Hence immediate opt-out remains minmax optimal in the richer problem.

The result shows that choice overload is not an artifact of binary rewards. Even when rewards can take a continuum of values, a sufficiently large homogeneous menu makes search unattractive.

\subsection{Dynamic Reoptimization}\label{subsec:dynamic-reoptimization}

It is well known that a choice based on the minmax regret criterion can lead to \emph{dynamic inconsistency}, arising from two potential sources (see \citet{hayashi2011context} for details): (1) new information may change the regret-maximizing scenario guiding the DM's choice; (2) regret-based preferences are opportunity dependent---removing some alternatives can change the optimal choice among the remaining ones.

In the homogeneous benchmark, this issue takes a simple form. When global profitability holds so that search is worthwhile, the commitment solution randomizes once at the outset; conditional on opening the first box, the DM plans to search exhaustively until she finds the prize or exhausts the menu. If, after an empty draw, she is allowed to reoptimize, she instead views the remaining problem as a fresh \((n-1)\)-box problem. The saddle-point belief for this continuation problem is not the Bayesian update of the original \(n\)-box saddle-point belief: it assigns less probability to the remaining prize states, or equivalently more probability to the no-prize state. Thus, as the DM eliminates options without finding a prize, she becomes increasingly concerned that no prize exists and has an incentive to re-randomize rather than carry out the original exhaustive-search plan.

Let us consider a \emph{sophisticated} DM who anticipates this future behavior but cannot commit to a plan. Her problem can be viewed as an \emph{intrapersonal game}, in which the DM's current self chooses an action at the current stage, taking future selves' continuation behavior as given. For each set \(N\subseteq[n]\) of remaining boxes, let
\[
\Omega_N
:=
\{\omega\in\Omega:u_i=0\text{ for all }i\notin N\}
\]
denote the set of states consistent with having reached \(N\) without discovering a high reward. Let \(u_N(a,\omega)\) denote the continuation payoff, with past search costs sunk, from following strategy \(a\) in the subproblem with remaining set \(N\).

\begin{defn}\label{def:intra}
A strategy \(a=(a_N)_{N\in\mathcal N}\) is an \textbf{intrapersonal equilibrium} if, for each set \(N\in\mathcal N\) of remaining boxes,
\[
a_N\in
\arg\min_{\tilde a_N\in\Delta(N\cup\{0\})}
\max_{\omega\in\Omega_N}
\Big[
u_N(a^*(\omega),\omega)
-
u_N\big((\tilde a_N,(a_{N'})_{N'\subsetneq N}),\omega\big)
\Big].
\]
\end{defn}

An intrapersonal equilibrium can equivalently be interpreted as a \emph{sequential saddle point}: at every continuation problem, the current self chooses a stage action that minimizes continuation regret against Nature's worst-case state, taking as given the strategy followed by future selves.

\begin{theorem}
\label{thm:corr}
Assume the DM cannot commit and search costs are homogeneous. There exists an intrapersonal equilibrium and an integer cutoff
\[
\bar n^{nc}\leq \bar{n}
\]
such that, at any history with \(m\) unopened boxes:
\begin{enumerate}
\item if \(m\leq \bar n^{nc}\), the DM randomizes between opening a uniformly selected remaining box and opting out;
\item if \(m>\bar n^{nc}\), the DM opts out with probability one.
\end{enumerate}
\end{theorem}

The cutoff structure in \cref{thm:corr} reflects the tension between the current self's incentive to search and her anticipation that future selves will reoptimize. To see this, consider first a naive DM who fails to foresee future incentives to reoptimize. Such a DM would adopt the \(n\)-box commitment plan at the initial node, initiating search with probability \(Q^*(n)\), but would then renege after each empty draw and solve the smaller continuation problem anew, using probabilities \(Q^*(n-1),Q^*(n-2)\) and so on, rather than following the original exhaustive-search plan. A sophisticated DM anticipates this future tendency to under-search. To preserve her future chances of finding the prize, she may therefore choose a higher initial search probability than under the corresponding commitment plan. This force can make the initial search probability non-monotone in the number of options, in contrast to the commitment comparative static in \cref{thm:cs-adding-option}. The large-menu conclusion is nevertheless unchanged: once the menu is sufficiently large, even the sophisticated DM opts out with probability one.

The cutoff \(\bar{n}^{nc}\) is weakly below the commitment cutoff $\bar n$ because the inability to commit makes initiating search less valuable. Indeed, choice overload has two sources now. The first is the baseline needle-in-a-haystack problem: adding options expands the set of possible locations in which the prize may be hidden. The second is intrapersonal: without commitment, the DM cannot ensure that future selves will carry out the search plan that made the initial search worthwhile.

We provide an explicit characterization of the equilibrium search probabilities and regret levels in \cref{sec:intrapersonal-eq}.

\subsection{Independent and Identically Distributed Rewards}\label{subsec:iid-rewards}

The baseline model allows Nature to choose any distribution over reward assignments, including arbitrary correlation across boxes. We now consider a more structured ambiguity set. Rewards are independent and identically distributed: each box contains the high reward with probability \(p\in[0,1]\), independently across boxes, but the DM does not know the value of \(p\). This is a robust analogue of the classical \citet{Weitzman1979} environment.

By symmetry, conditional on searching, the DM can be taken to select an unopened box uniformly at random. Her dynamic strategy is therefore described by a vector
\[
(\alpha_{m})_{m\in\{1,\ldots,n\}},
\]
where \(\alpha_{m}\) is the probability of continuing search when \(m\) boxes remain and no high reward has yet been found. Given \(p\), the minimized current-stage regret can be written as
\begin{equation}\label{eq:indep-commitment}
R_{m}(p)
:=
\min_{\alpha_m\in[0,1]}\ \Big[(1-\alpha_{m})
\bigl(1-(1-p)^{m}\bigr)(\uh-c)
+
\alpha_{m}(1-p)\bigl(c+R_{m-1}(p)\bigr)\Big],
\end{equation}
with \(R_0(p)=0\). The first term is the regret from stopping: if at least one of the remaining boxes contains the high reward, the oracle would have opened such a box and obtained \(\uh-c\). The second term is the regret from searching and drawing a low-reward box: the DM pays the cost \(c\) and then faces the \((m-1)\)-box continuation problem.

The robustly optimal continuation rule is characterized as follows.

\begin{theorem}
\label{thm:ind}
Assume rewards are independent and identically distributed and search costs are homogeneous. Conditional on not having found a high reward and on there being \(m\in\{1,\ldots,n\}\) unopened boxes, the DM opens a uniformly selected remaining box with probability
\[
\alpha_{m}^*
=
\frac{m(\uh-c)^{m}}
{(m-1)(\uh-c)^{m}+\uh^{m}}.
\]
Nature's worst-case success probability is
\[
\hat p=\frac{c}{\uh},
\]
and the associated regret is
\[
R_{m}^*
=
\left(
1-\left(\frac{\uh-c}{\uh}\right)^{m}
\right)(\uh-c).
\]
The described strategy also constitutes the unique intrapersonal equilibrium.
\end{theorem}

\begin{figure}[htb]
    \centering
    \includegraphics[width=0.65\textwidth]{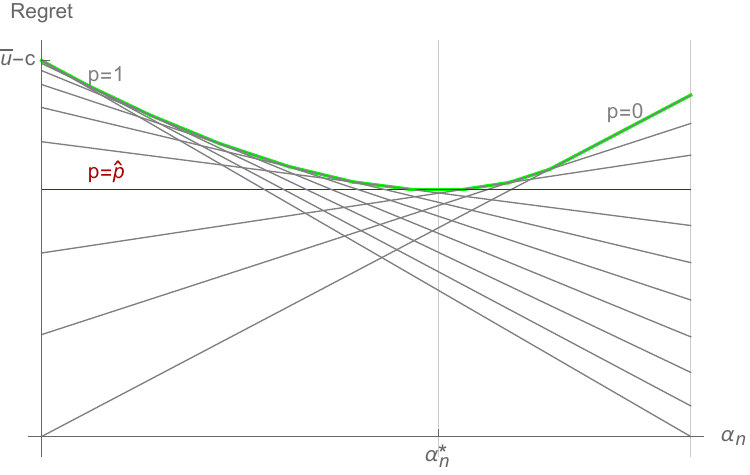}
    \caption{Saddle point with \(n=3\) unopened boxes; \(\uh=1\), \(c=0.3\).}
    \label{fig:iid-saddle}
\end{figure}

The DM's regret-minimizing search behavior is illustrated in \Cref{fig:iid-saddle}.
For each fixed \(p\), the expression in \eqref{eq:indep-commitment} is linear in \(\alpha_{m}\). For each fixed \(\alpha_{m}\), Nature chooses the success probability \(p\) that maximizes regret. The DM's worst-case regret is therefore the upper envelope of these linear functions, and the optimal search probability minimizes this envelope. In contrast to the baseline model with arbitrary correlation, the implementation of the optimal rule does not require commitment: the rule described in \cref{thm:ind} not only minimizes ex-ante regret but also constitutes an intrapersonal equilibrium.

The next result shows that, even when Nature is restricted to assigning rewards independently and identically—thereby effectively ruling out the possibility of hiding a single prize—adding options discourages the DM from initiating search. 

\begin{corollary}
\label{co:ind}
Consider the solution described in \cref{thm:ind}.
\begin{enumerate}
\item The probability of search \(\alpha_n^*\) is strictly decreasing in \(n\), with
\[
\lim_{n\to\infty}\alpha_n^*=0.
\]
\item The associated regret \(R_n^*\) is strictly increasing in \(n\), with
\[
\lim_{n\to\infty}R_n^*=\uh-c.
\]
\end{enumerate}
\end{corollary}

As in the setting with arbitrarily correlated rewards, the model predicts choice overload. This demonstrates that, although the single-prize construction in the baseline model provides a transparent benchmark, it is not essential to the result: also with i.i.d. rewards, larger menus discourage the DM from initiating search. While a larger menu raises the probability that some option is valuable, it also increases the number of unsuccessful draws the DM may face before finding one. Under minmax regret, the latter force dominates in the limit: search probabilities converge to zero, and regret converges to the loss from missing a high-reward box altogether.

\section{Conclusion}\label{sec:conclusion}

This paper revisits the classic Pandora's box problem by studying a DM who confronts ambiguity about the value of her options and aims to minimize her maximum anticipated regret. By doing so, we provide a tractable framework for analyzing sequential search in complex environments and offer a new theoretical foundation for the well-documented phenomenon of choice overload.

Our analysis demonstrates that a regret-minimizing agent's willingness to search is fundamentally shaped by the size of the choice set and the cost profile. The DM's worst-case belief centers on a ``hidden treasure'' scenario, which generates a stark prediction: the DM's propensity to search falls when new options are added to the choice set.

Crucially, our framework allows us to identify the precise mechanism driving this disengagement: selection error, or the fear of investing search effort in the wrong options. As the choice environment becomes more symmetric, the DM facing sufficiently many options becomes paralyzed by the ambiguity of where to start. This diagnosis points directly to potential remedies in \textit{assortment design}: platforms may mitigate overload by deliberately breaking the symmetry of the menu---for instance, by offering recommendations or introducing cost heterogeneity. By \emph{steering} the DM toward an ``easy" path, such designs alleviate the regret associated with selection error and restore the incentive to explore.

\appendix

\section{Proofs}
\subsection{Proof of \cref{lem:hc-one-prize}}
Fix \(a\in\Delta(A)\). Let \(\omega\) be a state with at least one prize, and let
\[
i^*(\omega)\in \arg\min\{c_i:\omega_i=\uh\}
\]
be the cheapest prize location. Let \(\omega'\) be the state that keeps only this prize:
\[
\omega'_{i^*(\omega)}=\uh,
\qquad
\omega'_j=0\quad\text{for all }j\neq i^*(\omega).
\]
The oracle payoff is the same in the two states, because the cheapest prize location is the same:
\[
u(a^*(\omega'),\omega')=u(a^*(\omega),\omega).
\]

It remains to compare the DM's payoff. First, consider a pure plan \(b\in A\). Since the set of prize boxes in \(\omega'\) is a subset of the set of prize boxes in \(\omega\), the plan \(b\) cannot find a prize earlier in \(\omega'\) than in \(\omega\). More precisely, up to the first time a prize is found in \(\omega\), the histories in \(\omega\) and \(\omega'\) coincide unless \(b\) opens a prize box that was removed in the construction of \(\omega'\). In that case, \(b\) stops in state \(\omega\), while in state \(\omega'\) it has not found a prize and may only continue searching or stop later. Hence, relative to \(\omega\), the reward obtained by \(b\) in \(\omega'\) weakly falls and the total search cost weakly rises. Therefore, $u(b,\omega')\leq u(b,\omega).$ Taking expectations over the pure plans according to the mixed rule \(a\) gives
\[
u(a,\omega')\leq u(a,\omega).
\]
Together with equality of the oracle payoffs, this implies
\[
u(a^*(\omega'),\omega')-u(a,\omega')
\geq
u(a^*(\omega),\omega)-u(a,\omega).
\]

Now let \(\pi\in\Delta(\Omega)\) be any regret-maximizing distribution. Construct a new distribution by leaving unchanged the probability assigned to states with no prize or exactly one prize, and by moving the probability assigned to each state with several prizes to the one-prize state constructed above. The preceding inequality shows that this operation weakly increases expected regret. Since \(\pi\) was already regret-maximizing, the new distribution is also regret-maximizing. By construction, it assigns positive probability only to states with at most one prize.
\qed

\subsection{Proof of \cref{lem:hc-best-reply}}

For fixed \(\mathbf p\), the oracle's expected payoff is independent of the DM's rule. Hence the DM's best replies to \(\mathbf p\) are exactly the rules that maximize the DM's expected payoff under \(\mathbf p\). Since this payoff is linear in the DM's mixed rule, it suffices to consider pure rules.

Under the reduced state space, a pure rule is identified with the finite list of boxes it opens along the history in which no prize is found. If a prize is found, the rule stops. After a set \(S\subset[n]\) has been opened and found empty, the posterior is
\begin{equation}\label{eq:hc-posterior}
\Pr(i\mid S)=
\frac{p_i}{p_0+\sum_{\ell\notin S}p_\ell}
\qquad\text{for every }i\notin S.
\end{equation}

\noindent\textbf{Step 1:} Fix such a history \(S\) and compare two continuation plans that differ only by swapping the next two unopened boxes \(i\) and \(j\). The two plans coincide after both boxes have been opened and found empty. Thus the conditional expected payoff difference from opening \(i\) before \(j\), rather than \(j\) before \(i\), is
\begin{equation}\label{eq:hc-swap}
\Pr(i\mid S)c_j-\Pr(j\mid S)c_i
=
\frac{p_i c_j-p_j c_i}{p_0+\sum_{\ell\notin S}p_\ell}.
\end{equation}
Hence opening \(i\) before \(j\) is weakly better if and only if
\[
\frac{p_i}{c_i}\geq \frac{p_j}{c_j}.
\]
Repeated adjacent swaps therefore produce an optimal rule that opens boxes in weakly decreasing order of \(p_i/c_i\) before stopping.\\

\noindent\textbf{Step 2:} We next show the all-or-nothing property for ties. Consider a maximal class \(B\) of boxes with a common positive ratio
\[
\frac{p_i}{c_i}=\lambda>0
\qquad\text{for all }i\in B.
\]
Suppose, toward a contradiction, that a pure best reply opens some but not all boxes in \(B\). Let \(k\in B\) be the last opened box from \(B\), and let \(\ell\in B\) be an unopened box. By Step 1,  no box $j$ with a higher ratio $\frac{p_j}{c_j}>\lambda$ can be opened after \(k\). Similarly, no box $m$ with a lower ratio $\frac{p_m}{c_m}<\lambda$ can be opened after \(k\), since \(\ell\) is left unopened. Hence the rule stops after \(k\) if no prize is found.

Let \(D>0\) be the probability of reaching \(k\). Since deleting \(k\) cannot improve the rule,
\[
\frac{p_k}{D}\cdot\uh-c_k\geq 0\quad\Longleftrightarrow\quad\lambda\uh\geq D.
\]
Since adding \(\ell\) after \(k\) cannot improve the rule,
\[
\frac{p_\ell}{D-p_k}\cdot \uh-c_\ell\leq 0\quad\Longleftrightarrow\quad\lambda\uh\leq D-p_k.
\]
This is impossible because \(p_k>0\). Therefore, for every positive-ratio class, a pure best reply opens either all boxes in the class or none of them.
\qed

\subsection{Proof of \cref{lem:hc-monotone-ratio}}

Suppose, toward a contradiction, that
\[
\frac{p_i}{c_i}<\frac{p_j}{c_j}
\qquad\text{for some }i<j.
\]
Then \(p_j>0\), so state \(j\) lies in Nature's support.

Let \(b\) be any pure rule in the support of the DM's mixed strategy. Since \((a,\mathbf p)\) is a saddle point, every such \(b\) is a best reply to \(\mathbf p\). By \cref{lem:hc-best-reply}, \(b\) cannot open \(i\) while leaving \(j\) unopened, and, if it opens both boxes, it opens \(j\) before \(i\). Therefore the payoff of \(b\) in state \(j\) is weakly larger than its payoff in state \(i\): either \(b\) never opens \(i\), in which case state \(j\) is weakly better for the DM, or \(b\) opens \(j\) before \(i\), in which case the prize is found weakly earlier in state \(j\). Averaging over the DM's mixed strategy gives
\[
u(a,j)\geq u(a,i).
\]
Next, since \(c_i<c_j\), the oracle payoff is strictly larger in state \(i\):
\[
u(a^*(i),i)=\uh-c_i>\uh-c_j=u(a^*(j),j).
\]
Thus
\[
u(a^*(i),i)-u(a,i)
>
u(a^*(j),j)-u(a,j).
\]
Nature assigns positive probability to state \(j\), but moving that probability from state \(j\) to state \(i\) would strictly increase expected regret against \(a\). This contradicts the optimality of \(\mathbf p\) at a saddle point.
\qed

\subsection{Proof of \cref{thm:hc-block-structure}}

Let \((a,\mathbf p)\) be a saddle point of the reduced game, where \(a\) is the DM's mixed strategy. Write
\[
\lambda_i:=\frac{p_i}{c_i}.
\]
By \cref{lem:hc-monotone-ratio}, the sequence \((\lambda_i)_{i=1}^n\) is weakly decreasing.

There is at least one active prize state: if, instead, Nature were to put all probability on state \(0\), the DM's unique best reply would be to stop immediately, and Nature could obtain strictly positive regret by assigning the prize to box \(1\), since \(c_1<\uh\). Let
\[
m:=\max\{i\in[n]:p_i>0\}.
\]
If \(p_j>0\) and \(i<j\), then $\lambda_i\geq \lambda_j>0,$ so \(p_i>0\). Hence the active boxes form an initial segment \([m]\). Partition \([m]\) into maximal contiguous intervals on which \(\lambda_i\) is constant, and denote them by
\[
B^{(1)},B^{(2)},\ldots,B^{(K)}.
\]
Let \(\lambda^{(k)}\) be the common value of \(p_i/c_i\) on \(B^{(k)}\). Since \((\lambda_i)\) is weakly decreasing and the blocks are maximal,
\[
\lambda^{(1)}>\lambda^{(2)}>\cdots>\lambda^{(K)}.
\]
Since \(a\) is a best reply to \(\mathbf p\), every pure rule $b$ in its support is itself a best reply to \(\mathbf p\). By \cref{lem:hc-best-reply}, \(b\) opens boxes in weakly decreasing order of \(p_i/c_i\) and opens either all boxes or no boxes from each positive-ratio class. On the active set \([m]\), these positive-ratio classes are precisely the blocks $B^{(1)},B^{(2)},\ldots,B^{(K)}.$ Therefore every pure rule in the support of \(a\) searches the active blocks in their order and, for each block, searches either all boxes in the block or none of them.
\qed

\subsection{Accounting Identity and Target Vector}\label{app:accounting-identity}

\begin{lemma}[Accounting identity]\label{lem:accounting}
Fix a nonempty set $S\subseteq[n]$ and an exhaustive order
$\pi=(\pi_1,\ldots,\pi_m)$ of $S$. Let
\[
E^\pi_{\pi_k}:=\sum_{r=1}^k c_{\pi_r}
\]
be the cumulative search cost incurred under this order when the prize is in box $\pi_k$. Then
\begin{equation}\label{eq:accounting-appendix}
\sum_{i\in S} c_iE^\pi_i
=
\frac{C(S)^2+\Sigma(S)}{2}
=
C(S)\Phi(S).
\end{equation}
Consequently, if the prize is distributed over $S$ according to $\Pr(i\mid i\in S)=c_i/C(S)$, the expected cumulative cost of locating the prize under any exhaustive order is $\Phi(S)$.
\end{lemma}

\begin{proof}
For the order $\pi$, we have
\[
\sum_{i\in S}c_iE^\pi_i
=
\sum_{k=1}^m c_{\pi_k}\sum_{r=1}^k c_{\pi_r}.
\]
The right-hand side contains each square term $c_i^2$ once and, for each unordered pair $\{i,j\}$, contains exactly one product $c_ic_j$, namely the product corresponding to whichever of $i$ and $j$ is searched later. Hence
\[
\sum_{i\in S}c_iE^\pi_i
=
\sum_{i\in S}c_i^2+
\sum_{\{i,j\}\subseteq S}c_ic_j
=
\frac{1}{2}\left[
\left(\sum_{i\in S}c_i\right)^2+
\sum_{i\in S}c_i^2
\right]
=
\frac{C(S)^2+\Sigma(S)}{2}.
\]
Dividing by $C(S)$ gives the expected cumulative cost under the cost-proportional distribution.
\end{proof}

\paragraph{Target vector:} Fix block $B^{(k)}$ and consider equation \eqref{eq:within-block-ind}, guaranteeing regret equalization between any two boxes contained in $B^{(k)}$. Condition \eqref{eq:within-block-ind} pins down $E_i$ up to an additive constant $a^{(k)}$:
\begin{equation}\label{eq:affine-target}
E_i=\frac{c_i}{Q^{(k)}}+a^{(k)},
\qquad i\in B^{(k)}.
\end{equation}
The constant $a^{(k)}$ is determined by the previous accounting identity. Namely, any lottery over exhaustive orders of $B^{(k)}$ must satisfy
\begin{equation}\label{eq:accounting}
\sum_{i\in B^{(k)}} c_iE_i
=
C(B^{(k)})\Phi(B^{(k)}).
\end{equation}
Substituting \eqref{eq:affine-target} into \eqref{eq:accounting} gives
\[
a^{(k)}
=
\Phi(B^{(k)})-
\frac{\Sigma(B^{(k)})}{Q^{(k)}C(B^{(k)})}.
\]
Finally, substituting for $a^{(k)}$ yields target vector \eqref{eq:target-vector-main}.

\subsection{Proof of \cref{thm:hc-concave-tail}: Cases and Reach Probabilities}\label{app:hc-cases-reach}

The proof constructs Nature's distribution and the within-tail search lottery and verifies their implementability and optimality. Because these verifications are lengthy and largely algebraic, we relegate them to Online Appendix \ref{app:proof-concave-tail} but, for later reference, record the case distinction and reach probabilities here. Throughout, write $L_{\ell:n}:=\Sigma_{\ell:n}/C_{\ell:n}$.

\paragraph{Case A (\(p_0>0\)):}
This case applies if \(\ell^*=1\), or if \(\ell^*\ge2\) and
\(\uh\le\bar u_{\ell^*-1:n}\). The DM randomizes between entering the tail and stopping. Her reach probabilities of the tail and last singleton block are
\begin{equation}\label{eq:reach-tail-A}
Q^{(\ell^*)}=Q^T_A(\ell^*),\quad\text{and for $\ell^*\geq2$,}\quad Q^{(\ell^*-1)}=Q_A^{-T}(\ell^*),
\end{equation}
where
\begin{eqnarray*}
Q^T_A(\ell):=\frac{\uh-L_{\ell:n}}
{\uh+C_{\ell:n}-\Phi([\ell,n])},\quad\text{and for $\ell\geq2$,}\quad Q_A^{-T}(\ell):=\frac{\uh-c_{\ell-1}-Q_A^{T}(\ell) C_{\ell:n}}{\uh}.
\end{eqnarray*}
For \(\ell^*\geq3\), set the remaining reach probabilities recursively
\begin{equation}\label{eq:reach-recursion}
Q^{(i)}
=
\frac{c_{i+1}-c_i+Q^{(i+1)}(\uh-c_{i+1})}{\uh}, \quad i=1,\ldots,\ell^*-2.
\end{equation}

\paragraph{Case B (\(p_0=0\)):}
This case applies if \(\ell^*\ge2\) and
\(\uh>\bar u_{\ell^*-1:n}\). By the minimality of \(\ell^*\) and genericity,
these conditions imply \(\uh<\Phi([\ell^*-1,n])\). Conditional on reaching the tail, the DM searches it with probability one. The reach probabilities of the tail and last singleton block are
\begin{equation}\label{eq:reach-tail-B}
Q^{(\ell^*)}=Q^{(\ell^*-1)}=Q_B^{T}(\ell^*),
\end{equation}
where
\[
Q_B^{T}(\ell):=
\frac{2(\Sigma_{\ell:n}-c_{\ell-1}C_{\ell:n})}
{C_{\ell:n}^2+\Sigma_{\ell:n}}.
\]
For \(i=1,\ldots,\ell^*-2\), set $Q^{(i)}$ as in \eqref{eq:reach-recursion}.

\subsection{Proof of \cref{thm:cs-adding-option}}

Let \(a\) and \(\tilde a\) be the saddle-point search rules selected by
\cref{thm:hc-concave-tail} in the original and expanded problems, and let
\(R^*\) and \(\tilde R^{*}\) be the corresponding saddle-point regrets. Let
\(\ell\) and \(\tilde\ell\) denote the associated tail cutoffs. Since the new box is
inserted after box \(1\), the cost of the first box is still \(c_1\). 

First, we argue that $\tilde R^{*}\ge R^*.$ By choosing the saddle-point strategy of the original problem, thereby ignoring the newly added box, Nature can guarantee a regret of $R^*$. The saddle-point regret in the expanded problem must therefore be weakly larger.

Crucially, in both saddle-point constructions, state \(1\) is active. This implies that the regret in state $1$ under the optimal search rule equals the respective saddle-point regret $R^*$ and $\tilde R^*$:
\[
R^*=R^{(1)}=\uh-c_1-u(a,1),\qquad \tilde R^*=\tilde R^{(1)}=\uh-c_1-u(\tilde a,1).
\]
Therefore
\begin{equation}\label{eq:cs-state1-payoff-order}
u(\tilde a,1)
=
\uh-c_1-\tilde R^{*}
\le
\uh-c_1-R^*
=
u(a,1).
\end{equation}

\medskip
\noindent\emph{Case 1: \(\tilde\ell>1\).} Then box \(1\) is the first singleton block in the expanded problem. Hence
\[
u(\tilde a,1)=\tilde Q^{(1)}(\uh-c_1).
\]
In the original problem, conditional on state \(1\), the DM obtains at most
\(\uh-c_1\) whenever she initiates search and obtains zero if she does not
initiate. Thus
\[
u(a,1)\le Q^{(1)}(\uh-c_1).
\]
Combining this inequality with \eqref{eq:cs-state1-payoff-order} gives
\[
\tilde Q^{(1)}(\uh-c_1)\le Q^{(1)}(\uh-c_1).
\]
Since \(\uh>c_1\), we obtain \(\tilde Q^{(1)}\le Q^{(1)}\).

\medskip
\noindent\emph{Case 2: \(\tilde\ell=1\).}
Define the initiation probability that equalizes the regret in state $0$ with single-$[n]$-block search for the original menu by
\[
Q_{[n]}
:=Q_A^T(1)=
\frac{\uh-L_{1:n}}{\uh+C_{1:n}-\Phi([n])}.
\]
Let $[\tilde n]:=[n]\cup\{\mathrm{new}\}$ denote the expanded set. Since \(\tilde\ell=1\), the expanded problem is a one-block problem, so
\[
\tilde Q^{(1)}=Q_{[\tilde n]}
:=
\frac{\uh-L_{1:\tilde n}}
{\uh+C_{1:n}+c_{\mathrm{new}}-\Phi([\tilde n])},
\qquad\text{with }\quad
L_{1:\tilde n}=\frac{\Sigma_{1:n}+c_{\mathrm{new}}^2}{C_{1:n}+c_{\mathrm{new}}}.
\]
Using \(\Phi([n])=(C_{1:n}+L_{1:n})/2\) and
\(\Phi([\tilde n])=(C_{1:n}+c_{\mathrm{new}}+L_{1:\tilde n})/2\), direct simplification gives
\begin{equation}\label{eq:cs-qprime-qhat-algebra}
Q_{[\tilde n]}\le Q_{[n]}
\quad\Longleftrightarrow\quad
\uh\ge
\frac{2C_{1:n}L_{1:n}-c_{\mathrm{new}}(C_{1:n}-L_{1:n})}
{C_{1:n}-L_{1:n}+2c_{\mathrm{new}}}.
\end{equation}
The condition \(\tilde\ell=1\) also implies
\(\uh\ge \bar u_{1:\tilde n}\). Another direct simplification gives
\[
\bar u_{1:\tilde n}
-
\frac{2C_{1:n}L_{1:n}-c_{\mathrm{new}}(C_{1:n}-L_{1:n})}
{C_{1:n}-L_{1:n}+2c_{\mathrm{new}}}
=
\frac{
(c_{\mathrm{new}}-c_1)(C_{1:n}^2+C_{1:n}L_{1:n}+2C_{1:n}c_{\mathrm{new}}+2c_{\mathrm{new}}^2)
}{
C_{1:n}(C_{1:n}-L_{1:n}+2c_{\mathrm{new}})
}
>0,
\]
because \(c_{\mathrm{new}}>c_1\). Hence \eqref{eq:cs-qprime-qhat-algebra} holds, and therefore
\begin{equation}\label{eq:cs-qprime-below-qhat}
\tilde{Q}^{(1)}\le Q_{[n]}.
\end{equation}

It remains to compare \(Q_{[n]}\) with the original \(Q^{(1)}\). If the original
problem also has \(\ell=1\), then the original construction is the one-block
construction, so \(Q^{(1)}=Q_{[n]}\). Suppose instead that \(\ell>1\). Since \(\tilde \ell=1\), we have
\(\uh>\Phi([\tilde n])\). The function \(\Phi\) is increasing under inclusion, so
\(\uh>\Phi([n])\). Consider then the following feasible distribution for Nature in
the original problem:
\[
\check p_0
=
\frac{\uh-\Phi([n])}{\uh+C_{1:n}-\Phi([n])},\qquad \check p_i
=
\frac{c_i}{\uh+C_{1:n}-\Phi([n])},
\qquad i=1,\ldots,n.
\]
All ratios \(\check p_i/c_i\) are equal. By \cref{lem:hc-best-reply}, the DM
either searches the whole block \([n]\) or stops. Stopping gives payoff zero,
and exhaustive search also gives payoff zero under \(\check{\mathbf p}\), since
\begin{eqnarray*}
&&\sum_{i\in[n]}\check p_i\bigl(\uh-\Phi([n])\bigr)
-
\check p_0C_{1:n}\\
&=&
\frac{C_{1:n}}{\uh+C_{1:n}-\Phi([n])}\bigl(\uh-\Phi([n])\bigr)
-
\frac{\uh-\Phi([n])}{\uh+C_{1:n}-\Phi([n])}C_{1:n}=0.
\end{eqnarray*}
Thus Nature can guarantee expected regret
\[
\sum_{i=1}^n \check p_i(\uh-c_i)
=
\frac{\uh C_{1:n}-\Sigma_{1:n}}{\uh+C_{1:n}-\Phi([n])}
=
Q_{[n]}C_{1:n}.
\]
Therefore
\begin{equation}\label{eq:cs-value-lower-bound}
R^*\ge Q_{[n]}C_{1:n}.
\end{equation}
Because \(\uh>\Phi([n])\), the original concave-tail construction is in the
positive-\(p_0\) case. Hence state \(0\) is active and
\[
R^*
=
R_0
=
\sum_{i=1}^n c_i\Pr_a(\text{box }i\text{ is opened in state }0).
\]
Since every box is opened with probability at most \(Q^{(1)}\), we have 
\[
R^*\le Q^{(1)} C_{1:n}.
\]
Together with \eqref{eq:cs-value-lower-bound}, this gives $Q_{[n]}\le Q^{(1)}.$ Combining with \eqref{eq:cs-qprime-below-qhat}, we obtain
\[
\tilde Q^{(1)}=Q_{[\tilde n]}\leq Q_{[n]}\le Q^{(1)},
\]
completing the proof.
\qed

\subsection{Proof of \cref{thm:cs-flatter-costs}}

For \(t\in[0,1]\), set
\[
c_i(t):=c_i+t(\tilde c_i-c_i),
\qquad i=1,\ldots,n,
\]
and let \(Q^{(1)}(t)\) denote the probability of initiating search in the
saddle-point construction of \cref{thm:hc-concave-tail} for the cost profile
\(c(t)\).  Thus
\[
Q^{(1)}(0)=Q^{(1)},
\qquad
Q^{(1)}(1)=\tilde Q^{(1)}.
\]
Writing \(d_i:=\tilde c_i-c_i\), the assumptions of the theorem imply
\[
d_1\geq d_2\geq\cdots\geq d_n\geq0,
\]
and every \(c(t)\) is strictly increasing and concave. We use the following lemma.  Its proof is given in
\cref{app:proof-flattening-inequalities}.

\begin{lemma}
\label{lem:cs-flattening-inequalities}
For \(t\in[0,1]\), let $c_i(t):=c_i+t(\tilde c_i-c_i)$. In the
following statements, all cost-dependent expressions are evaluated at
\(c(t)\).
\begin{enumerate}
\item[(i)]
For every \(\ell\),
\[
\frac{d\Phi([\ell,n])}{dt}\geq0,
\qquad
\uh\geq\bar u_{\ell:n}
\quad\Longrightarrow\quad
\frac{dQ_A^T(\ell)}{dt}\leq0.
\]
For \(\ell\geq2\),
\[
\uh\geq\Phi([\ell,n])
\quad\Longrightarrow\quad
\frac{dQ_A^{-T}(\ell)}{dt}\leq0,
\qquad
\frac{dQ_B^T(\ell)}{dt}\leq0.
\]

\item[(ii)]
Implementability cutoffs are nested:
\[
\bar u_{\ell:n}\leq\bar u_{\ell-1:n},
\qquad \ell=2,\ldots,n.
\]

\item[(iii)]
If \(\ell\geq2\) and
\(\uh=\bar u_{\ell-1:n}\), then
\[
Q_A^T(\ell)=Q_A^{-T}(\ell)=Q_B^T(\ell).
\]
For \(\ell=1,\ldots,n-1\), if
\(\uh=\bar u_{\ell:n}\), then
\[
Q_A^T(\ell)=Q_B^T(\ell+1).
\]
If instead
\(\uh=\Phi([\ell,n])\geq\bar u_{\ell:n}\), then
\[
Q_A^T(\ell)\geq Q_B^T(\ell+1).
\]
Finally,
\[
Q_B^T(\ell)\geq Q_B^T(\ell+1),
\qquad \ell=2,\ldots,n-1.
\]
\end{enumerate}
\end{lemma}

\paragraph{Intervals with a fixed construction.}
Consider an interval on which the tail \([\ell,n]\) and the case of the
construction are fixed.  In Case A, part~(i) of
\cref{lem:cs-flattening-inequalities} shows that \(Q_A^T(\ell)\), and also
\(Q_A^{-T}(\ell)\) when \(\ell\geq2\), are weakly decreasing.  In Case B,
the same part shows that \(Q_B^T(\ell)\) is weakly decreasing.  Since
\[
Q^{(\ell-1)}=Q^{(\ell)}=Q_B^T(\ell)
\]
in Case B, the boundary reach probability is weakly decreasing in either
case. Moving backward through the singleton blocks, differentiation of
\eqref{eq:reach-recursion} gives
\[
\frac{dQ^{(i)}}{dt}
=
\frac{
d_{i+1}-d_i
+
\frac{dQ^{(i+1)}}{dt}(\uh-c_{i+1})
-
Q^{(i+1)}d_{i+1}
}{\uh}.
\]
Thus \(dQ^{(i)}/dt\leq0\) whenever
\(dQ^{(i+1)}/dt\leq0\).  Backward induction shows that
\(Q^{(1)}(t)\) is weakly decreasing on every interval on which the
construction is fixed.

\paragraph{Switching points.}
It remains to show that a change of construction cannot create an upward
jump.  First, if the case changes while the tail remains \([\ell,n]\), the
switch occurs at \(\uh=\bar u_{\ell-1:n}\).  Part~(iii) of
\cref{lem:cs-flattening-inequalities} gives $Q_A^T(\ell)=Q_A^{-T}(\ell)=Q_B^T(\ell)$,
so the boundary value is continuous.

Second, suppose the tail expands from \([\ell+1,n]\) to \([\ell,n]\).
Part~(i) of the lemma shows that \(\Phi([\ell,n])\) is weakly increasing in
\(t\).  Hence the larger tail cannot become admissible because its
profitability condition becomes newly satisfied.  The switch must occur at $\uh=\bar u_{\ell:n}.$ At this value, part~(iii) gives $Q_B^T(\ell+1)=Q_A^T(\ell).$ Moreover, part~(ii) implies \(\uh\leq\bar u_{\ell-1:n}\) when \(\ell\geq2\); for \(\ell=1\), Case A applies by definition. Thus \(Q_A^T(\ell)\) is the new reach probability
of box \(\ell\), while \(Q_B^T(\ell+1)\) is its old reach probability.
There is therefore no upward jump.

Third, suppose the tail contracts from \([\ell,n]\) to
\([\ell+1,n]\).  If implementability is the binding condition, then
\(\uh=\bar u_{\ell:n}\).  Part~(ii) implies that the old construction is in
Case A when \(\ell\geq2\), while Case A applies by definition when
\(\ell=1\).  Part~(iii) again shows that the old and new reach probabilities
of box \(\ell\) agree. If instead profitability is the binding condition, then $\uh=\Phi([\ell,n]).$ When implementability binds simultaneously, the preceding argument applies. Otherwise \(\uh>\bar u_{\ell:n}\), so the construction with the smaller tail \([\ell+1,n]\) is in Case B.  If the old construction is in Case A,
part~(iii) gives $Q_A^T(\ell)\geq Q_B^T(\ell+1).$ If the old construction is in Case B, the final inequality in part~(iii) gives $Q_B^T(\ell)\geq Q_B^T(\ell+1).$ Thus a contraction of the tail cannot increase the reach probability of
box \(\ell\).

The recursion \eqref{eq:reach-recursion} is continuous and increasing in
\(Q^{(i+1)}\).  Hence the absence of an upward jump at the tail boundary
propagates through every preceding singleton block.  Nonadjacent changes
can be decomposed into adjacent ones, and simultaneous boundary equalities
follow from the same identities by taking one-sided limits.  Consequently,
\(Q^{(1)}(t)\) is weakly decreasing on \([0,1]\), and therefore
\[
\tilde Q^{(1)}
=Q^{(1)}(1)
\leq Q^{(1)}(0)
=Q^{(1)}.
\]
\qed

\subsection{Proof of \cref{thm:hc-flat-limit}}

For \(k=1,\ldots,n\), write \(T_k:=[n-k+1,n]\) for the tail of size \(k\).
As \(\nu\downarrow0\),
\[
C^\nu(T_k)\to kc,
\qquad
\Sigma^\nu(T_k)\to kc^2,
\qquad
\Phi^\nu(T_k)\to\frac{k+1}{2}c.
\]
Moreover,
\[
\bar u^\nu_{n-k+1:n}\to c.
\]
For \(k=1\), this is immediate from the singleton definition. For \(k\ge2\), it follows from
\[
\frac{\Sigma^\nu(T_k)}{C^\nu(T_k)}-c^\nu_{n-k+1}\to0
\]
while
\[
C^\nu(T_k)^2-\Sigma^\nu(T_k)\to k(k-1)c^2>0.
\]

\paragraph{Part (i):}
If \(\uh>\frac{n+1}{2}c\), then, for all sufficiently small \(\nu\),
\[
\uh>\Phi^\nu([n])
\qquad\text{and}\qquad
\uh\ge \bar u^\nu_{1:n}.
\]
Hence the whole set is feasible in \eqref{eq:hc-ell-star-main}, so $\ell_\nu=1$ eventually. The one-block formula in the construction of
\cref{thm:hc-concave-tail} gives
\[
Q_\nu^{(1)}
=
\frac{\uh-\Sigma^\nu_{1:n}/C^\nu_{1:n}}
{\uh+C^\nu_{1:n}-\Phi^\nu([n])}.
\]
Taking limits yields
\[
Q_\nu^{(1)}
\to
\frac{\uh-c}
{\uh+nc-\frac{n+1}{2}c}
=
\frac{\uh-c}{\uh+\frac{n-1}{2}c}.
\]

\paragraph{Part (ii):} Now suppose \(\uh<\frac{n+1}{2}c\). Then \(\bar n<n\). Set
\[
m:=n-\bar n+1.
\]
For every tail \(T_k\) with \(k>\bar n\),
\[
\frac{k+1}{2}c>\uh.
\]
Therefore, for all sufficiently small \(\nu\),
\[
\Phi^\nu(T_k)>\uh,
\qquad k>\bar n.
\]
No tail of size larger than \(\bar n\) can then satisfy the profitability
condition in \eqref{eq:hc-ell-star-main}. Hence $\ell_\nu\ge m$ eventually.

We now use the reach-probability bounds from the construction in
\cref{thm:hc-concave-tail}. In both Case A and Case B, every singleton block
before the tail satisfies
\[
0\le Q_\nu^{(i)}
\le
\frac{c^\nu_{i+1}-c^\nu_i}{c^\nu_{i+1}},
\qquad i<\ell_\nu.
\]
This is exactly the backward-induction bound used to prove feasibility of the
singleton-front recursion. Since \(c^\nu_{i+1}-c^\nu_i\to0\), the right-hand side
converges to zero.

Thus, if \(i\le m\) and \(i<\ell_\nu\), then \(Q_\nu^{(i)}\to0\). It remains only to
handle the possible boundary case \(i=m=\ell_\nu\). In that case the enlarged tail
\([m-1,n]\) has size \(\bar n+1\), and
\[
\bar u^\nu_{m-1:n}\to c<\uh.
\]
Thus, for all sufficiently small \(\nu\), the larger tail fails because it is
not profitable: $\uh<\Phi^\nu([m-1,n]).$ The construction is therefore Case B. The concavity bound for the tail-reach probability \eqref{eq:reach-tail-B} gives
\[
0\le Q_\nu^{(m)}
\le
\frac{c^\nu_m-c^\nu_{m-1}}{c^\nu_m}
\to0.
\]
Therefore $Q_\nu^{(k)}\to0$ for every block index $k=1,\ldots,m=n-\bar n+1.$
\qed

\bibliographystyle{apalike}
\bibliography{Library}

\newpage
\setcounter{page}{1}

\linespread{1.15}

\section{Online Appendix}

\subsection{Proof of \cref{thm:global-profitability}}

We first record the auxiliary fact used in the proof.

\begin{lemma}\label{lem:hc-guaranteed-prize}
Fix a nonempty set \(S\subseteq[n]\) and suppose the prize is known to lie in \(S\), with conditional prior \(\mathbf p\in\Delta(S)\). If
\[
\uh>\Phi(S),
\]
then opting out is not a best reply.
\end{lemma}

\begin{proof}
Let \(\pi=(\pi_1,\ldots,\pi_m)\) be an exhaustive order of \(S\) that minimizes the expected opening cost. By the adjacent-swap argument in \eqref{eq:hc-swap}, we may choose \(\pi\) so that
\[
\frac{p_{\pi_1}}{c_{\pi_1}}
\geq
\frac{p_{\pi_2}}{c_{\pi_2}}
\geq
\cdots
\geq
\frac{p_{\pi_m}}{c_{\pi_m}}.
\]
Write
\[
C_k^\pi:=\sum_{t=1}^k c_{\pi_t},
\qquad k=1,\ldots,m.
\]
If the DM searches \(S\) exhaustively in this order, the expected opening cost is
\[
\sum_{k=1}^m p_{\pi_k}C_k^\pi.
\]
For each \(k\), let \(P_k:=\sum_{t=1}^k p_{\pi_t}\). Since the ratios \(p_{\pi_t}/c_{\pi_t}\) are weakly decreasing,
\[
\frac{P_k}{C_k^\pi}
=
\frac{\sum_{t=1}^k c_{\pi_t}(p_{\pi_t}/c_{\pi_t})}
{\sum_{t=1}^k c_{\pi_t}}
\geq
\frac{\sum_{t=1}^m c_{\pi_t}(p_{\pi_t}/c_{\pi_t})}
{\sum_{t=1}^m c_{\pi_t}}
=
\frac{1}{C(S)}.
\]
Hence \(P_k\geq C_k^\pi/C(S)\). Summation by parts gives
\[
\sum_{k=1}^m p_{\pi_k}C_k^\pi
=
C(S)-\sum_{k=1}^{m-1}P_k c_{\pi_{k+1}}
\leq
C(S)-\sum_{k=1}^{m-1}\frac{C_k^\pi}{C(S)}c_{\pi_{k+1}}
=
\sum_{k=1}^m \frac{c_{\pi_k}}{C(S)}C_k^\pi.
\]
By \Cref{lem:accounting}, the last term is equal to \(\Phi(S)\). Therefore the exhaustive search order \(\pi\) yields expected payoff at least
\[
\uh-\Phi(S)>0.
\]
Since opting out yields payoff \(0\), opting out is not a best reply.
\end{proof}

\begin{proof}[Proof of \cref{thm:global-profitability}]

For a fixed search rule \(a\), write
\[
R_i:=u(a^*(i),i)-u(a,i),
\qquad i=0,1,\ldots,n.
\]
At a saddle point, every state in Nature's support maximizes \(R_i\).

A useful preliminary observation is that \(\Phi\) is strictly increasing under inclusion. Indeed, if \(S\neq\emptyset\) and \(j\notin S\), then
\[
\Phi(S\cup\{j\})-\Phi(S)
=
\frac{c_j\bigl(C(S)^2+2c_jC(S)-\Sigma(S)\bigr)}
{2C(S)(C(S)+c_j)}
>0,
\]
because \(C(S)^2\geq \Sigma(S)\).

\medskip

\noindent\emph{Claim 1: Full support if \(\uh>\Phi([n])\).}
Fix a saddle point \((a,\mathbf p)\). By \cref{thm:hc-block-structure}, the active prize states form an initial segment \([m]\), partitioned into active blocks $B^{(1)},\ldots,B^{(K)}.$

We first show that \(p_0>0\). Suppose instead that \(p_0=0\). Then the DM knows that the prize lies in \([m]\). Since \(\Phi\) is increasing under inclusion,
\[
\uh>\Phi([n])\geq \Phi([m]).
\]
By \cref{lem:hc-guaranteed-prize}, opting out is not a best reply. Hence, every pure rule in the support of \(a\) enters the first active block \(B^{(1)}\). By \cref{thm:hc-block-structure}, it searches \(B^{(1)}\) exhaustively, so $Q^{(1)}=1$ and
\[
R^{(1)}
=
\Phi(B^{(1)})-\frac{\Sigma(B^{(1)})}{C(B^{(1)})}
=
\frac{C(B^{(1)})^2-\Sigma(B^{(1)})}{2C(B^{(1)})}
<
C(B^{(1)}).
\]
In state \(0\), however, every pure rule in the support of \(a\) opens every box in \(B^{(1)}\) before it can stop. Therefore
\[
R^{(0)}\geq C(B^{(1)})>R^{(1)},
\]
contradicting Nature's optimality. Thus \(p_0>0\).

Next, suppose \(m<n\). Take any \(j>m\). Since \(p_j=0\) and \(p_0>0\), opening \(j\) at any no-prize history that can arise under state \(0\) strictly lowers the DM's expected payoff against \(\mathbf p\). Hence no pure best reply in the support of \(a\) opens \(j\). Therefore the DM's payoff is the same in states \(j\) and \(0\): $u(a,j)=u(a,0).$ But the oracle payoff in state \(j\) is strictly larger than in state \(0\), so
\[
R_j
=
R_0+(\uh-c_j)
>
R_0.
\]
Since \(p_0>0\), state \(0\) is in Nature's support, and this again contradicts Nature's optimality. Hence \(m=n\), so every prize state is in Nature's support.

It remains only to show that the DM inspects every box with positive probability. Suppose some box \(j\) is opened with probability zero under \(a\). Since \(p_0>0\) and \(p_j>0\), both states \(0\) and \(j\) are in Nature's support. As box \(j\) is never opened, the DM's payoff is the same in states \(j\) and \(0\). Thus
\[
R_j=R_0+(\uh-c_j)>R_0,
\]
contradicting Nature's optimality. Therefore every box is inspected with positive probability.

\medskip

\noindent\emph{Claim 2: At least two active blocks if \(\uh<\Phi([n])\).}
Suppose, toward a contradiction, that there is a saddle point with a single active block \([m]\).

If \(m=n\), then entering and exhaustively searching the unique active block yields expected payoff
\[
(1-p_0)\bigl(\uh-\Phi([n])\bigr)-p_0C([n])<0.
\]
By \cref{thm:hc-block-structure}, every pure rule in the support of \(a\) either searches the whole block or does not enter it. Since stopping gives payoff \(0\), every best reply stops immediately. Against immediate stopping, Nature's unique best reply is state \(1\), contradicting \(m=n\). Hence \(m<n\).

We now show that \(p_0=0\). Suppose instead that \(p_0>0\). Since \(p_{m+1}=0\), opening box \(m+1\) at any no-prize history that can arise under state \(0\) strictly lowers the DM's expected payoff against \(\mathbf p\). Hence no pure best reply in the support of \(a\) opens \(m+1\). Therefore
\[
u(a,m+1)=u(a,0),
\]
while
\[
u(a^*(m+1),m+1)=\uh-c_{m+1}>0=u(a^*(0),0).
\]
Thus
\[
R_{m+1}=R_0+(\uh-c_{m+1})>R_0,
\]
contradicting Nature's optimality. Hence \(p_0=0\).

With \(p_0=0\), the DM knows that the prize lies in the unique active block \([m]\). Exhaustively searching this block yields expected payoff
\[
\uh-\Phi([m]).
\]
By the genericity condition \eqref{eq:genericity}, this expression is nonzero.

If \(\uh-\Phi([m])<0\), then every best reply stops immediately. Against immediate stopping, Nature's unique best reply is state \(1\). Hence we would have \(m=1\). But then
\[
\uh-\Phi([1])=\uh-c_1>0,
\]
a contradiction.

If \(\uh-\Phi([m])>0\), then every pure rule in the support of \(a\) searches \([m]\) exhaustively. As above,
\[
R^{(1)}
=
\Phi([m])-\frac{\Sigma([m])}{C([m])}
=
\frac{C([m])^2-\Sigma([m])}{2C([m])}
<
C([m]).
\]
In state \(0\), the DM pays at least \(C([m])\), so $R^{(0)}>R^{(1)}$, contradicting Nature's optimality. Therefore, a saddle point cannot have a single active block. Since every saddle point has at least one active block, it must have at least two active blocks.
\end{proof}

\subsection{Proof of \cref{lem:core-implementability}}

\begin{proof}
Fix a block \(B\). For a deterministic order \(\pi=(\pi_1,\ldots,\pi_m)\) of \(B\), let
\[
E_{\pi_k}^{\pi}:=\sum_{t=1}^k c_{\pi_t}
\]
be the cumulative cost incurred inside the block when the prize is in box \(\pi_k\), and write
\[
x_i^\pi:=c_iE_i^\pi .
\]

\noindent\emph{Necessity:} Fix a subset \(S\subseteq B\). For any deterministic order \(\pi\), the sum $\sum_{i\in S}x_i^\pi$ is at least the corresponding weighted cumulative cost obtained by searching only the boxes in \(S\), in the order induced by \(\pi\). By \Cref{lem:accounting}, applied to the set \(S\), this latter quantity is
\[
\Phi(S)C(S)=\frac{C(S)^2+\Sigma(S)}{2}.
\]
Thus
\[
\sum_{i\in S}x_i^\pi
\geq
\frac{C(S)^2+\Sigma(S)}{2}.
\]
For \(S=B\), there are no outside boxes, so the inequality holds with equality. Taking expectations over any lottery over permutations gives \eqref{eq:core-inequality-main}, with equality at \(S=B\).\\

\noindent\emph{Sufficiency:} Define
\[
v(S):=\frac{C(S)^2+\Sigma(S)}{2},
\qquad S\subseteq B.
\]
The set of vectors \(x=(x_i)_{i\in B}\) satisfying
\[
\sum_{i\in S}x_i\geq v(S)
\quad\text{for all }S\subseteq B,
\qquad
\sum_{i\in B}x_i=v(B),
\]
is the core of the cooperative game \(v\). The game \(v\) is supermodular: for \(S\subseteq T\subseteq B\) and \(i\notin T\),
\[
v(S\cup\{i\})-v(S)
=
c_i\bigl(C(S)+c_i\bigr)
\leq
c_i\bigl(C(T)+c_i\bigr)
=
v(T\cup\{i\})-v(T).
\]
Hence \(v\) is a convex game. By the Shapley--Ichiishi characterization of convex games, equivalently Edmonds' characterization of base polyhedra of supermodular functions, the core is the convex hull of the marginal vectors associated with permutations of \(B\); see \citet{ichiishi1981super,edmonds2003submodular}.

For a permutation \(\pi=(\pi_1,\ldots,\pi_m)\), let \(S_k:=\{\pi_1,\ldots,\pi_k\}\), with \(S_0=\emptyset\). The associated marginal vector is
\[
v(S_k)-v(S_{k-1})
=
\frac{C(S_k)^2+\Sigma(S_k)}{2}
-
\frac{C(S_{k-1})^2+\Sigma(S_{k-1})}{2}
=
c_{\pi_k}\sum_{t=1}^k c_{\pi_t}
=
x_{\pi_k}^\pi .
\]
Thus the marginal vectors of \(v\) are exactly the cost-weighted cumulative-expenditure vectors generated by deterministic search orders.

Now suppose \(E\) satisfies \eqref{eq:core-inequality-main}, with equality at \(S=B\), and set \(x_i=c_iE_i\). Then \(x\) lies in the core of \(v\). Therefore there exist weights \((\sigma_\pi)_\pi\), with \(\sigma_\pi\geq0\) and \(\sum_\pi\sigma_\pi=1\), such that
\[
x_i=\sum_\pi \sigma_\pi x_i^\pi
\qquad\text{for every }i\in B.
\]
Since \(c_i>0\), this is equivalent to
\[
E_i=\sum_\pi \sigma_\pi E_i^\pi
\qquad\text{for every }i\in B.
\]
The lottery that chooses order \(\pi\) with probability \(\sigma_\pi\) implements \(E\).
\end{proof}

\subsection{Proof of \cref{prop:prefix-sufficiency}}

\begin{proof}
List the elements of \(B\) in increasing cost order as
\[
b_1<b_2<\cdots<b_m,
\]
and define the prefix sets
\[
S_k:=\{b_1,\ldots,b_k\},
\qquad k=0,1,\ldots,m,
\]
with \(S_0=\emptyset\) and \(S_m=B\).  Let \(Q\in(0,1]\) be the reach probability of the block and set
\[
\beta:=
\Phi(B)-\frac{\Sigma(B)}{Q\,C(B)}.
\]
By \eqref{eq:target-vector-main},
\[
E_i=\frac{c_i}{Q}+\beta
\qquad\text{for every }i\in B.
\]
Hence, for every \(S\subseteq B\),
\[
\sum_{i\in S}c_iE_i
=
\frac{\Sigma(S)}{Q}+\beta C(S).
\]
Letting
\[
f(S):=
\left(\frac1Q-\frac12\right)\Sigma(S)
+\beta C(S)
-\frac{C(S)^2}{2},
\]
inequality \eqref{eq:core-inequality-main} is thus equivalent to
\[
f(S)\geq 0.
\]
Moreover, the target vector satisfies equality at the full block: $f(B)=0$. Also, clearly \(f(\emptyset)=0\).

The necessity of the prefix conditions is immediate, since every prefix \(S_k\) is a subset of \(B\). We prove sufficiency. For \(s\in[0,C(B)]\), define the piecewise-linear function \(\psi\) by
\[
\psi(s):=
\Sigma(S_{k-1})+c_{b_k}\bigl(s-C(S_{k-1})\bigr)
\qquad
\text{whenever }s\in[C(S_{k-1}),C(S_k)].
\]
Thus \(\psi\) is the polygonal line connecting the prefix points
\[
(C(S_0),\Sigma(S_0)),(C(S_1),\Sigma(S_1)),\ldots,(C(S_m),\Sigma(S_m)).
\]
We claim that, for every subset \(S\subseteq B\),
\[
\Sigma(S)\geq \psi(C(S)).
\]
To see this, fix \(s\in[0,C(B)]\) and consider the relaxed problem
\[
\min_{0\leq z_\ell\leq 1}
\sum_{\ell=1}^m z_\ell c_{b_\ell}^2
\qquad
\text{subject to}
\qquad
\sum_{\ell=1}^m z_\ell c_{b_\ell}=s.
\]
Since the cost per unit of the constraint is $c_{b_\ell}^2/c_{b_\ell}=c_{b_\ell},$ and \(c_{b_1}<\cdots<c_{b_m}\), the minimizing solution fills the cheapest boxes first. Thus, if \(s\in[C(S_{k-1}),C(S_k)]\), the minimum is attained at
\[
z_\ell=1\quad(\ell<k),
\qquad
z_k=\frac{s-C(S_{k-1})}{c_{b_k}},
\qquad
z_\ell=0\quad(\ell>k),
\]
and the minimum value is exactly
\[
\Sigma(S_{k-1})+c_{b_k}(s-C(S_{k-1}))=\psi(s).
\]
Every subset \(S\subseteq B\) corresponds to a feasible \(0\)-\(1\) vector in this relaxed problem with \(s=C(S)\). Hence
\[
\Sigma(S)\geq \psi(C(S)).
\]
Now define
\[
H(s):=\left(\frac1Q-\frac12\right)\psi(s)+\beta s-\frac{s^2}{2}.
\]
Since \(\frac1Q-\frac12\geq0\), the preceding inequality implies that, for every \(S\subseteq B\),
\[
f(S)
=
\left(\frac1Q-\frac12\right)\Sigma(S)+\beta C(S)-\frac{C(S)^2}{2}
\geq
\left(\frac1Q-\frac12\right)\psi(C(S))+\beta C(S)-\frac{C(S)^2}{2}
=
H(C(S)).
\]
On each interval \([C(S_{k-1}),C(S_k)]\), the function \(\psi\) is affine. Therefore \(H\) is concave on each such interval. Hence, for every \(s\in[C(S_{k-1}),C(S_k)]\),
\[
H(s)\geq \min\{H(C(S_{k-1})),H(C(S_k))\}.
\]
But
\[
H(C(S_k))
=
\left(\frac1Q-\frac12\right)\Sigma(S_k)+\beta C(S_k)-\frac{C(S_k)^2}{2}
=
f(S_k)
\qquad
\text{for every }k=0,1,\ldots,m.
\]
Thus, if \(f(S_k)\geq0\) for every prefix \(S_k\), then \(H(s)\geq0\) for every \(s\in[0,C(B)]\). Consequently, for every subset \(S\subseteq B\),
\[
f(S)\geq H(C(S))\geq0.
\]
This proves \eqref{eq:core-inequality-main} for all subsets \(S\subseteq B\).
\end{proof}

\subsection{Proof of \cref{lem:hc-concave-block-single-prefix}}

\begin{proof}
Let \(T=[\ell,n]\), and let \(Q^T\in(0,1]\) be the reach probability of the tail. If \(T\) is a singleton, implementability is automatic and there are no proper prefix constraints; the displayed threshold condition is void. Hence assume \(\ell<n\).

By \cref{prop:prefix-sufficiency}, within-block implementability is equivalent to the core inequalities for the prefixes \([\ell,k]\), \(k=\ell,\ldots,n\). For a proper prefix \([\ell,k]\), substituting the target vector from \eqref{eq:target-vector-main} into \eqref{eq:core-inequality-main} gives
\[
Q^T\left[
C([\ell,k])\bigl(C(T)-C([\ell,k])\bigr)
+\frac{C([\ell,k])}{C(T)}\Sigma(T)-\Sigma([\ell,k])
\right]
\ge
2\left[
\frac{C([\ell,k])}{C(T)}\Sigma(T)-\Sigma([\ell,k])
\right].
\]
Since \([\ell,k]\) is a proper prefix and costs are increasing, this is equivalent to
\begin{equation}\label{eq:hc-concave-prefix-secant-minimal}
\frac{\Sigma(T)/C(T)-\Sigma([\ell,k])/C([\ell,k])}
{C(T)-C([\ell,k])}
\le
\frac{Q^T}{2-Q^T}.
\end{equation}
We show that the left-hand side of \eqref{eq:hc-concave-prefix-secant-minimal} is largest at \(k=\ell\). Consider the polygonal line through the points
\[
\left(C([\ell,k]),\frac{\Sigma([\ell,k])}{C([\ell,k])}\right),
\qquad k=\ell,\ldots,n.
\]
It is enough to show that this line is concave. The slope of the segment from the point indexed by \(k-1\) to the point indexed by \(k\), for \(k=\ell+1,\ldots,n\), is
\[
\frac{\Sigma([\ell,k])/C([\ell,k])-\Sigma([\ell,k-1])/C([\ell,k-1])}{c_k}
=
\frac{c_kC([\ell,k-1])-\Sigma([\ell,k-1])}
{C([\ell,k-1])C([\ell,k])}.
\]
For \(k=\ell+1,\ldots,n-1\), the difference between two consecutive such slopes is
\[
\frac{
(c_k+c_{k+1})\bigl(c_kC([\ell,k-1])-\Sigma([\ell,k-1])\bigr)
-(c_{k+1}-c_k)C([\ell,k-1])C([\ell,k])
}
{C([\ell,k-1])C([\ell,k])C([\ell,k+1])}.
\]
The denominator is positive. For the numerator, first note the telescoping identity
\[
c_kC([\ell,k-1])-\Sigma([\ell,k-1])
=
\sum_{q=\ell+1}^k (c_q-c_{q-1})C([\ell,q-1]).
\]
By concavity, \(c_q-c_{q-1}\ge c_{k+1}-c_k\) for every \(q\le k\). Hence the numerator is at least
\[
(c_{k+1}-c_k)\left[
(c_k+c_{k+1})\sum_{q=\ell}^{k-1}C([\ell,q])
-C([\ell,k-1])C([\ell,k])
\right].
\]
The expression in brackets is nonnegative. Indeed,
\[
C([\ell,k-1])^2
=\sum_{s,t=\ell}^{k-1}c_sc_t
\le
c_k\sum_{s,t=\ell}^{k-1}\min\{c_s,c_t\}
=
c_k\sum_{t=\ell}^{k-1}(2(k-t)-1)c_t.
\]
Adding \(c_kC([\ell,k-1])\) to both sides gives
\[
C([\ell,k-1])C([\ell,k])
\le
2c_k\sum_{t=\ell}^{k-1}(k-t)c_t
=
2c_k\sum_{q=\ell}^{k-1}C([\ell,q])
\le
(c_k+c_{k+1})\sum_{q=\ell}^{k-1}C([\ell,q]).
\]
Thus consecutive slopes are weakly decreasing, so the polygonal line is concave.

For a concave polygonal line, the secant slope from any point to the fixed right endpoint is a weighted average of the adjacent slopes to its right. Since those adjacent slopes are weakly decreasing, these secant slopes are weakly decreasing as the left endpoint moves to the right. Therefore the left-hand side of \eqref{eq:hc-concave-prefix-secant-minimal} is maximal at the first prefix \(k=\ell\). Hence, if the core inequality holds for \(\{\ell\}\), it holds for every prefix, and \cref{prop:prefix-sufficiency} gives implementability. The converse is immediate because \(\{\ell\}\) is itself a prefix.

Finally, the prefix condition for \(\{\ell\}\) is
\[
c_\ell\left(
\frac{c_\ell}{Q^T}
+\Phi(T)
-\frac{\Sigma(T)}{Q^TC(T)}
\right)
\ge c_\ell^2,
\]
which is equivalent to
\[
Q^T
\ge
\frac{\Sigma(T)/C(T)-c_\ell}{\Phi(T)-c_\ell}.
\]
This is \eqref{eq:block-prefix-threshold-main}.
\end{proof}

\subsection{Proof of \cref{thm:hc-concave-tail}: Construction and Verification}\label{app:proof-concave-tail}

\begin{proof}
The proof has two cases. Case A is the positive-\(p_0\) construction, used when either \(\ell^*=1\) or \(\uh\le \bar u_{\ell^*-1:n}\). In this case the DM is indifferent between entering the tail and stopping. Case B is the zero-\(p_0\) construction, used when \(\ell^*\ge2\) and \(\uh>\bar u_{\ell^*-1:n}\). Then minimality of \(\ell^*\) and genericity imply \(\uh<\Phi([\ell^*-1,n])\), so the enlarged tail \([\ell^*-1,n]\) is not profitable even though its implementability threshold is passed.

We first record the elementary facts. For \(\ell<n\), direct simplification gives
\begin{equation}\label{eq:hc-QA-impl-proof}
Q_A^T(\ell)-
\frac{L_{\ell:n}-c_\ell}{\Phi([\ell,n])-c_\ell}
=
\frac{2(C_{\ell:n}-L_{\ell:n})(\uh-\bar u_{\ell:n})}
{(C_{\ell:n}+L_{\ell:n}-2c_\ell)
 (C_{\ell:n}-L_{\ell:n}+2\uh)}.
\end{equation}
Thus the Case-A tail-reach probability satisfies the single-prefix condition
in \cref{lem:hc-concave-block-single-prefix} if and only if
\(\uh\geq\bar u_{\ell:n}\).  If \(\ell=n\), implementability is automatic.

For \(\ell\geq2\), three further simplifications of
\eqref{eq:hc-tail-cutoff-main} give
\begin{equation}\label{eq:hc-QA-zero-bound-difference-proof}
\frac{\uh-c_{\ell-1}}{\uh+C_{\ell:n}}-Q_A^T(\ell)
=
\frac{C_{\ell:n}-L_{\ell:n}+2c_{\ell-1}}
{(\uh+C_{\ell:n})(C_{\ell:n}-L_{\ell:n}+2\uh)}
(\bar u_{\ell-1:n}-\uh),
\end{equation}
\begin{equation}\label{eq:hc-QA-QB-difference-proof}
\frac{\uh-c_{\ell-1}-Q_A^T(\ell)C_{\ell:n}}{\uh}
-Q_B^T(\ell)
=
\frac{2(\uh-\Phi([\ell,n]))
 (C_{\ell:n}-L_{\ell:n}+2c_{\ell-1})}
{\uh(C_{\ell:n}+L_{\ell:n})
 (C_{\ell:n}-L_{\ell:n}+2\uh)}
(\uh-\bar u_{\ell-1:n}),
\end{equation}
and
\begin{equation}\label{eq:hc-QB-zero-bound-difference-proof}
\frac{\uh-c_{\ell-1}}{\uh+C_{\ell:n}}-Q_B^T(\ell)
=
\frac{C_{\ell:n}-L_{\ell:n}+2c_{\ell-1}}
{(C_{\ell:n}+L_{\ell:n})(\uh+C_{\ell:n})}
(\uh-\bar u_{\ell-1:n}).
\end{equation}
Consequently,
\begin{equation}\label{eq:hc-caseA-order-proof}
\uh\leq\bar u_{\ell-1:n}
\ \text{and}\ 
\uh>\Phi([\ell,n])
\quad\Longrightarrow\quad
Q_A^T(\ell)
\leq
\frac{\uh-c_{\ell-1}-Q_A^T(\ell)C_{\ell:n}}{\uh}
\leq Q_B^T(\ell),
\end{equation}
whereas
\begin{equation}\label{eq:hc-caseB-zero-bound-proof}
\uh\geq\bar u_{\ell-1:n}
\quad\Longrightarrow\quad
Q_B^T(\ell)
\leq
\frac{\uh-c_{\ell-1}}{\uh+C_{\ell:n}}
\leq Q_A^T(\ell).
\end{equation}
In particular, at \(\uh=\bar u_{\ell-1:n}\),
\begin{equation}\label{eq:hc-case-boundary-equality-proof}
Q_A^T(\ell)=Q_A^{-T}(\ell)=Q_B^T(\ell).
\end{equation}
Concavity also implies
\begin{equation}\label{eq:hc-QB-concavity-bound-proof}
Q_B^T(\ell)
\leq
\frac{c_\ell-c_{\ell-1}}{c_\ell},
\qquad \ell=2,\ldots,n.
\end{equation}
Indeed,
\[
Q_B^T(\ell)
=
\frac{\sum_{i=\ell}^n c_i(c_i-c_{\ell-1})}
{\sum_{i=\ell}^n c_iC([\ell,i])}
\]
is a weighted average of
\[
\frac{c_i-c_{\ell-1}}{C([\ell,i])},
\qquad i=\ell,\ldots,n,
\]
with positive weights \(c_iC([\ell,i])\).  These ratios decrease with \(i\).
For \(i<n\), concavity and strict monotonicity give
\[
c_i-c_{\ell-1}
\geq(i-\ell+1)(c_{i+1}-c_i),
\qquad
C([\ell,i])<(i-\ell+1)c_{i+1},
\]
and hence
\[
(c_i-c_{\ell-1})c_{i+1}
>
(c_{i+1}-c_i)C([\ell,i]).
\]
This is equivalent to the desired monotonicity, so the weighted average is
bounded above by its first term.

\medskip
\noindent\textbf{Case A: \(\ell^*=1\), or \(\ell^*\ge2\) and \(\uh\le\bar u_{\ell^*-1:n}\).}

\noindent\emph{Construction.}
Write $\ell:=\ell^*$. Nature chooses
\[
p_i=\frac{c_i}{\uh}\prod_{j<i}\left(1-\frac{c_j}{\uh}\right),
\qquad i=1,\ldots,\ell-1,
\]
\[
p_i=
\prod_{j<\ell}\left(1-\frac{c_j}{\uh}\right)
\frac{c_i}{\uh+C_{\ell:n}-\Phi([\ell,n])},
\qquad i=\ell,\ldots,n,
\]
and
\[
p_0=
\prod_{j<\ell}\left(1-\frac{c_j}{\uh}\right)
\frac{\uh-\Phi([\ell,n])}{\uh+C_{\ell:n}-\Phi([\ell,n])}.
\]
The ratios \(p_i/c_i\) are strictly decreasing on the singleton part and constant on \([\ell,n]\). If \(\ell>1\), the boundary comparison is
\[
\frac{p_{\ell-1}}{c_{\ell-1}}
=
\frac{\prod_{j<\ell}(1-c_j/\uh)}{\uh-c_{\ell-1}}
>
\frac{\prod_{j<\ell}(1-c_j/\uh)}
{\uh+C_{\ell:n}-\Phi([\ell,n])}
=\frac{p_i}{c_i},
\quad i\ge\ell.
\]
Thus Nature's active prize states are \([n]\), with blocks
\[
\{1\},\ldots,\{\ell-1\},[\ell,n].
\]
The DM enters the tail with reach probability 
\[
Q_A^T(\ell)=\frac{\uh-L_{\ell:n}}
{\uh+C_{\ell:n}-\Phi([\ell,n])}.
\] 
Conditional on entering the tail, she uses any lottery over permutations implementing
\[
E_i^A=
\frac{c_i}{Q_A^T(\ell)}
+
\Phi([\ell,n])
-
\frac{\Sigma_{\ell:n}}{Q_A^T(\ell) C_{\ell:n}},
\qquad i=\ell,\ldots,n.
\]
This target is implementable by \cref{lem:hc-concave-block-single-prefix} and \eqref{eq:hc-QA-impl-proof}, because \(\ell\) satisfies \eqref{eq:hc-ell-star-main}. If \(\ell>1\), set
\[
Q^{(\ell-1)}=Q_A^{-T}(\ell)=\frac{\uh-c_{\ell-1}-Q_A^T(\ell) C_{\ell:n}}{\uh},
\]
and recursively
\[
Q^{(i)}=
\frac{c_{i+1}-c_i+Q^{(i+1)}(\uh-c_{i+1})}{\uh},
\qquad i=1,\ldots,\ell-2.
\]
By \eqref{eq:hc-caseA-order-proof} and \eqref{eq:hc-QB-concavity-bound-proof},
\[
0<Q_A^T(\ell)\le Q_A^{-T}(\ell)\le Q_B^T(\ell)
\le \frac{c_{\ell}-c_{\ell-1}}{c_{\ell}}.
\]
A backward induction using concavity then gives
\[
1>Q^{(1)}\ge Q^{(2)}\ge\cdots\ge Q^{(\ell-1)}\ge Q^{(\ell)}>0.
\]
Indeed, if \(Q^{(i+1)}\le(c_{i+1}-c_i)/c_{i+1}\), then
\[
Q^{(i)}-Q^{(i+1)}
=\frac{c_{i+1}-c_i-Q^{(i+1)}c_{i+1}}{\uh}\ge0,
\]
and
\[
Q^{(i)}\le\frac{c_{i+1}-c_i}{c_{i+1}}
\le\frac{c_i-c_{i-1}}{c_i},
\quad i\ge2.
\]
Hence the reach probabilities are feasible: after an unsuccessful search of block \(i\), the DM continues to the next block with probability \(Q^{(i+1)}/Q^{(i)}\). If \(\ell=1\), there are no singleton blocks and feasibility only requires \(0<Q_A^T(1)<1\), which follows from \(\uh>\Phi([1,n])\).\\

\noindent\emph{Verification.}
Let
\[
\Pi_i:=p_0+\sum_{j=i}^n p_j.
\]
At the tail, the payoff from entering and exhaustively searching is
\[
\frac{C_{\ell:n}}{\uh+C_{\ell:n}-\Phi([\ell,n])}
\bigl(\uh-\Phi([\ell,n])\bigr)
-
\frac{\uh-\Phi([\ell,n])}{\uh+C_{\ell:n}-\Phi([\ell,n])}
C_{\ell:n}
=0.
\]
For each singleton \(i<\ell\),
\[
\frac{p_i}{\Pi_i}=\frac{c_i}{\uh},
\qquad
\frac{\Pi_{i+1}}{\Pi_i}=1-\frac{c_i}{\uh}.
\]
Thus, if the continuation value is zero, opening box \(i\) also has value zero. Backward induction from the tail shows that the DM is indifferent at every block-start history at which she randomizes. Together with the ratio ordering and \cref{lem:hc-best-reply}, the constructed search rule is a best reply to Nature's distribution.

It remains to check Nature's optimality. For \(i<\ell\),
\begin{eqnarray}\label{regret-singleton}
R_i=(1-Q^{(i)})\uh+\sum_{k=1}^iQ^{(k)}c_k-c_i,
\end{eqnarray}
and the recursion for the \(Q^{(i)}\)'s gives \(R_1=\cdots=R_{\ell-1}\). For \(j\in[\ell,n]\),
\[
R_j=(1-Q_A^T(\ell))\uh+\sum_{k=1}^{\ell-1}Q^{(k)}c_k+Q_A^T(\ell)E_j^A-c_j.
\]
The target formula gives
\[
Q_A^T(\ell)E_j^A-c_j
=
Q_A^T(\ell)\Phi([\ell,n])-L_{\ell:n},
\]
which is independent of \(j\). Moreover, the definition of \(Q_A^T(\ell)\) is equivalent to
\[
Q_A^T(\ell)C_{\ell:n}
=
(1-Q_A^T(\ell))\uh+Q_A^T(\ell)\Phi([\ell,n])-L_{\ell:n}.
\]
Hence every tail state has the same regret as state \(0\), whose regret is
\[
R_0=\sum_{k=1}^{\ell-1}Q^{(k)}c_k+Q_A^T(\ell)C_{\ell:n}.
\]
If \(\ell>1\), the definition of \(Q^{(\ell-1)}=Q_A^{-T}(\ell)\) gives
\[
(1-Q_A^{-T}(\ell))\uh-c_{\ell-1}=Q_A^T(\ell)C_{\ell:n},
\]
so \(R_{\ell-1}=R_0\). Therefore all states \(0,1,\ldots,n\) yield the same regret, and Nature is optimal.

\medskip
\noindent\textbf{Case B: \(\ell^*\ge2\) and \(\uh>\bar u_{\ell^*-1:n}\).} Write again $\ell:=\ell^*$. By minimality of \(\ell\), the tail \([\ell-1,n]\) cannot satisfy both inequalities in \eqref{eq:hc-ell-star-main}. Since \(\uh>\bar u_{\ell-1:n}\), genericity implies
\[
\uh<\Phi([\ell-1,n]).
\]

\noindent\emph{Construction.}
Nature chooses \(p_0=0\) and
\[
p_i=\frac{c_i}{\uh}\prod_{j<i}\left(1-\frac{c_j}{\uh}\right),
\qquad i=1,\ldots,\ell-2,
\]
\[
p_{\ell-1}
=
\prod_{j<\ell-1}\left(1-\frac{c_j}{\uh}\right)
\frac{\Phi([\ell,n])-(\uh-c_{\ell-1})}{\Phi([\ell,n])},
\]
and
\[
p_i=
\prod_{j<\ell-1}\left(1-\frac{c_j}{\uh}\right)
\frac{\uh-c_{\ell-1}}{\Phi([\ell,n])}\frac{c_i}{C_{\ell:n}},
\qquad i=\ell,\ldots,n.
\]
These probabilities are positive because
\[
\Phi([\ell,n])<\uh<\Phi([\ell-1,n])<\Phi([\ell,n])+c_{\ell-1}.
\]
The last inequality follows from
\[
\Phi([\ell-1,n])
=
\frac{C_{\ell:n}\Phi([\ell,n])+c_{\ell-1}(C_{\ell:n}+c_{\ell-1})}
{C_{\ell:n}+c_{\ell-1}}
<\Phi([\ell,n])+c_{\ell-1}.
\]
The ratios \(p_i/c_i\) are strictly decreasing for \(i<\ell-1\) and constant on \([\ell,n]\). The boundary comparison with the tail is equivalent to
\[
\uh<\Phi([\ell-1,n]),
\]
so \(p_{\ell-1}/c_{\ell-1}>p_i/c_i\) for \(i\ge\ell\). If \(\ell>2\), the comparison between boxes \(\ell-2\) and \(\ell-1\) follows by writing \(t:=\uh-\Phi([\ell,n])\in(0,c_{\ell-1})\): it is equivalent to
\[
t^2+t\bigl(\Phi([\ell,n])-c_{\ell-2}-c_{\ell-1}\bigr)
+c_{\ell-2}c_{\ell-1}>0.
\]
This is immediate if \(\Phi([\ell,n])\ge c_{\ell-2}+c_{\ell-1}\). Otherwise the quadratic is minimized at an interior point, and its minimum is larger than
\[
c_{\ell-2}c_{\ell-1}-\frac{c_{\ell-2}^2}{4}>0,
\]
because \(\Phi([\ell,n])>c_{\ell-1}\). Hence the active blocks are again
\[
\{1\},\ldots,\{\ell-1\},[\ell,n].
\]
The DM enters the tail and the last singleton block with reach probability 
\[
Q_B^T(\ell)=\frac{2(\Sigma_{\ell:n}-c_{\ell-1}C_{\ell:n})}
{C_{\ell:n}^2+\Sigma_{\ell:n}}.
\]
We thus set \(Q^{(\ell)}=Q^{(\ell-1)}=Q_B^T(\ell)\). For \(i=1,\ldots,\ell-2\), set recursively
\[
Q^{(i)}=
\frac{c_{i+1}-c_i+Q^{(i+1)}(\uh-c_{i+1})}{\uh},
\]
as before. Conditional on entering the tail, she uses any lottery over permutations implementing
\[
E_i^B=\frac{c_i-c_{\ell-1}}{Q_B^T(\ell)},
\qquad i=\ell,\ldots,n.
\]
This is the target vector in \eqref{eq:target-vector-main}, since
\[
Q_B^T(\ell)\Phi([\ell,n])=L_{\ell:n}-c_{\ell-1}.
\]
If \(\ell<n\), the single-prefix condition in \cref{lem:hc-concave-block-single-prefix} is equivalent to
\[
Q_B^T(\ell)\le\frac{c_\ell-c_{\ell-1}}{c_\ell},
\]
which holds by \eqref{eq:hc-QB-concavity-bound-proof}; if \(\ell=n\), implementability is automatic. The same backward induction as in Case A gives
\[
1>Q^{(1)}\ge\cdots\ge Q^{(\ell-1)}=Q^{(\ell)}>0,
\]
so the reach probabilities are feasible.\\

\noindent\emph{Verification.}
At the tail, the residual belief is proportional to costs and the no-prize state has zero probability, so exhaustive search has value
\[
\uh-\Phi([\ell,n])>0.
\]
At the boundary singleton,
\[
\frac{p_{\ell-1}}{\sum_{j=\ell-1}^n p_j}
=
\frac{\Phi([\ell,n])-(\uh-c_{\ell-1})}{\Phi([\ell,n])},
\qquad
\frac{\sum_{j=\ell}^n p_j}{\sum_{j=\ell-1}^n p_j}
=
\frac{\uh-c_{\ell-1}}{\Phi([\ell,n])}.
\]
Therefore opening box \(\ell-1\) has value
\[
\frac{\Phi([\ell,n])-(\uh-c_{\ell-1})}{\Phi([\ell,n])}
(\uh-c_{\ell-1})
+
\frac{\uh-c_{\ell-1}}{\Phi([\ell,n])}
\bigl(\uh-\Phi([\ell,n])-c_{\ell-1}\bigr)
=0.
\]
For every earlier singleton, the same calculation as in Case A gives value zero. Thus the DM is indifferent exactly at the singleton blocks where she randomizes and strictly willing to search the tail once it is reached. By the ratio ordering and \cref{lem:hc-best-reply}, she is best replying.

For \(i<\ell\), $R_i$ is given by \eqref{regret-singleton} and the recursion gives \(R_1=\cdots=R_{\ell-1}\). For \(j\in[\ell,n]\),
\[
R_j
=(1-Q_B^T(\ell))\uh+\sum_{k=1}^{\ell-1}Q^{(k)}c_k+Q_B^T(\ell)E_j^B-c_j
=(1-Q_B^T(\ell))\uh+\sum_{k=1}^{\ell-1}Q^{(k)}c_k-c_{\ell-1}
=R_{\ell-1},
\]
where the last equality uses \(Q^{(\ell-1)}=Q_B^T(\ell)\). Thus all active prize states have the same regret. The only off-support state is \(0\), with regret
\[
R_0=\sum_{k=1}^{\ell-1}Q^{(k)}c_k+Q_B^T(\ell)C_{\ell:n}.
\]
By \eqref{eq:hc-caseB-zero-bound-proof},
\[
R_{\ell-1}-R_0
=
\uh-c_{\ell-1}-Q_B^T(\ell)(\uh+C_{\ell:n})
\ge0.
\]
Hence Nature has no profitable deviation to state \(0\).
\end{proof}

\subsection{Proof of \cref{lem:cs-flattening-inequalities}}\label{app:proof-flattening-inequalities}

\begin{proof}

\noindent\textbf{Part (i).}
Fix \(\ell\), suppress dependence on \(t\), and use dots for derivatives
with respect to \(t\). Differentiation gives
\[
\dot C_{\ell:n}
=\sum_{i=\ell}^n d_i,
\qquad
\dot L_{\ell:n}
=
\frac{1}{C_{\ell:n}}
\sum_{i=\ell}^n d_i(2c_i-L_{\ell:n}).
\]
Consequently,
\[
2\dot\Phi([\ell,n])
=
\sum_{i=\ell}^n d_i
\left(
1+\frac{2c_i-L_{\ell:n}}{C_{\ell:n}}
\right)
\geq0,
\]
since \(C_{\ell:n}-L_{\ell:n}+2c_i>0\). For the Case-A tail probability,
\[
-\frac{(2\uh+C_{\ell:n}-L_{\ell:n})^2}{2}
\dot Q_A^T(\ell)
=
\sum_{i=\ell}^n d_i
\left[
\uh-L_{\ell:n}
+
\frac{\uh+C_{\ell:n}}{C_{\ell:n}}
(2c_i-L_{\ell:n})
\right].
\]
The bracket is increasing in \(i\). If \(\ell<n\), its value at
\(i=\ell\) is increasing in \(\uh\), and at
\(\uh=\bar u_{\ell:n}\) equals
\[
\frac{c_\ell}{C_{\ell:n}}
\left(
C_{\ell:n}-L_{\ell:n}+2c_\ell
+
\frac{4C_{\ell:n}(L_{\ell:n}-c_\ell)}
{C_{\ell:n}-L_{\ell:n}}
\right)
\geq0.
\]
Thus \(\dot Q_A^T(\ell)\leq0\) whenever
\(\uh\geq\bar u_{\ell:n}\). For \(\ell=n\), this follows directly from
\(Q_A^T(n)=1-c_n/\uh\). Now suppose \(\ell\geq2\). We first claim that
\[
\dot L_{\ell:n}\leq d_{\ell-1}.
\]
Indeed,
\begin{align*}
C_{\ell:n}\bigl(d_{\ell-1}-\dot L_{\ell:n}\bigr)
={}&
d_{\ell-1}
\bigl((n-\ell+1)L_{\ell:n}-C_{\ell:n}\bigr)\\
&+
\sum_{i=\ell}^n
(d_{\ell-1}-d_i)(2c_i-L_{\ell:n}).
\end{align*}
The first term is nonnegative by Cauchy--Schwarz. Summation by parts gives
\[
\sum_{i=\ell}^n(d_{\ell-1}-d_i)(2c_i-L_{\ell:n})
=
(d_{\ell-1}-d_\ell)
\sum_{i=\ell}^n(2c_i-L_{\ell:n})
+
\sum_{j=\ell+1}^n(d_{j-1}-d_j)
\sum_{i=j}^n(2c_i-L_{\ell:n}).
\]
All coefficients are nonnegative. Moreover, concavity and
\(L_{\ell:n}\leq c_n\) imply
\[
\sum_{i=j}^n(2c_i-L_{\ell:n})
\geq
(n-j+1)(c_j+c_n-L_{\ell:n})>0.
\]
This proves the claim. Hence
\begin{align*}
\frac{(C_{\ell:n}+L_{\ell:n})^2}{2}
\dot Q_B^T(\ell)
={}&
(C_{\ell:n}+c_{\ell-1})\dot L_{\ell:n}
-d_{\ell-1}(C_{\ell:n}+L_{\ell:n})-(L_{\ell:n}-c_{\ell-1})\dot C_{\ell:n}\\
\leq{}&
-(L_{\ell:n}-c_{\ell-1})
\bigl(d_{\ell-1}+\dot C_{\ell:n}\bigr)
\leq0.
\end{align*}
It remains to consider \(Q_A^{-T}(\ell)\). For
\(k=\ell,\ldots,n\), define
\begin{align*}
H_k(\uh):={}&
(2\uh+C_{\ell:n}-L_{\ell:n})^2+2(k-\ell+1)(2\uh-L_{\ell:n})(\uh-L_{\ell:n})\\
&-2(\uh+C_{\ell:n})
\bigl(2C([\ell,k])-(k-\ell+1)L_{\ell:n}\bigr).
\end{align*}
Differentiation followed by summation by parts yields
\begin{align*}
&-\uh(2\uh+C_{\ell:n}-L_{\ell:n})^2
\dot Q_A^{-T}(\ell)\\
&\quad=
(d_{\ell-1}-d_\ell)
(2\uh+C_{\ell:n}-L_{\ell:n})^2+
\sum_{k=\ell}^{n-1}(d_k-d_{k+1})H_k(\uh)
+d_nH_n(\uh).
\end{align*}
All coefficients on the right are nonnegative. At
\(\uh=\Phi([\ell,n])\),
\[
H_k(\Phi)
=
4C_{\ell:n}^2
+(k-\ell+1)(C_{\ell:n}+L_{\ell:n})^2
-2(3C_{\ell:n}+L_{\ell:n})C([\ell,k]).
\]
Because the costs are increasing,
\[
\frac{C([\ell,k])}{k-\ell+1}
\leq
\frac{C_{\ell:n}}{n-\ell+1}.
\]
Therefore
\begin{align*}
H_k(\Phi)
\geq{}&
4C_{\ell:n}^2
+(k-\ell+1)
\left[
(C_{\ell:n}+L_{\ell:n})^2
-\frac{
2C_{\ell:n}(3C_{\ell:n}+L_{\ell:n})
}{n-\ell+1}
\right].
\end{align*}
If the expression in square brackets is nonnegative, this lower bound is
immediate. Otherwise, replacing \(k-\ell+1\) by \(n-\ell+1\) gives
\[
H_k(\Phi)
\geq
(C_{\ell:n}+L_{\ell:n})
\left[
(n-\ell+1)(C_{\ell:n}+L_{\ell:n})
-2C_{\ell:n}
\right]
\geq0.
\]
Finally,
\[
H_k'(\Phi)
=
4\bigl((k-\ell+3)C_{\ell:n}-C([\ell,k])\bigr)>0,
\qquad
H_k''(\uh)=8(k-\ell+2)>0.
\]
Thus \(H_k(\uh)\geq0\) whenever
\(\uh\geq\Phi([\ell,n])\). The summation-by-parts decomposition then
implies
\[
\dot Q_A^{-T}(\ell)\leq0.
\]

\paragraph{Part (ii):}
For \(\ell=n\), the inequality $\bar u_{\ell:n}\leq\bar u_{\ell-1:n}$ follows from
\[
\bar u_{n:n}=c_n,
\qquad
\bar u_{n-1:n}=\frac{c_n^2}{c_{n-1}}>c_n.
\]
For \(\ell<n\), clearing positive denominators reduces the desired inequality
to
\begin{align*}
&(c_\ell-c_{\ell-1})C_{\ell+1:n}^2
+2c_\ell^2C_{\ell+1:n}
-(c_{\ell-1}+c_\ell)\Sigma_{\ell+1:n}
\geq0.
\end{align*}
The left-hand side divided by $2$ is
\[
\sum_{j=1}^{n-\ell}c_{\ell+j}
\left[
c_\ell^2-c_{\ell-1}c_{\ell+j}
+(c_\ell-c_{\ell-1})
\sum_{r=1}^{j-1}c_{\ell+r}
\right].
\]
Concavity gives
\[
c_{\ell+j}\leq c_\ell+j(c_\ell-c_{\ell-1}),
\qquad
\sum_{r=1}^{j-1}c_{\ell+r}\geq(j-1)c_\ell.
\]
Each bracket is therefore at least
\(j(c_\ell-c_{\ell-1})^2>0\), proving part~(ii).

\paragraph{Part (iii):}
The first boundary identity is
\eqref{eq:hc-case-boundary-equality-proof}; the second follows by direct
substitution. At a profitability boundary,
\begin{align*}
\left.
\bigl(Q_A^T(\ell)-Q_B^T(\ell+1)\bigr)
\right|_{\uh=\Phi([\ell,n])}
\ =\ 
\frac{C_{\ell+1:n}-L_{\ell+1:n}+2c_\ell}
{(C_{\ell+1:n}+L_{\ell+1:n})C_{\ell:n}}
\bigl(\Phi([\ell,n])-\bar u_{\ell:n}\bigr).
\end{align*}
This is nonnegative whenever \(\Phi([\ell,n])\geq\bar u_{\ell:n}\). Finally, for \(\ell=2,\ldots,n-1\),
\begin{align*}
Q_B^T(\ell)-Q_B^T(\ell+1)
={}&
\frac{2C_{\ell:n}c_\ell}
{C_{\ell+1:n}^2+C_{\ell+1:n}L_{\ell+1:n}
 +2C_{\ell+1:n}c_\ell+2c_\ell^2}
\\
&\times
\left(
\frac{c_\ell-c_{\ell-1}}{c_\ell}
-Q_B^T(\ell+1)
\right).
\end{align*}
To sign the final factor, note that inequality \eqref{eq:hc-QB-concavity-bound-proof},
together with concavity, gives
\[
Q_B^T(\ell+1)
\leq
\frac{c_{\ell+1}-c_\ell}{c_{\ell+1}}
\leq
\frac{c_\ell-c_{\ell-1}}{c_\ell}.
\]
Hence \(Q_B^T(\ell)\geq Q_B^T(\ell+1)\), completing the proof of
part~(iii).
\end{proof}

\subsection{Beyond Binary Rewards: Details}

\subsubsection{Proof of \cref{lem:rich-symmetric}}

\begin{proof}
Let \((a,F)\) be a commitment saddle point. For any permutation \(\pi\) of box labels, let \((a^\pi,F^\pi)\) be the correspondingly relabeled strategy and distribution. Homogeneous costs and the symmetric ambiguity set imply that \((a^\pi,F^\pi)\) is also a saddle point with the same value. By bilinearity of expected regret, averaging over all permutations gives another saddle point,
\[
\bar a:=\frac{1}{n!}\sum_\pi a^\pi,
\qquad
\bar F:=\frac{1}{n!}\sum_\pi F^\pi.
\]
The distribution \(\bar F\) is exchangeable. The averaged rule \(\bar a\) is invariant to relabeling of unopened boxes; hence, at every history, all unopened boxes receive the same probability. Conditional on continuing search, the rule therefore selects each unopened box with equal probability.
\end{proof}

\subsubsection{The $2$-Box Case}\label{sec:beyond-n2}

The DM faces two boxes with rewards arbitrarily distributed on $[0,\uh]^2$. Her strategy consists of an initial search probability $\alpha_2(0)$ and a continuation rule $\alpha_1:[0,\uh]\to[0,1]$. More specifically, a pure strategy specifies whether to initiate search and, conditional on opening the first box and observing reward $v\in[0,\uh]$, whether to stop or continue. Regarding the latter, it is without loss to focus on cutoff strategies, given which the DM stops if the first reward is above a threshold. To hedge against ambiguity, she may randomize both in the initial decision and in the cutoff, with $\alpha_1(v)$ specifying the probability that the cutoff is above $v$, or equivalently the probability that the DM opens the second box after observing $v$.

The following proposition establishes the qualitative characteristics of the optimal search rule. To state the result, let $\alpha_2^{\{0,\uh\}}$ and $R_2^{\{0,\uh\}}$ denote the initial search probability and associated regret in the $2$-box binary-support benchmark: for $\uh\geq3/2c$,
\[
\alpha_2^{\{0,\uh\}}
=
\frac{\uh-c}{\uh+\frac{1}{2}c},
\qquad
R_2^{\{0,\uh\}}
=
\frac{2c(\uh-c)}{\uh+\frac{1}{2}c}
=
\frac{4(\uh-c)c}{2\uh+c};
\]
for $c<\uh<3/2$, $\alpha_2^{\{0,\uh\}}=0$ and $R_2^{\{0,\uh\}}=\uh-c$.

\begin{proposition}\label{prop:2-general}
Let \(\Omega=[0,\uh]^n\) with \(n=2\).  The saddle-point
regret with arbitrary rewards coincides with its binary-support counterpart
if and only if
\[
    \uh\leq4c.
\]
More precisely:
\begin{enumerate}
\item If \(c<\uh\leq\frac32c\), the DM opts out with probability one and
\[
    R_2^*=R_2^{\{0,\uh\}}=\uh-c.
\]
\item If \(\frac32c<\uh\leq4c\), the binary-support strategy, extended to
intermediate rewards, is a saddle-point strategy:
\[
    \alpha_2^*(0)=
    \frac{\uh-c}{\uh+\frac12c},
    \qquad
    \alpha_1^*(v)=\mathbf 1_{\{v<\uh-c\}}.
\]
Its regret is
\[
    R_2^*=R_2^{\{0,\uh\}}
    =\frac{2c(\uh-c)}{\uh+\frac12c}.
\]
\end{enumerate}
\end{proposition}

\begin{proof}
For an unordered reward pair \(\{v,u\}\), where \(0\leq v\leq u\leq\uh\), the DM's regret is
\begin{equation}\label{eq:two-general-regret}
G_{\alpha_1,\alpha_2}(v,u)
=(u-c)_+
-\alpha_2(0)\left[\left(\frac{u+v}{2}-c\right)
+\frac{1}{2}\bigl(\alpha_1(v)(u-v-c)-\alpha_1(u)c\bigr)\right].
\end{equation}
Since binary-support distributions are feasible for Nature in the present
problem,
\begin{equation}\label{eq:two-general-binary-lower}
    R_2^*\geq R_2^{\{0,\uh\}}.
\end{equation}
It therefore suffices to construct matching upper bounds when
\(\uh\leq4c\) (\emph{if}), and a strict lower bound when \(\uh>4c\) (\emph{only if}).

\paragraph{The range \(c<\uh\leq3c/2\).}
If the DM opts out, her regret is at most \(\uh-c\), since this is the
largest payoff available to the oracle.  Together with
\eqref{eq:two-general-binary-lower}, this proves part 1.

\paragraph{The range \(3c/2<\uh\leq4c\).}
Consider the strategy
\[
    \alpha_2(0)
    =
    \frac{\uh-c}{\uh+\frac12c},
    \qquad
    \alpha_1(v)
    =
    \mathbf 1_{\{v<\uh-c\}}.
\]
Notice that
\[
    \frac{1-\alpha_2(0)}{\alpha_2(0)}
    =
    \frac{3c}{2(\uh-c)}.
\]
We distinguish three cases.
\begin{enumerate}
\item If \(v,u<\uh-c\), the DM opens the second box after either possible
first-box observation.  When this case is nonempty, regret is
\begin{align*}
G_{\alpha_1,\alpha_2}(v,u)
&=
\bigl(1-\alpha_2(0)\bigr)(u-c)
+\alpha_2(0)c\\
&\leq
\alpha_2(0)
\left[
\frac{3c(\uh-2c)}{2(\uh-c)}+c
\right]\\
&\leq
2c\alpha_2(0),
\end{align*}
where the last inequality is equivalent to \(\uh\leq4c\).

\item If \(v<\uh-c\leq u\), the DM continues after observing \(v\) and stops
after observing \(u\).  Therefore,
\begin{align*}
G_{\alpha_1,\alpha_2}(v,u)
&=
\bigl(1-\alpha_2(0)\bigr)(u-c)
+\frac12\alpha_2(0)c\\
&\leq
\bigl(1-\alpha_2(0)\bigr)(\uh-c)
+\frac12\alpha_2(0)c\\
&=
2c\alpha_2(0).
\end{align*}

\item If \(\uh-c\leq v\leq u\), the DM stops after either possible
first-box observation.  Hence
\begin{align*}
G_{\alpha_1,\alpha_2}(v,u)
&=
\bigl(1-\alpha_2(0)\bigr)(u-c)
+\frac12\alpha_2(0)(u-v)\\
&\leq
\bigl(1-\alpha_2(0)\bigr)(\uh-c)
+\frac12\alpha_2(0)c\\
&=
2c\alpha_2(0).
\end{align*}
\end{enumerate}

Thus every reward pair generates regret at most \(2c\alpha_2(0)\).
Since
\[
    2c\alpha_2(0)
    =
    \frac{2c(\uh-c)}{\uh+\frac12c}
    =
    R_2^{\{0,\uh\}},
\]
the lower bound in \eqref{eq:two-general-binary-lower} is attained.

\paragraph{The range \(\uh>4c\).} We exhibit a distribution that gives Nature strictly more than the binary-support value.  Define
\begin{equation}\label{eq:two-general-eta}
    \eta
    :=
    \frac{
        3(\uh-c)
        -
        \sqrt{7(\uh-c)^2+6c(\uh-c)}
    }{2},
    \qquad
    \hat v
    :=
    \uh-c-\eta.
\end{equation}
Nature assigns positive probabilities to
\[
    \{0,0\},
    \qquad
    \{0,\hat v\},
    \qquad
    \{0,\uh\},
    \qquad
    \{\hat v,\uh\}.
\]
The probabilities sum to one and are chosen in the proportions
\begin{equation}\label{eq:two-general-weights}
\begin{split}
\pi_{00}:\pi_{0\hat v}:\pi_{0\uh}:\pi_{\hat v\uh}
={}&
\Big[
(\hat v-c)\eta
+(\uh-c)(2\hat v-c)
\Big]
:
2c\eta
:
2c(2\hat v-c)
:
2c^2.
\end{split}
\end{equation}
Consider an arbitrary strategy \((\alpha_1,\alpha_2)\).  Its expected
regret under this distribution is affine in
\[
    \alpha_2(0),\qquad
    \alpha_2(0)\alpha_1(0),\qquad
    \alpha_2(0)\alpha_1(\hat v),\qquad
    \alpha_2(0)\alpha_1(\uh).
\]
The probabilities in \eqref{eq:two-general-weights} satisfy
\begin{align*}
c\pi_{00}
+\frac{c-\hat v}{2}\pi_{0\hat v}
+\frac{c-\uh}{2}\pi_{0\uh}
&=0\\
\frac c2\pi_{0\hat v}
+\frac{c+\hat v-\uh}{2}\pi_{\hat v\uh}
&=0\\
c\pi_{00}
+\left(c-\frac{\hat v}{2}\right)\pi_{0\hat v}
+\left(c-\frac{\uh}{2}\right)\pi_{0\uh}
+\left(c-\frac{\uh+\hat v}{2}\right)\pi_{\hat v\uh}
&=0.
\end{align*}
The first equality eliminates the coefficient on
\(\alpha_2(0)\alpha_1(0)\), the second eliminates the coefficient on
\(\alpha_2(0)\alpha_1(\hat v)\), and the third eliminates the coefficient
on \(\alpha_2(0)\).  The remaining coefficient on
\(\alpha_2(0)\alpha_1(\uh)\) is
\[
    \frac c2\bigl(\pi_{0\uh}+\pi_{\hat v\uh}\bigr)>0.
\]
Consequently, the DM minimizes expected regret by setting
\(\alpha_1(\uh)=0\), and Nature guarantees expected regret
\begin{align}
&\pi_{0\hat v}(\hat v-c)
+\bigl(\pi_{0\uh}+\pi_{\hat v\uh}\bigr)(\uh-c)
=
\frac{2c(2\eta+\uh)}{2\eta+\uh+2c}.
\label{eq:two-general-strict-lower}
\end{align}
This lower bound is strictly larger than the binary-support regret:
\begin{align}
&
\frac{2c(2\eta+\uh)}{2\eta+\uh+2c}
-
R_2^{\{0,\uh\}}=
\frac{
    2c^2(6\eta-\uh+4c)
}{
    (2\eta+\uh+2c)(2\uh+c)
}
>0.
\label{eq:two-general-strict-gap}
\end{align}
Indeed,
\[
6\eta-\uh+4c
=
8(\uh-c)+3c
-
3\sqrt{7(\uh-c)^2+6c(\uh-c)}
>0,
\]
because
\[
\bigl(8(\uh-c)+3c\bigr)^2
-
9\bigl(7(\uh-c)^2+6c(\uh-c)\bigr)
=
(\uh-4c)^2
>0.
\]
It follows that
\[
    R_2^*
    \geq
    \frac{2c(2\eta+\uh)}{2\eta+\uh+2c}
    >
    R_2^{\{0,\uh\}},
\]
which establishes the claim.
\end{proof}

\Cref{prop:2-general} shows that when the range of possible rewards is sufficiently small, $\uh \leq 4c$, our earlier characterization applies: at the saddle point, Nature optimally selects a distribution concentrated on the bounds of the support. In this case, the binary-support search strategy, suitably extended, remains robustly optimal. 

If, on the other hand, the range is large, $\uh > 4c$, Nature can strictly increase regret by choosing a distribution with interior support and the robustly optimal search rule entails randomization over acceptance thresholds. By contrast, if the DM were to adopt a deterministic cutoff strategy, Nature could generate excessive regret for the DM by assigning a reward slightly below the cutoff to one box and zero to the other, thereby inducing the DM to search too long, or by assigning a reward slightly above the cutoff to one box and $\uh$ to the other, thereby inducing premature stopping. To illustrate this, we provide an explicit saddle-point construction for the case
\begin{equation}\label{eq:two-general-first-nonbinary-range}
    4c<\uh\leq(20+6\sqrt{10})c.
\end{equation}
Retain \(\eta\) and \(\hat v\) from
\eqref{eq:two-general-eta}.  The optimal initial-search probability is
\begin{equation}\label{eq:two-general-explicit-initial}
\alpha_2^*(0)
=
\frac{
    2(\uh-c)(2\eta+c)
}{
    2(\uh-c)(2\eta+c)
    +c(\uh-c)-\eta^2
}.
\end{equation}
Conditional on \(\uh>4c\), this probability lies in $[0,1]$ precisely on the range in \eqref{eq:two-general-first-nonbinary-range}.  The DM's continuation rule is
\begin{equation}\label{eq:two-general-explicit-rule}
\alpha_1^*(v)
=
\begin{cases}
\displaystyle
1-
\frac{
    \eta^2
}{
    (2\eta+c)(\uh-c-v)
},
&
\displaystyle
0\leq v<
\uh-c-\frac{\eta^2}{2\eta+c},
\\[10pt]
0,
&
\displaystyle
\uh-c-\frac{\eta^2}{2\eta+c}
\leq v\leq\uh.
\end{cases}
\end{equation}
Thus the qualitative departure from the binary rule is that the DM randomizes over a positive interval of cutoffs.  Even after
observing zero, the DM stops with positive probability:
\[
1-\alpha_1^*(0)
=
\frac{
    \eta^2
}{
    (2\eta+c)(\uh-c)
}
>0.
\]
Nature uses the four reward pairs and the probabilities specified in
\eqref{eq:two-general-weights}.  These four pairs have a transparent role.
The state \(\{0,0\}\) disciplines excessive search, while
\(\{0,\uh\}\) disciplines excessive opting out.  The two states involving
\(\hat v\) make the DM indifferent at the interior reward: after observing
\(\hat v\), she cannot tell whether the unopened box contains zero or
\(\uh\).

The saddle-point regret is
\begin{equation}\label{eq:two-general-explicit-value}
R_2^*
=
\alpha_2^*(0)c
\left(
    2-
    \frac{
        \eta^2
    }{
        (2\eta+c)(\uh-c)
    }
\right)
=
\frac{
    2c(2\eta+\uh)
}{
    2\eta+\uh+2c
}.
\end{equation}
By \eqref{eq:two-general-strict-gap}, this value is strictly larger than
the binary-support regret.

\paragraph{Verification of the saddle point.}
For later use, note that the definition of \(\eta\) implies
\begin{equation}\label{eq:two-general-identities}
(\uh-c)^2
-6(\uh-c)\eta
+2\eta^2
-3c(\uh-c)
=0,
\end{equation}
and the proposed initial-search probability satisfies
\begin{align}
\frac{1-\alpha_2^*(0)}{\alpha_2^*(0)}(\uh-c)
+
\frac12
\left(
    c+\frac{\eta^2}{2\eta+c}
\right)
=
2c-
\frac{
    c\eta^2
}{
    (2\eta+c)(\uh-c)
}.
\label{eq:two-general-value-identity}
\end{align}
\begin{enumerate} 
\item Consider first
\[
    v\geq
    \uh-c-\frac{\eta^2}{2\eta+c}.
\]
The DM stops after either possible first-box observation.  Therefore,
\begin{align*}
\frac{
    G_{\alpha_1^*,\alpha_2^*}(v,u)
}{
    \alpha_2^*(0)
}
&=
\frac{1-\alpha_2^*(0)}{\alpha_2^*(0)}(u-c)
+\frac{u-v}{2}\\
&\leq
\frac{1-\alpha_2^*(0)}{\alpha_2^*(0)}(\uh-c)
+\frac12
\left(
    c+\frac{\eta^2}{2\eta+c}
\right)\\
&=
2c-
\frac{
    c\eta^2
}{
    (2\eta+c)(\uh-c)
},
\end{align*}
where the final equality follows from \eqref{eq:two-general-value-identity}. 
\item Consider next
\[
v<
\uh-c-\frac{\eta^2}{2\eta+c}
\leq u.
\]
Direct substitution of the continuation rule gives
\begin{align*}
&
\frac{
    G_{\alpha_1^*,\alpha_2^*}(v,\uh)
    -
    G_{\alpha_1^*,\alpha_2^*}(v,u)
}{
    \alpha_2^*(0)
}
\\
&\qquad=
(\uh-u)
\left[
\frac{1-\alpha_2^*(0)}{\alpha_2^*(0)}
+
\frac{
    \eta^2
}{
    2(2\eta+c)(\uh-c-v)
}
\right]
\geq0.
\end{align*}
Thus, conditional on the lower reward \(v\), Nature cannot obtain more by
reducing the higher reward below \(\uh\).

\item Finally, consider
\[
    v\leq u<
    \uh-c-\frac{\eta^2}{2\eta+c}.
\]
Direct substitution gives
\begin{align*}
&
\frac{
    G_{\alpha_1^*,\alpha_2^*}(v,\uh)
    -
    G_{\alpha_1^*,\alpha_2^*}(v,u)
}{
    \alpha_2^*(0)
}
\\
&=
\frac{1-\alpha_2^*(0)}{\alpha_2^*(0)}(\uh-u)
\\
&\quad
+\frac12
\left[
\frac{
    c\eta^2
}{
    (2\eta+c)(\uh-c-u)
}
-c
+
\frac{
    \eta^2(\uh-u)
}{
    (2\eta+c)(\uh-c-v)
}
\right]
\\
&\geq
\frac{1-\alpha_2^*(0)}{\alpha_2^*(0)}(\uh-u)
\\
&\quad
+\frac12
\left[
\frac{
    c\eta^2
}{
    (2\eta+c)(\uh-c-u)
}
-c
+
\frac{
    \eta^2(\uh-u)
}{
    (2\eta+c)(\uh-c)
}
\right],
\end{align*}
where the inequality follows from \(v\geq0\). Viewed as a function of \(u\), the final expression is strictly convex.
Its unique stationary point is
\[
    u=\uh-c-\eta=\hat v,
\]
and \eqref{eq:two-general-identities} implies that its value at
\(u=\hat v\) is zero.  Consequently,
\[
    G_{\alpha_1^*,\alpha_2^*}(v,u)
    \leq
    G_{\alpha_1^*,\alpha_2^*}(v,\uh).
\]
Moreover,
\begin{align*}
G_{\alpha_1^*,\alpha_2^*}(v,\uh)
&=
\alpha_2^*(0)
\left(
    2c-
    \frac{
        c\eta^2
    }{
        (2\eta+c)(\uh-c)
    }
\right)=
R_2^*.
\end{align*}
\end{enumerate}

Thus no reward pair gives Nature more than the value in
\eqref{eq:two-general-explicit-value}.  Conversely, the coefficient
calculation in the proof of \Cref{prop:2-general} shows that Nature's
four-point distribution guarantees this value.  The two strategies are
therefore mutual best replies.

\subsection{Dynamic Reoptimization: Details}\label{sec:intrapersonal-eq}

This subsection provides the recursive characterization underlying \cref{thm:corr}. Consider first the case in which the DM has a single box remaining. This problem coincides with the one-box commitment problem and with the one-box independent-rewards case. Now suppose the DM has \(m\geq2\) boxes remaining. By homogeneity, and using the one-prize reduction, we can restrict attention to Nature assigning the prize uniformly across the remaining boxes with total probability \(P\in[0,1]\). Thus, with probability \(1-P\) there is no prize, and with probability \(P/m\) each remaining box is the unique prize box.

Let \(\tilde R_m(\alpha_m,P)\) denote the DM's continuation regret when she opens one of the \(m\) remaining boxes with probability \(\alpha_m\), conditional on not having found a high reward so far, and when future selves follow the equilibrium continuation strategy. If the DM opens a box and does not find the prize, the total conditional probability that a prize remains in the continuation problem is
\[
\pi_m(P):=\frac{(m-1)P}{m-P}.
\]
Hence, for \(m\geq2\),
\begin{eqnarray*}
\tilde R_m(\alpha_m,P)
&=&
(1-\alpha_m)P(\uh-c)
+
\alpha_m\left(1-\frac{P}{m}\right)
\left[
c+\tilde R_{m-1}(\alpha_{m-1}^*,\pi_m(P))
\right].
\end{eqnarray*}
In the active region characterized below, the continuation strategy \(\alpha_{m-1}^*\) equalizes regret across Nature's beliefs, so
\[
\tilde R_{m-1}(\alpha_{m-1}^*,P)=R_{m-1}^*
\qquad\text{for all }P\in[0,1].
\]
The recursion then reduces to a static minmax problem in \(\alpha_m\) and \(P\).

\begin{proposition}\label{prop:intrapersonal-eq}
Let \((c_1,\ldots,c_n)=(c,\ldots,c)\). There exists an intrapersonal equilibrium and an integer cutoff
\[
\bar{n}^{nc}\leq \frac{2\uh-c}{c}
\]
such that, conditional on not having opened a box with reward \(\uh\) and having \(m\) unopened boxes remaining, the DM opens a uniformly selected remaining box with probability \(\alpha_m^*\) and incurs continuation regret \(R_m^*\), where:
\begin{enumerate}
\item For all \(m\leq\bar{n}^{nc}\), \(\alpha_m^*\) and \(R_m^*\) are recursively defined by \(R_0^*=0\) and
\[
\alpha_m^*
=
\frac{m(\uh-c)}
{m(\uh-c)+c+R_{m-1}^*},
\qquad
R_m^*
=
\frac{m(\uh-c)(c+R_{m-1}^*)}
{m(\uh-c)+c+R_{m-1}^*}.
\]
In particular,
\[
\alpha_1^*=\frac{\uh-c}{\uh},
\qquad
R_1^*=\frac{(\uh-c)c}{\uh}.
\]
\item For all \(m>\bar{n}^{nc}\),
\[
\alpha_m^*=0,
\qquad
R_m^*=\uh-c.
\]
\end{enumerate}
\end{proposition}

\begin{proof}
For \(m=1\), the problem is the one-box problem:
\[
\tilde R_1(\alpha_1,P)
=
(1-\alpha_1)P(\uh-c)+\alpha_1(1-P)c.
\]
The minmax solution equalizes the regret from not opening the prize box and from opening an empty box, yielding
\[
\alpha_1^*=\frac{\uh-c}{\uh},
\qquad
R_1^*=\frac{(\uh-c)c}{\uh}.
\]

Now consider \(m\geq2\). Suppose that the \((m-1)\)-box continuation strategy has already been constructed and that, in the active region,
\[
\tilde R_{m-1}(\alpha_{m-1}^*,P)=R_{m-1}^*
\qquad
\text{for all }P\in[0,1].
\]
Then the \(m\)-box regret can be written as
\[
\tilde R_m(\alpha_m,P)
=
(1-\alpha_m)P(\uh-c)
+
\alpha_m\left(1-\frac{P}{m}\right)
\bigl(c+R_{m-1}^*\bigr).
\]
For a fixed \(\alpha_m\), this expression is linear in \(P\). The coefficient of \(P\) vanishes if
\[
(1-\alpha_m)(\uh-c)
=
\frac{\alpha_m}{m}\bigl(c+R_{m-1}^*\bigr),
\]
or equivalently
\[
\alpha_m
=
\frac{m(\uh-c)}{m(\uh-c)+c+R_{m-1}^*}.
\]
At this value, \(\tilde R_m(\alpha_m,P)\) is independent of \(P\), and the associated regret is
\[
R_m^*
=
\alpha_m^*\bigl(c+R_{m-1}^*\bigr)
=
\frac{m(\uh-c)(c+R_{m-1}^*)}
{m(\uh-c)+c+R_{m-1}^*}.
\]

Turning to the DM's optimality condition, for a fixed \(P\) the current self is indifferent between opening a box and opting out if
\[
P(\uh-c)
=
\left(1-\frac{P}{m}\right)
\bigl(c+R_{m-1}^*\bigr).
\]
Solving for \(P\) gives Nature's candidate saddle-point probability:
\[
P_m^*
=
\frac{m\bigl(c+R_{m-1}^*\bigr)}
{m(\uh-c)+c+R_{m-1}^*}.
\]
This probability is admissible if and only if \(P_m^*\leq1\), or
\begin{equation}\label{eq:bar-N}
c+R_{m-1}^*
\leq
\frac{m}{m-1}(\uh-c).
\end{equation}
Whenever \eqref{eq:bar-N} holds, the pair \((\alpha_m^*,P_m^*)\) is a saddle point of the \(m\)-box continuation problem, and the regret \(R_m^*\) is independent of \(P\). This proves the recursive characterization in the active region.

Let \(\bar{n}^{nc}\) be the largest integer for which the active construction is admissible. The bound
\[
\bar{n}^{nc}\leq \frac{2\uh-c}{c}
\]
follows from comparison with the homogeneous commitment problem. If \(m>(2\uh-c)/c\), then
\[
\uh<\frac{m+1}{2}c,
\]
so even under commitment the homogeneous binary-support problem is not globally profitable. Since lack of commitment cannot make the continuation value of search larger than under commitment, the no-commitment active construction cannot be admissible for such \(m\).

It remains to verify the opt-out region. Consider first \(m=\bar{n}^{nc}+1\). By definition of \(\bar{n}^{nc}\), condition \eqref{eq:bar-N} fails:
\[
c+R_{\bar{n}^{nc}}^*
>
\frac{\bar{n}^{nc}+1}{\bar{n}^{nc}}(\uh-c).
\]
Take Nature's belief to be \(P=1\). If the DM opts out, her regret is \(\uh-c\). If she opens a box, then with probability \(1-1/(\bar{n}^{nc}+1)\) she does not find the prize and continues with \(\bar{n}^{nc}\) boxes remaining. Hence opening is weakly worse than opting out exactly when
\[
\uh-c
\leq
\left(1-\frac{1}{\bar{n}^{nc}+1}\right)
\bigl(c+R_{\bar{n}^{nc}}^*\bigr),
\]
which is equivalent to the failure of \eqref{eq:bar-N} at \(m=\bar{n}^{nc}+1\). Thus \(\alpha_{\bar{n}^{nc}+1}^*=0\) is optimal against \(P=1\), and \(P=1\) is optimal for Nature against \(\alpha_{\bar{n}^{nc}+1}^*=0\). The resulting regret is \(R_{\bar{n}^{nc}+1}^*=\uh-c\).

For any \(m>\bar{n}^{nc}+1\), future selves also opt out after an unsuccessful search, so \(R_{m-1}^*=\uh-c\). Against \(P=1\), opening a box yields regret
\[
\left(1-\frac{1}{m}\right)\bigl(c+\uh-c\bigr)
=
\left(1-\frac{1}{m}\right)\uh.
\]
Opting out yields regret \(\uh-c\). Since the failure of \eqref{eq:bar-N} at \(\bar{n}^{nc}+1\), together with \(R_{\bar{n}^{nc}}^*\leq\uh-c\), implies \(\bar{n}^{nc}+1>\uh/c\), we have \(m>\uh/c\) for all \(m>\bar{n}^{nc}+1\). Therefore
\[
\uh-c
\leq
\left(1-\frac{1}{m}\right)\uh,
\]
so opting out remains optimal. This completes the construction of an intrapersonal equilibrium.
\end{proof}

\subsection{Independent and Identically Distributed Rewards: Details}

\subsubsection{Proof of \cref{thm:ind}}\label{app:proof-ind}

\begin{proof}
For a vector of continuation probabilities \(\bm \alpha_k=(\alpha_1,\ldots,\alpha_k)\), let
\(\mathcal R_k(\bm \alpha_k,p)\) denote the DM's regret when there are \(k\) unopened boxes and rewards are i.i.d. with success probability \(p\). We set \(\mathcal R_0(\emptyset,p)=0\). The recursion corresponding to \eqref{eq:indep-commitment} is
\begin{eqnarray}\label{eq:ind-proof-recursion}
\mathcal R_k((\bm \alpha_{k-1},\alpha_k),p)
=
(1-\alpha_k)\bigl(1-(1-p)^k\bigr)(\uh-c)+
\alpha_k(1-p)\bigl(c+\mathcal R_{k-1}(\bm \alpha_{k-1},p)\bigr).
\end{eqnarray}
With
\[
\alpha_k^*
=
\frac{k(\uh-c)^k}{(k-1)(\uh-c)^k+\uh^k},
\]
we prove by induction that, for every \(k\leq n\), $\bm\alpha^*_k=(\alpha_1^*,...,\alpha_k^*)$, together with \(\hat p=c/\uh\), forms a saddle point of the \(k\)-box problem, and that the associated regret is
\[
R_k^*
=
\left(1-\left(\frac{\uh-c}{\uh}\right)^k\right)(\uh-c).
\]
For \(k=1\), the problem is the one-box problem:
\[
\mathcal R_1(\alpha_1,p)
=
(1-\alpha_1)p(\uh-c)+\alpha_1(1-p)c.
\]
The saddle point equalizes the regret from not opening a prize box and from opening an empty box:
\[
\alpha_1^*=\frac{(\uh-c)}{(\uh-c)+c}=\frac{\uh-c}{\uh},
\qquad
\hat p=\frac{c}{\uh},
\qquad
R_1^*=\frac{(\uh-c)c}{\uh}.
\]
This coincides with the closed-form expressions above. Now suppose the claim holds for \(k-1\). Thus,
\[
\mathcal R_{k-1}(\bm \alpha_{k-1}^*,p)\leq R_{k-1}^*
\quad\text{for all }p\in[0,1],
\]
with equality at \(p=\hat p\), and
\[
\mathcal R_{k-1}(\bm \alpha_{k-1},\hat p)\geq R_{k-1}^*
\quad\text{for all }\bm \alpha_{k-1}.
\]
Given the equilibrium continuation strategy \(\bm \alpha_{k-1}^*\), define the auxiliary upper bound
\begin{eqnarray*}
\widehat{\mathcal R}_k(\alpha_k,p)
&:=&
(1-\alpha_k)\bigl(1-(1-p)^k\bigr)(\uh-c)
+
\alpha_k(1-p)\bigl(c+R_{k-1}^*\bigr).
\end{eqnarray*}
By the induction hypothesis,
\[
\mathcal R_k((\bm \alpha_{k-1}^*,\alpha_k),p)
\leq
\widehat{\mathcal R}_k(\alpha_k,p),
\]
with equality at \(p=\hat p\). We next choose \(\alpha_k\) so that \(\widehat{\mathcal R}_k(\alpha_k,p)\) is maximized at \(p=\hat p\). Differentiating with respect to \(p\) gives
\[
\frac{\partial \widehat{\mathcal R}_k(\alpha_k,p)}{\partial p}
=
(1-\alpha_k)k(1-p)^{k-1}(\uh-c)
-
\alpha_k\bigl(c+R_{k-1}^*\bigr).
\]
This derivative is decreasing in \(p\). Hence \(\hat p\) is a maximizer if the derivative vanishes at \(p=\hat p\), i.e.,
\[
(1-\alpha_k)k\left(\frac{\uh-c}{\uh}\right)^{k-1}(\uh-c)
=
\alpha_k\bigl(c+R_{k-1}^*\bigr).
\]
Substituting for \(R_{k-1}^*\), we have
\[
c+R_{k-1}^*
=
c+\left(1-\left(\frac{\uh-c}{\uh}\right)^{k-1}\right)(\uh-c)
=
\uh\left(1-\left(\frac{\uh-c}{\uh}\right)^k\right).
\]
Solving the indifference condition therefore yields
\[
\alpha_k^*
=
\frac{k(\uh-c)^k}{(k-1)(\uh-c)^k+\uh^k}.
\]
At \(p=\hat p\), the current self is indifferent between opening and stopping because
\begin{eqnarray*}
(1-\hat{p})(c+R^*_{k-1})&=&\left(\frac{\uh-c}{\uh}\right)\bigl(c+R_{k-1}^*\bigr)\\
&=&
\left(1-\left(\frac{\uh-c}{\uh}\right)^k\right)(\uh-c)\\
&=&\left(1-(1-\hat{p})^k\right)(\uh-c).
\end{eqnarray*}
Thus, for every \(\alpha_k\),
\[
\widehat{\mathcal R}_k(\alpha_k,\hat p)
=
\left(1-\left(\frac{\uh-c}{\uh}\right)^k\right)(\uh-c).
\]
In particular,
\[
\mathcal R_k(\bm \alpha_k^*,\hat p)
=
\widehat{\mathcal R}_k(\alpha_k^*,\hat p)
=
\left(1-\left(\frac{\uh-c}{\uh}\right)^k\right)(\uh-c)
=:R_k^*.
\]
We now verify the saddle-point inequalities. First, for any \(p\),
\[
\mathcal R_k(\bm \alpha_k^*,p)
\leq
\widehat{\mathcal R}_k(\alpha_k^*,p)
\leq
\widehat{\mathcal R}_k(\alpha_k^*,\hat p)
=
R_k^*,
\]
where the first inequality follows from the induction hypothesis and the second from the fact that \(\widehat{\mathcal R}_k(\alpha_k^*,p)\) is maximized at \(\hat p\). Hence Nature's best response to \(\bm \alpha_k^*\) is \(\hat p\).

Second, for any alternative commitment strategy \(\bm \alpha_k=(\alpha_1,\ldots,\alpha_k)\), the induction hypothesis implies
\[
\mathcal R_{k-1}(\bm \alpha_{k-1},\hat p)\geq R_{k-1}^*.
\]
Using \eqref{eq:ind-proof-recursion}, we obtain
\begin{eqnarray*}
\mathcal R_k(\bm \alpha_k,\hat p)
&\geq&
(1-\alpha_k)\left(1-\left(1-\hat{p}\right)^k\right)(\uh-c)
+
\alpha_k\left(1-\hat{p}\right)\bigl(c+R_{k-1}^*\bigr)=
R_k^*.
\end{eqnarray*}
Thus \(\bm \alpha_k^*\) is a best response to \(\hat p\). Combining the two inequalities,
\[
\mathcal R_k(\bm \alpha_k,\hat p)
\geq
\mathcal R_k(\bm \alpha_k^*,\hat p)
\geq
\mathcal R_k(\bm \alpha_k^*,p)
\qquad
\text{for all }\bm \alpha_k\text{ and }p.
\]
Therefore \((\bm \alpha_k^*,\hat p)\) is a saddle point of the \(k\)-box commitment problem.

The same induction also proves the intrapersonal claim. At any continuation problem with \(k\) boxes remaining, future selves use \(\bm \alpha_{k-1}^*\). The preceding argument shows that the current self's minmax problem has a saddle point \((\alpha_k^*,\hat p)\). Hence the strategy vector \((\alpha_1^*,\ldots,\alpha_n^*)\) is an intrapersonal equilibrium.

Finally, the intrapersonal equilibrium is unique in terms of search probabilities. By induction, \(\mathcal R_{k-1}(\bm \alpha_{k-1}^*,p)\) is differentiable at \(p=\hat p\) and has derivative zero there. Therefore, in the \(k\)-box continuation problem, the derivative of
\(\mathcal R_k((\bm \alpha_{k-1}^*,\alpha_k),p)\) at \(p=\hat p\) is zero if and only if \(\alpha_k=\alpha_k^*\). If \(\alpha_k\neq \alpha_k^*\), Nature can perturb \(p\) locally around \(\hat p\) and raise regret above \(R_k^*\). Thus no other continuation probability can be minmax optimal.
\end{proof}

\subsubsection{Proof of \cref{co:ind}}\label{app:proof-co-ind}

\begin{proof}
By \cref{thm:ind},
\[
\alpha_n^*
=
\frac{n(\uh-c)^n}{(n-1)(\uh-c)^n+\uh^n},
\qquad
R_n^*
=
(1-\left(\frac{\uh-c}{\uh}\right)^n)(\uh-c).
\]
The limits follow immediately:
\[
\lim_{n\to\infty}\alpha_n^*=0,
\qquad
\lim_{n\to\infty}R_n^*=\uh-c.
\]
Moreover, \(R_n^*\) is strictly increasing in \(n\), since \(\left(\frac{\uh-c}{\uh}\right)^n\) is strictly decreasing.

It remains to show that \(\alpha_n^*\) is strictly decreasing in \(n\). Treating \(n\) as a real variable, differentiating the closed-form expression for $\alpha_n^*$ gives
\[
\frac{\partial \alpha_n^*}{\partial n}
=
\frac{
(\uh-c)^n\uh^n
\left[
1-\left(\frac{\uh-c}{\uh}\right)^n
+
\ln\left(\left(\frac{\uh-c}{\uh}\right)^n\right)
\right]
}
{
\left((n-1)(\uh-c)^n+\uh^n\right)^2
}.
\]
Let \(x=\left(\frac{\uh-c}{\uh}\right)^n\in(0,1)\). Since
\[
1-x+\ln x<0
\qquad\text{for all }x\in(0,1),
\]
we have \(\partial \alpha_n^*/\partial n<0\). Hence \(\alpha_n^*\) is strictly decreasing in \(n\).
\end{proof}

\end{document}